\documentclass{svjour3}

\usepackage{graphicx}
\usepackage{amsmath,epsfig,rotating,amssymb}

\begin{document}

\newcommand{\msol}{\hbox{\,${\rm M}_\odot$}}
\newcommand{\Msun}{\mbox{M$_{\scriptsize \odot}$}}
\newcommand{\Lsun}{\mbox{L$_{\scriptsize \odot}$}} 
\newcommand{\lsol}{\hbox{\,${\rm L}_\odot$}}
\newcommand{\etal}{{\it{et al.~}}}
\newcommand{\pdf}{{\it{pdf}}}
\newcommand{\ie}{{\it{i.e.~}}}
\newcommand{\eg}{{\it{e.g.~}}}
\newcommand{\ang}{\mbox{\AA}}
\newcommand{\degrees}{$^{\circ}$}
\newcommand{\HI}{\mbox{H\,{\sc i}}}
\newcommand{\HII}{\mbox{H\,{\sc ii}}}
\newcommand{\OI}{\mbox{[O\,{\sc i}]}}
\newcommand{\NeII}{\mbox{Ne\,{\sc ii}}}
\newcommand{\FeII}{\mbox{Fe\,{\sc ii}}}
\newcommand{\NeIII}{\mbox{Ne\,{\sc iii}}}
\newcommand{\ArII}{\mbox{Ar\,{\sc ii}}}
\newcommand{\ArIII}{\mbox{Ar\,{\sc iii}}}   
\newcommand{\OIII}{\mbox{O\,{\sc iii}}}
\newcommand{\SII}{\mbox{S\,{\sc ii}}}
\newcommand{\SIII}{\mbox{S\,{\sc iii}}}
\newcommand{\SIV}{\mbox{S\,{\sc iv}}}
\newcommand{\CII}{\mbox{[C\,{\sc ii}]}}
\newcommand{\CI}{\mbox{[C\,{\sc i}]}}
\newcommand{\NII}{\mbox{N\,{\sc ii}}}   
\def\pc{\hbox{\,pc}}
\def\mum{\hbox{\,$\mu$m}}
\def\mumd{\hbox{\,$\mu$m$^{-2}$}}
\def\cm{\hbox{\,cm}}
\def\s{\hbox{\,s}}
\def\h{\hbox{\,h}}
\def\sec{\hbox{\,sec}}
\def\min{\hbox {\,min}}
\def\smu{\hbox{\,s$^{-1}$}}
\def\smd{\hbox{\,s$^{-2}$}}
\def\yr{\hbox{\,yr}}
\def\yrmu{\hbox{\,yr$^{-1}$}}
\def\My{\hbox{\,My}}
\def\Mymu{\hbox{\,My$^{-1}$}}
\def\K{\hbox{\,K}}
\def\pcmu{\hbox{\,pc$^{-1}$}}
\def\pcmd{\hbox{\,pc$^{-2}$}}   
\def\kms{\hbox{\,km\,s$^{-1}$}}
\def\kpc{\hbox{\,kpc}}
\def\cms{\hbox{\,cm\,s$^{-1}$}}
\def\erg{\hbox{\,erg}}
\def\cmpd{\hbox{\,cm$^2$}}
\def\fluxi{\hbox{\,erg\,\,s$^{-1}$\,cm$^{-2}$}}
\def\flux{\hbox{\,erg\,\,s$^{-1}$\,cm$^{-2}$\,sr$^{-1}$}}
\def\fluxpm{\hbox{\,erg\,\,s$^{-1}$\,cm$^{-2}$\,\mum$^{-1}$\,sr$^{-1}$}}
\def\mjy{\hbox{\,mJy}}
\def\megjysr{\hbox{\,MJy/sr}}
\def\jy{\hbox{\,Jy}}
\def\ghz{\hbox{\,GHz}}
\def \aa{A\&A}
\def\cc{\ifmmode{\,{\rm cm}^{-3}}\else{$\,{\rm cm}^{-3}$}\fi}
\def\cq{\ifmmode{\,{\rm cm}^{-2}}\else{$\,{\rm cm}^{-2}$}\fi}
\def\mic{\ifmmode{\,\mu{\rm m}}\else{$\mu$m}\fi}
\def\eccs{\ifmmode{\,{\rm erg}\,{\rm cm}^{-3} {\rm s}^{-1}}\else{$\,{\rm
erg}\,{\rm cm}^{-3} {\rm s}^{-1}$}\fi}
\def\ecqs{\ifmmode{\,{\rm erg}\,{\rm cm}^{-2}\,{\rm s}^{-1}\,{\rm
sr}^{-1}}\else{$\,{\rm erg}\,{\rm cm}^{-2}\,{\rm s}^{-1}\,{\rm sr}^{-1}$}\fi}
\def\deg{\ifmmode{^{\circ}}\else{$^{\circ}$}\fi} 
\def\pc{\ifmmode{\,{\rm pc}}\else{$\,{\rm pc}$}\fi} 
\def\kms{\ifmmode{\,{\rm km}\,{\rm s}^{-1}}\else{km s$^{-1}$}\fi} 
\def\kmspc{\ifmmode{\,{\rm km}\,{\rm s}^{-1}\,{\rm pc}^{-1}}\else{kms$^{-1}$ pc$^{-1}$}\fi} 
\def\MJysr{\ifmmode{\,{\rm MJy\,sr}^{-1}}\else{$\,{\rm MJy\,sr}^{-1}$}\fi} 
\def\Kkms{\ifmmode{\,{\rm K\,km\,s}^{-1}}\else{$\,{\rm K\,km\,s}^{-1}$}\fi} 
\def\CO{\ifmmode{\rm CO}\else{$\rm CO$}\fi} 
\def\twCO{\ifmmode{\rm ^{12}CO}\else{$\rm^{12}CO$}\fi} 
\def\thCO{\ifmmode{\rm ^{13}CO}\else{$\rm^{13}CO$}\fi} 
\def \Cp{\ifmmode{\rm C^+}\else{$\rm C^+$}\fi} 
\def \CHp{\ifmmode{\rm CH^+}\else{$\rm CH^+$}\fi}
\def \SHp{\ifmmode{\rm SH^+}\else{$\rm SH^+$}\fi}
\def \thCHp{\ifmmode{\rm ^{13}CH^+}\else{$\rm ^{13}CH^+$}\fi}
\def \twCHp{\ifmmode{\rm ^{12}CH^+}\else{$\rm ^{12}CH^+$}\fi}
\def \CHtp{\ifmmode{\rm CH_2^+}\else{$\rm CH_2^+$}\fi} 
\def\CHthp{\ifmmode{\rm CH_3^+}\else{$\rm CH_3^+$}\fi} 
\def\Hthp{\ifmmode{\rm H_3^+}\else{$\rm H_3^+$}\fi} 
\def \HCOp{\ifmmode{\rm HCO^+}\else{$\rm HCO^+$}\fi} 
\def \HthOp{\ifmmode{\rm H_3O^+}\else{$\rm H_3O^+$}\fi}
\def \OHp{\ifmmode{\rm OH^+}\else{$\rm OH^+$}\fi}  
\def \HtwOp{\ifmmode{\rm H_2O^+}\else{$\rm H_2O^+$}\fi} 
\def \HCfiN{\ifmmode{\rm HC_5N}\else{$\rm HC_5N$}\fi} 
\def\wat{\ifmmode{\rm H_2O}\else{$\rm H_2O$}\fi} 
\def\amm{\ifmmode{\rm NH_3}\else{$\rm NH_3$}\fi} 
\def \oxy{\ifmmode{\rm O_2}\else{$\rm O_2$}\fi} 
\def \HH{\ifmmode{\rm H_2}\else{$\rm H_2$}\fi}
\def \Jone{\ifmmode{\rm {(J=1--0)}}\else{{(J=1--0)}}\fi} 
\def\Jtwo{\ifmmode{\rm {(J=2--1)}}\else{{(J=2--1)}}\fi} 
\def\Jthr{\ifmmode{\rm {(J=3--2)}}\else{{(J=3--2)}}\fi} 
\def\Jfou{\ifmmode{\rm {(J=4--3)}}\else{{(J=4--3)}}\fi} 
\def\Jfo{\ifmmode{\rm {J=4--3}}\else{{J=4--3}}\fi} 
\def \Jon{\ifmmode{\rm {J=1--0}}\else{{J=1--0}}\fi} 
\def \Jtw{\ifmmode{\rm {J=2--1}}\else{{J=2--1}}\fi} 
\def \Jth{\ifmmode{\rm {J=3--2}}\else{{J=3--2}}\fi} 
\def\Jfi{\ifmmode{\rm {J=4--3}}\else{{J=4--3}}\fi} 
\def \Ta{\ifmmode{\rm T_A}\else{$\rm T_A$}\fi} 
\def \Tas{\ifmmode{\rm T_A^*}\else{$\rm T_A^*$}\fi} 
\def \Tmb{\ifmmode{\rm T_{mb}}\else{$\rm T_{mb}$}\fi} 
\def \Tr{\ifmmode{\rm T_r}\else{$\rm T_r$}\fi} 
\def \Trs{\ifmmode{\rm T_r^*}\else{$\rm T_r^*$}\fi}
\def \Icent{\ifmmode{I(100 \mu{\rm m})}\else{$I(100 \mu{\rm m})$}\fi}

\title{Turbulent molecular clouds}

\author{
 Patrick Hennebelle \and Edith Falgarone  }
\institute{Patrick Hennebelle \at  Laboratoire AIM, 
Paris-Saclay, CEA/IRFU/SAp - CNRS - Universit\'e Paris Diderot, 91191, 
Gif-sur-Yvette Cedex, France\\
\email{patrick.hennebelle@cea.fr}
\and
Edith Falgarone \at 
LERMA, CNRS, Ecole Normale Sup\'erieure \& Observatoire de Paris, 
24 rue Lhomond, 75231 Paris Cedex, France\\
\email{edith.falgarone@ens.fr}
}

\date{Received: September 15, 2012 /Accepted : October 10, 2012} 

\maketitle

\begin{abstract} Stars form within molecular clouds but our understanding of
  this fundamental process remains hampered by the complexity of the
  physics that drives their evolution.  We review our observational
  and theoretical knowledge of molecular clouds trying to confront the
  two approaches wherever possible.  After a broad presentation of the
  cold interstellar medium and molecular clouds, we emphasize the
  dynamical processes with special focus to turbulence and its impact
  on cloud evolution.  We then review our knowledge of the velocity,
  density and magnetic fields. We end by openings towards new
  chemistry models and the links between molecular cloud structure and 
  star--formation rates.   
\keywords{  Instabilities  \and  Interstellar  medium:
kinematics and dynamics -- structure -- clouds \and Star: formation} 
\end{abstract}

\section{Introduction: bridging theory and observations}

In the Galaxy and all spiral galaxies, molecular clouds are the sites
of star--formation. The roots of star--formation are therefore to be
sought in the physics of these clouds.  We review our current
understanding of the physics of molecular clouds both from a
theoretical and observational point of view, trying to systematically
confront the two approaches.  Such a confrontation turns out to be
more difficult a task than anticipated because on the observational
and theoretical sides, the concept of molecular clouds relies on
different foundations: in the former, molecular clouds are by
definition structures detected in molecular line emission, mostly that
of CO because \HH\ is not easily accessible to observations; in the
latter, molecular clouds are structures denser and more shielded from
the UV field than atomic and diffuse gas.  In this review, we will see
that molecular gas is not necessarily dense and shielded from UV
photons, that gas with only traces of \HH\ can harbor large amounts of
molecules other than \HH, and that CO, the traditional and powerful
tracer of molecular gas in the universe, is indeed a complex tracer.
Another difficulty on bridging observations and theory stems from the
projections of the observations (line--of--sight velocities, densities
providing only column densities), not to mention those involved in the
observations of magnetic fields.  These projections make local
quantities hard to derive, the inversion of the observables relying on
assumptions (homogeneity, geometry) and theoretical knowledge
(collisional excitation cross--sections of molecules, interstellar
chemistry, line radiative transfer, ...).

Observations of molecular clouds in the Galaxy now cover an impressive
range of size scales, from a few 100 pc to the milli--pc scale.  
The less controversial facts are: {\it (i)} scaling laws are
found to relate the masses, sizes and internal velocity dispersions of
clouds across this range of scales, hence the physics of molecular
clouds involves all of them; {\it (ii)} molecular clouds, as traced by
CO, include self--gravitating and non self--gravitating structures; {\it
  (iii)} to be understood, the interplay of gravity and turbulence has
to be put on even larger size scales (galactic dynamics, galaxy
interactions, ...) as shown by new observational capabilities
(Swinbank et al. 2010, Herrera et al. 2012).

Significant theoretical progress has been accomplished during the
last decades. This has been largely, though not exclusively, permitted
by the development of numerical simulations which, thanks to the
increase of computing power but also to the use of new numerical
schemes (such as adaptive mesh refinement) can now cover a broad range
of scales.  In parallel, much effort has been made to treat the
important physical processes involving magnetic fields, radiative
transfer and chemistry, which are now commonly incorporated
into numerical simulations.

In spite of these achievements, we are still far from a proper
modeling of the molecular clouds.  This is because the dynamics of
scale than can be tackled in the simulations remains far too
restricted. As an example, while the effective Reynolds number (see
definition in Section 3.1) of the numerical simulations is certainly
not larger than $\simeq 10^4$, its value within molecular clouds is
larger than $10^6--10^7$. Moreover, molecular clouds, as we will argue
later on, are not isolated entities, but are dynamically linked to
their environment. This makes the limitation of the range of scales
treated in simulations even more severe, since ideally one should
treat not only the molecular clouds but also the diffuse interstellar
medium in which they form and are embedded.

While the broad range of scales is certainly a severe issue, the
complexity of the physical processes is the other outstanding
difficulty that interstellar medium (ISM) studies are facing.  It is
well known, for instance, that in the ISM, thermal, kinetic, magnetic,
radiative and cosmic ray energies are all on the order of $\simeq$1 eV
cm$^{-3}$ (within one order of magnitude).  This energy equipartition
suggests that these processes are coupled to each other.  Adding
self-gravity for the densest parts of the interstellar gas, this makes
six {\it fields of knowledge}, which are interacting in the cloud
evolution and the star--formation process.

Given this complexity, it is not a surprise that outstanding resilient
questions remain, among which:\\ 
-- Are molecular clouds well described
by turbulent density fluctuations (\eg\ Ballesteros--Paredes et
al. 1999)? \\ 
-- Is the observed hierarchy the outcome of gravitational
fragmentation or of large scale dynamics imposed by the galactic
environment (\eg\ Chi\`eze 1987, Koyama \& Ostriker 2009)?\\ 
-- Is gravity feeding turbulence throughout the hierarchy (\eg\ Field et
al. 2008)?\\ 
-- What is the nature of the non--thermal pressure? How
does it act (\eg\ Bonazzola et al. 1987)? \\ 
-- What is the evolutionary path between atomic gas and giant molecular clouds
(\eg\  Koda et al. 2009)? \\ 
-- Are molecules a prerequisite to star
formation (\eg\ Glover \& Clark 2012)?\\ 
-- What is the role played by stellar feedback 
in the evolution of molecular clouds (\eg\ Blitz \& Shu
1980)? \\

A complete and detailed overview of the subject is beyond the scope of
these pages and we had to make choices. We put special emphasis on the
dynamic processes and present the different theories directly used to
interpret relevant observations. When possible, we try to give some
hints of how the theoretical relations are inferred and their physical
significance.  Numerical simulations are discussed, particularly when
their results can be confronted to theory.  However, given the large
number of relevant works in this area, our review is by no means
exhaustive. The same applies to the observations, more attention have
been given to observations from which statistics are inferred and then
confronted to theories.  We also stress that, while the primary focus
of the present review is not star--formation, we will address closely
related topics, such as the origin of the core mass function and the
derivation of the star--formation rate, because they are integral parts
of molecular cloud physics. Other comprehensive references on closely
related topics can be found in Mac Low \& Klessen (2004), Elmegreen \&
Scalo (2004), Scalo \& Elmegreen (2004), McKee \& Ostriker (2007) and 
Kennicutt \& Evans (2012).

The review is divided in seven sections.  Section 2 provides a broad
overview of our observational knowledge of molecular clouds in the
Galaxy and presents a plausible formation scenario. Section 3 is
dedicated to the turbulent velocity field and its impact on the cloud
physics. The density structure and mass distribution are discussed in
Section 4 with subsections devoted to filaments and dense
cores. Section 5 focuses on magnetic fields.  Section 6 presents
openings and stress the key role played by turbulence.  In the last
section (section 7) conjectures and questions left open are discussed.

\section{Overview: the broad concept of molecular clouds}

\subsection{Traditional tracers of molecular clouds}
Molecular clouds have been discovered in the 1970s via their emission
in the J=1--0 transition of CO at $\nu=115$GHz (Wilson et al. 1970,
Penzias et al. 1972).  The beauty of this discovery lies in the fact
that two lines (those of the two main isotopologues $^{12}$CO and
$^{13}$CO) and the knowledge of the CO spectroscopy were sufficient to
infer that the line emitting gas in dark clouds (\ie\ those seen as
dark islands on visible maps of the sky) was cold ($T_K<10$K) and
dense ($n>10^3$\cc) and that the CO abundance relative to \HH\ was
low, $[\rm CO]/[\HH] <10^{-4}$. This was a powerful discovery: a trace
molecule was found to be a strong line emitter, opening the
possibility of surveys.  The first CO surveys of the Galactic plane
(Solomon et al. 1987, Dame et al. 1987) brought to light the existence
of CO-bright and massive entities, called the giant molecular clouds
(GMC), most of them harboring star forming regions.

More sensitive and fully sampled surveys (Dame et al. 2001) revealed a
weaker, so called ``cold'' CO emission bridging the CO-bright islands
of the GMCs in space and velocity, covering the inter--arm regions and
extending beyond the galactic plane.  A faint thick molecular disk was
discovered, with a scale height ($\sim$ 250 pc) comparable to that of
the HI layer in the inner Galaxy (Dame \& Thaddeus 1994).
Moreover, the mass comprised in the ``cold'' CO component was
estimated to be larger than that comprised in all the GMCs, a result
that in the 1980s raised a long and vivid controversy.  At the same
epoch, HI superclouds were found as very large scale structures in the
Galaxy (several 100 pc), each comprising several GMCs (Elmegreen \&
Elmegreen 1987) setting the large scale environment of what was then
understood as molecular clouds.  Since then, several $^{12}$CO(1--0)
(May et al. 1988, 1993, Sanders et al. 1986, Clemens et al. 1986) and
\thCO(1--0) (Jackson et al. 2006) surveys have further broadened the
view of molecular gas and clouds throughout the Galaxy, beyond the
neighborhood of star--forming regions.

Individual GMCs (size $\sim 50$ pc) are now resolved in nearby
external galaxies (\eg\ Gratier et al. 2010, Egusa et al. 2011, Koda
et al. 2009). Recent results obtained in \twCO(1--0) in the
grand--design spiral galaxy M51 are shown in Fig. 1. At low resolution
(4'' or $\sim 200$ pc), GMC associations follow the spiral arms and at
higher resolution (0.7'' or 30 pc) GMCs are resolved structures: some
are also found between the arms (Egusa et al. 2011). Interestingly, 
the most massive GMCs are located downstream of the spiral arms, in
close association with star forming regions, suggesting a time
evolution towards star--formation across the arms.  Koda et al. (2009)
find that GMCs and their \HH\ molecules are not dissociated upon
leaving the spiral arms, which supports a long lifetime for molecular
clouds and molecular gas in spiral galaxies.  For comparison, the
\twCO(1--0) column density map (resolution 0.03 pc) of the nearby
Taurus--Auriga complex, over the same scale of 30 pc, is shown in
Fig. 1 (Goldsmith et al. 2008): it reveals the complex structure of
the gas seen in \twCO(1--0), at all scales. Unexpectedly, this
complexity extends down to milliparsec scales (Falgarone et al. 1991,
2009).

Another observational technique has proven very powerful at revealing
massive clouds in the central regions of the Galaxy: shadows on the
bright mid--IR and near--IR stellar (and small particle emission)
backgrounds of the inner Galaxy. The infrared dark clouds (IRDCs)
(P\'erault et al. 1996, Egan et al. 1998, Hennebelle et al. 2001)
appear as massive entities of large column density ($N_{\rm H}$ up to
$10^{23}$ \cq) immersed in large clouds of lower extinction containing
10 times more mass (Kanulainen et al. 2011). They harbor small but
massive cores (Rathborne et al. 2010, Peretto \& Fuller 2010) that are
occasionally associated with massive young stellar objects (YSOs). In
spite of their large mass fractions at high column density, IRDCs, on
average, have a low star--forming activity.

\begin{figure}
  \includegraphics[width=0.27\textwidth,angle=0]{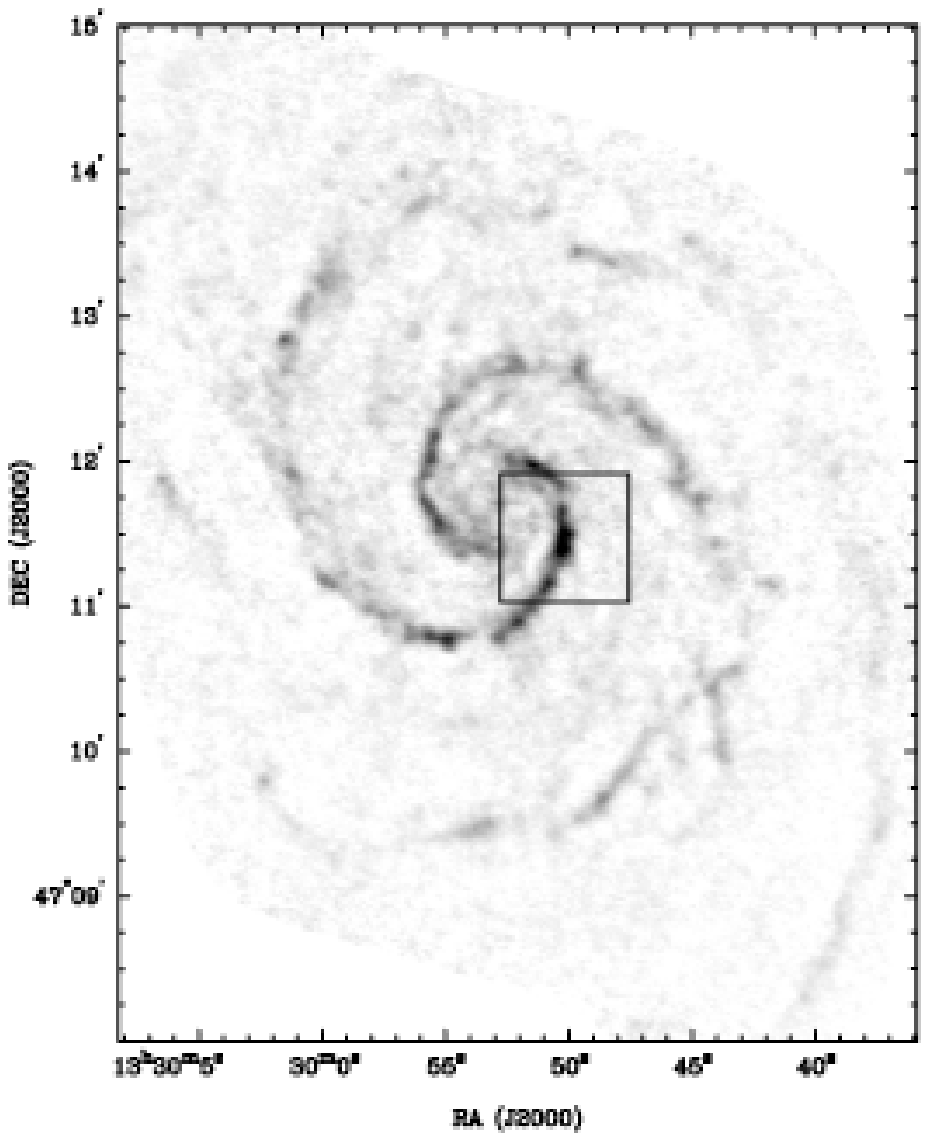}
  \includegraphics[width=0.3\textwidth,angle=0]{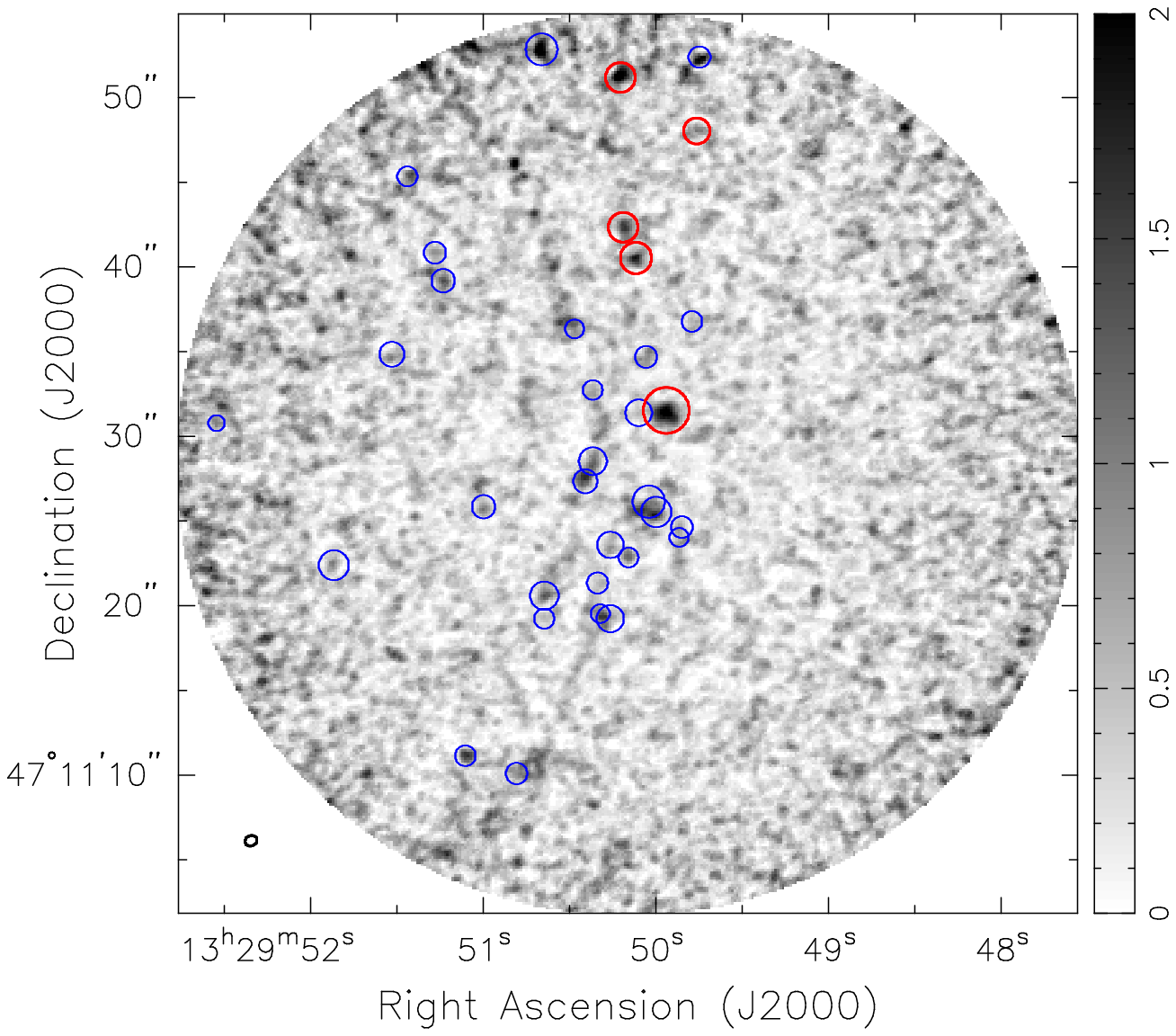}
  \includegraphics[width=0.35\textwidth,angle=0]{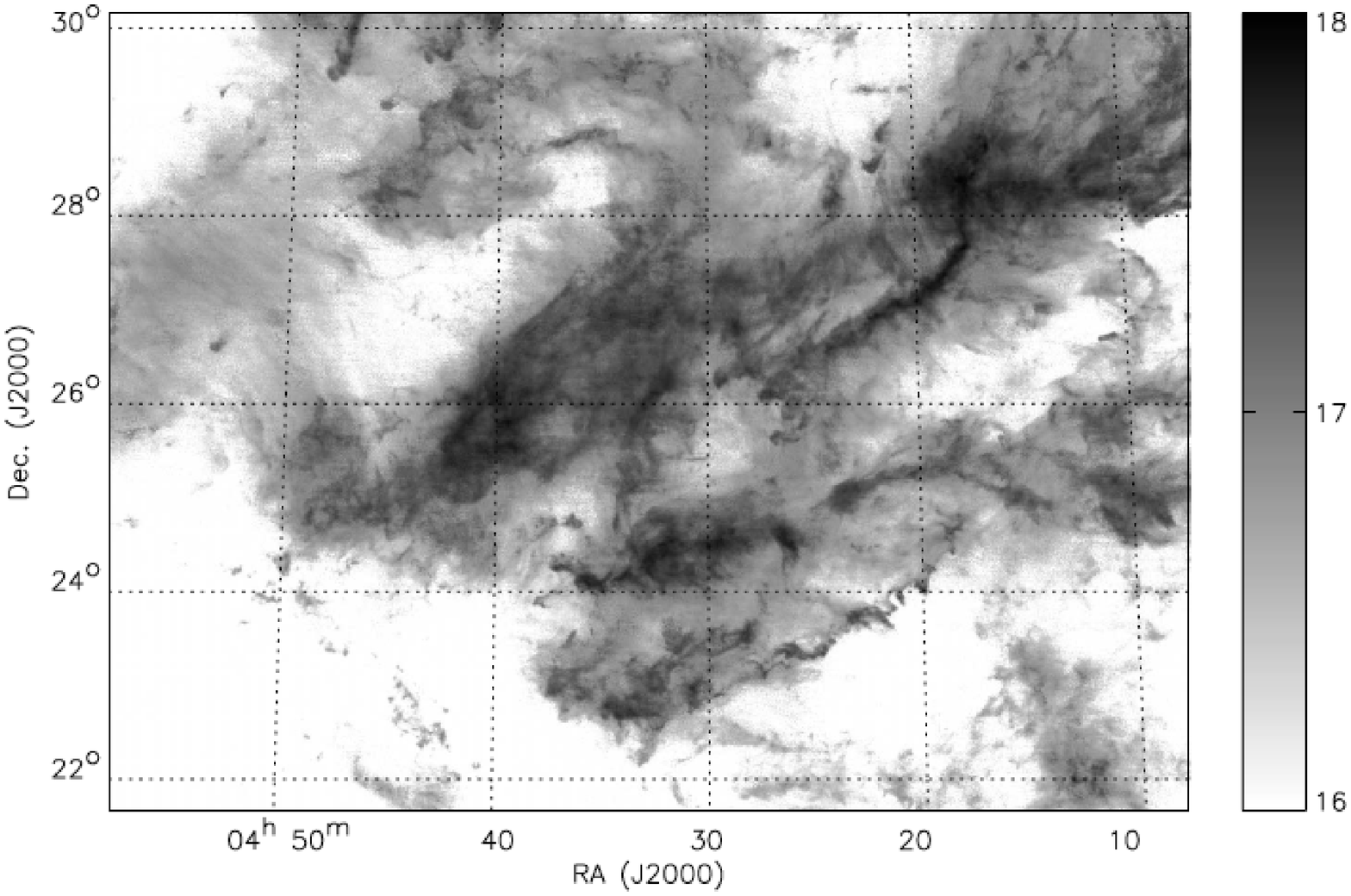}
  \caption{\twCO(1-0) integrated intensity maps of M51 at a resolution
of $\sim 200$ pc {\it (left)} and 30 pc {\it (center)} within the
square field shown in left panel (from Egusa et al. 2011). \twCO(1-0)
column density map of the Taurus-Auriga complex over the same size
$\sim 30$ pc, at a resolution of $\sim 0.03$ pc (Goldsmith et
al. 2008) {\it (right)}.  The intensity scale is logarithmic and shown
in the wedge. In the white regions, $N({\rm CO}) = 7.5 \times 10^{15}$
\cq, a factor $\sim 100$ smaller than the maxima.}
\label{COtaurus}
\end{figure}

\subsection{Diffuse molecular gas as a part of molecular clouds}

Beyond the above surveys of molecular clouds, the high sensitivity and
broad dynamic coverage of most recent observations of CO line emission
(including those of the all--sky {\it Planck} survey) and absorption
with the {\it Hubble Space Telescope} reveal the existence of CO
molecules in environments weakly shielded from the ambient UV--field.
This raises the issue of the CO molecules survival in these environments,
and of the nature of the gas component that CO emission is actually
tracing there. Because it is a CO--emitter and may be considered as the
birth--site of dense molecular clouds, this gas component called
``diffuse molecular gas'' deserves attention. The following sections
are dedicated to it.

On observational grounds, the definition of the diffuse neutral
ISM has evolved continuously since its
fortuitous discovery by Hartmann in 1904. It is now generally agreed
that it is a low density medium ($n_{\rm H}<$ a few $10^3$ \cc),
weakly shielded from the interstellar radiation field (ISRF) ($A_V<$ a
few mag), therefore comprising the diffuse atomic ($n_{\rm H} \sim
50$ \cc), diffuse molecular and translucent media, following the
terminology of Snow \& McCall (2006).  An even broader definition
includes the above plus the edges of molecular clouds as traced by
their \twCO($J$=1--0) emission, \ie\ all the gas that is not in dense
cores and/or star forming regions. Equivalently, the diffuse ISM can
be considered to be all material with total hydrogen column density less
than a few $\sim 10^{21}$ \cq\ at the parsec scale.  This unified
definition is motivated by the shapes of the probability distribution
functions (PDF) of the column density (inferred from dust extinction)
of nearby molecular clouds: they have a log--normal part from $A_V \sim
0.5$ mag up to $A_V \sim$ a few magnitudes, and a power--law tail at
higher column densities, characteristic of dense cores and filaments
present only in clouds actively forming stars (Kainulainen et
al. 2009).  The log--normal contribution that spans two orders of
magnitude in $N_{\rm H}$ corresponds to the diffuse ISM of the
molecular clouds.  The diffuse ISM, at large, may therefore be seen as
all the gas that does not belong to small--scale, self--gravitating
structures such as dense cores, dense filaments and star--forming
regions.  Note that the diffuse medium, as defined above, comprises a
significant mass fraction of molecular clouds.  In the Taurus complex,
at the 30~pc scale, Goldsmith et al. (2008) find that half the mass of
the complex is in regions having column densities less than $2 \times
10^{21}$ \cq.

 \subsubsection{The diffuse ISM: A two--phase medium}
The shape of the cooling function, dominated by \CII\ line emission,
makes the diffuse ISM thermally bistable at the pressure and
metallicity conditions prevailing in the Solar Neighborhood (Field et
al. 1969, Wolfire et al. 1995, 2003), \ie\ at a distance of less than
$\simeq$1kpc from the Sun.  The balance between heating and radiative
cooling processes generates two stable phases at very different
temperatures and densities, $T\sim 100$ K and $n_{\rm H}\sim 30$
\cc\ for the cold neutral medium (CNM), $T\sim 8000$ K and $n_{\rm H}
\sim 0.3$ \cc\ for the warm neutral medium (WNM). However, a broad
range of kinetic temperatures, pervading the thermally unstable range,
is usually inferred from UV absorption lines sampling the diffuse ISM
(e.g. Fitzpatrick \& Spitzer 1997), and recent sensitive HI
observations also probe temperatures in this range (Heiles \& Troland
2003, Begum \etal\ 2010).  The latter infer that about half the WNM is
at temperatures in the range 500 -- 5000 K, thus evolving between the
two stable phases.

\subsubsection{Pressure of the diffuse ISM}
Because it sits in the gravitational potential of the Galaxy and
extends up to $\sim$ 300 pc above the plane, the diffuse ISM has to be
supported by a pressure that is about 10 times its thermal pressure,
$P_{tot}=3 \times 10^{-12}$ dynes cm$^{-2}$, or $P_{tot}/k \sim 2
\times 10^4$ \cc\ K (Cox 2005).  The non-thermal contributions to the
total pressure are due to supersonic turbulence and magnetic fields,
in rough equipartition, as shown by measurements of magnetic field
intensity (Crutcher et al. 2010, see Section 5) and HI linewidths
(Haud \& Kalberla 2007).  The distribution of thermal pressures in the
Solar Neighborhood, inferred from [CI] fine--structure absorption
lines towards nearby stars (Jenkins \& Tripp 2011) peaks at about
$P_{th}/k \sim 3 \times 10^3$ \cc\ K, with large fluctuations, up to
several 10$^4$ \cc\ K, possibly due to expanding supernova
remnants. It is noteworthy that the total non--thermal pressure in the
Galactic plane is of the same order as the largest values of the
thermal pressure observed. This equality suggests that the
non--thermal energy eventually and occasionally dissipates into thermal
energy.

\subsubsection{An unexpected molecular richness}

\begin{figure}
\begin{minipage}{7cm}
  \includegraphics[width=0.7\textwidth,angle=0]{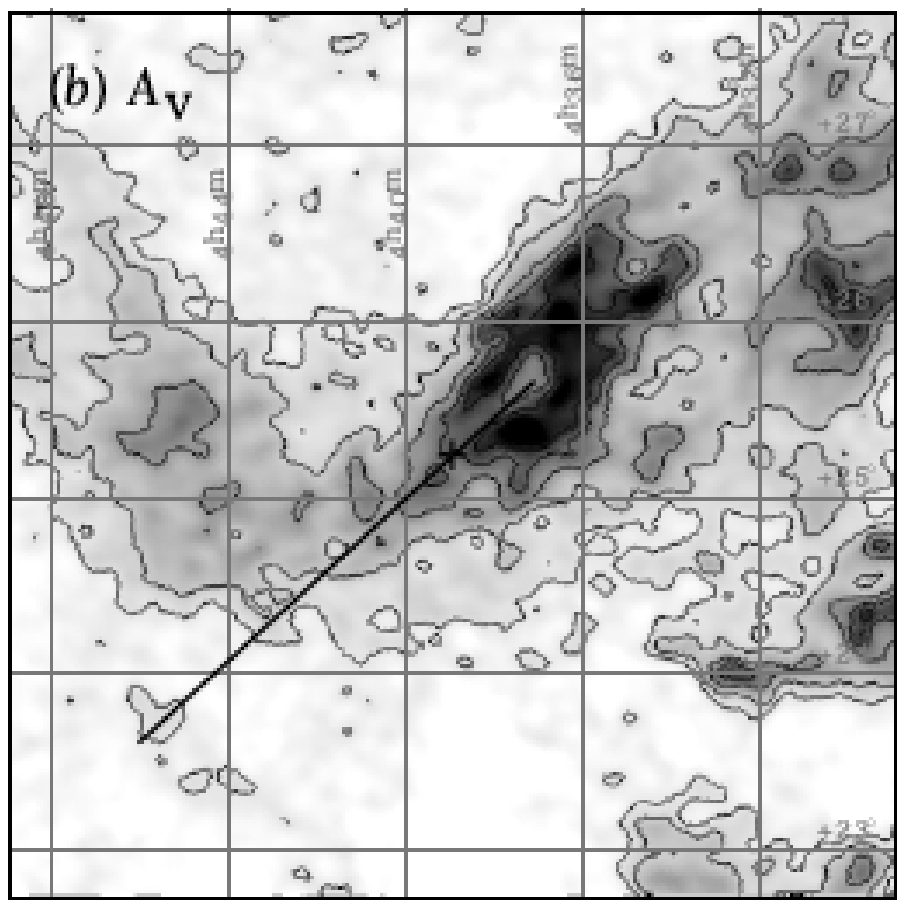}
\end{minipage}
\begin{minipage}{7cm}
  \includegraphics[width=0.7\textwidth,angle=0]{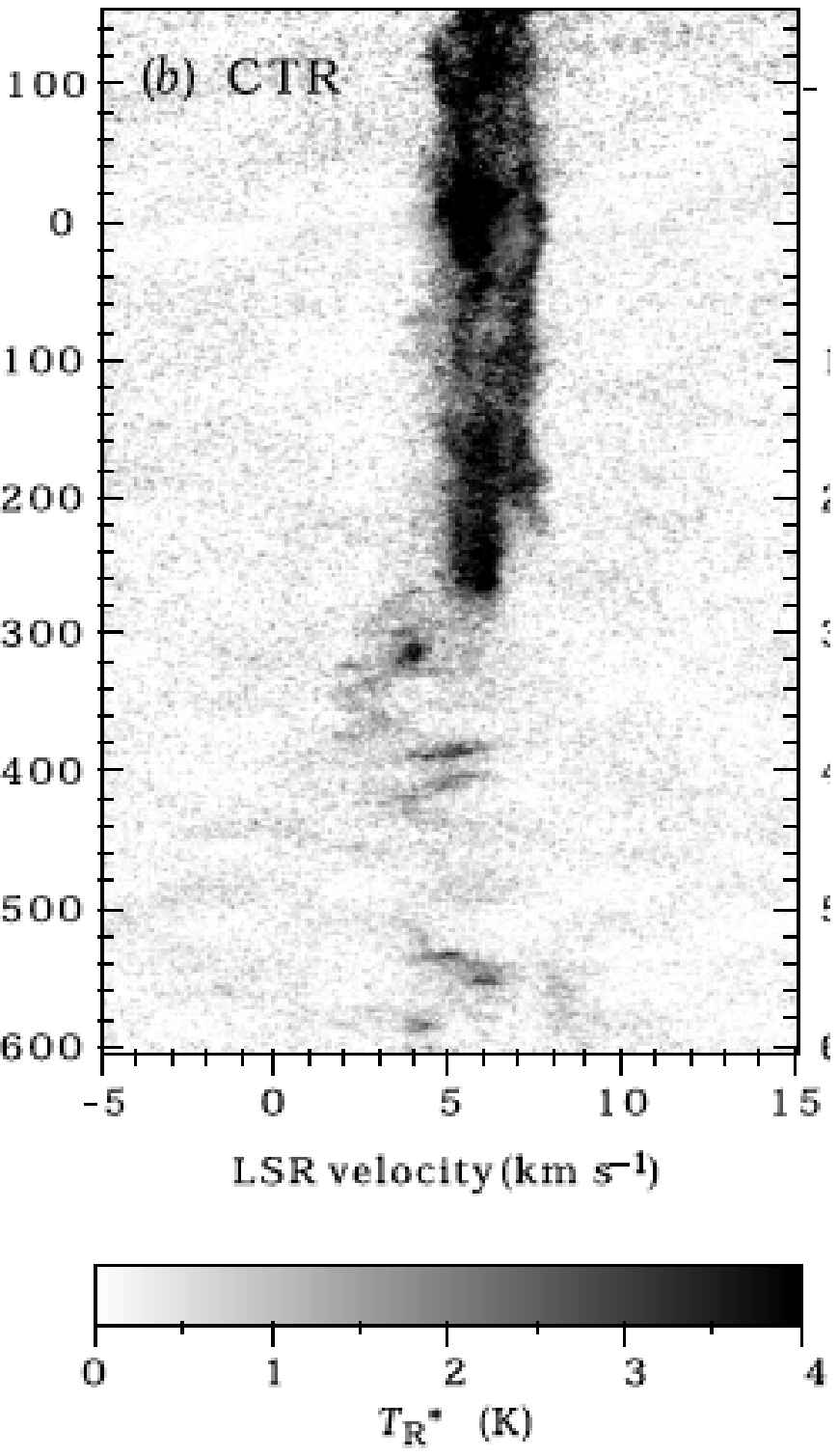}
\end{minipage}
  \caption{Illustration of the variability of the \twCO\Jone\ emission
along a cut across one edge of the Taurus molecular complex. (Left)
the extinction map showing the position of the cut. (Right) the
position-velocity diagram of the cut showing the sharp transition
between translucent gas of high velocity dispersion with tiny spots of
detected CO emission and the opaque part, with a much narrower
velocity dispersion (from Sakamoto \& Sunada 2003).}
\label{sakamoto}
\end{figure}

These large fluctuations of thermal pressure are a sign of a complex
dynamic evolution that may provide the missing important clues to
understand the origin of the remarkable molecular richness found in
this hostile medium, weakly shielded from UV photons (e.g. Liszt \&
Lucas 1998). In the case of CO, {\it (i)} column densities as low as a
few 10$^{12}$ \cq, smaller than that required for CO self--shielding
($N(\rm CO) \sim 10^{14}$ \cq, van Dishoeck \& Black 1988, Lee et al. 1996) 
have been observed in absorption in
unshielded environments where $N(\HH) < 3 \times 10^{19}$ \cq\
(Sonnentrucker et al. 2007); {\it (ii)} large fluctuations of the CO
emission of diffuse gas at small scale have been found that cannot be
ascribed to shielding fluctuations; {\it (iii)} both 
CO--dark and CO--overluminous diffuse molecular gas exist
with respect to the standard CO--to--\HH\ conversion factor (Liszt et
al. 2010, Liszt \& Pety 2012).
 
The sharp variability on arcminute scales of the CO emission of
diffuse gas is primarily due to CO chemistry, a point that cannot be
understood in the framework of UV--driven chemistry: small--scale gas
kinematics seems to be involved.  An illustration of that is shown in
Fig. \ref{sakamoto} from Sakamoto \& Sunada (2003).  The tiny spots of
CO emission in the unshielded part of the cut across the edge of a
molecular cloud are associated with large velocity variations.  The
authors suggest that these CO--rich structures may form through thermal
instability in shocked atomic gas (Koyama \& Inutsuka 2000, 2002) or
Kelvin--Helmholtz (KH) instability (Sakamoto 2002) although they favor
thermal instability, given the long timescale for the growth of the KH
instability.  Barriault et al. (2010) investigate the transition
between atomic and molecular gas in high-latitude clouds. They find
that there is no systematic coincidence between IR-excess (\ie\ FIR
emission in excess to that predicted from HI column density) and CO
emission and that CO appears where two HI velocity components merge or
where a large velocity--shear is present in the HI map.  These results
strengthen the possibility of a new CO formation route linked to the
dynamics of the HI gas (see Section 6.1.2).
 
Recently, the molecular richness of the diffuse medium has been
further illustrated by absorption spectroscopy in the submillimeter
domain against the dust emission of bright star--forming regions in the
Galaxy.  Small hydrides, the building blocks of IS chemistry, have
been detected with unexpected large abundances. Several species
provide unique clues to our understanding of the physics of the
diffuse ISM: the large excess of \OHp\ over \HtwOp\ implies that these
ions are formed in mostly atomic gas (Gerin et al. 2010, Neufeld et
al. 2010a); the ubiquitous HF is a powerful tracer of \HH\ in diffuse
gas because of the highly exothermic reaction of fluorine with
\HH\ (Neufeld et al. 2010b, Sonnentrucker et al. 2010); the large
abundances of \CHp\ and \SHp\,  both with highly endoenergetic
formation routes reveal the existence of local reservoirs of energy,
 exceeding by far that provided by UV--photons (Falgarone et al. 2010a,b, 
Godard et al. 2012). Last, the molecular fraction of the diffuse
medium $f_{\HH}= N(\HH)/(N({\rm H})+2N(\HH))$ where these hydrides are
observed is found to fluctuate by large factors (Godard et al. 2012),
from less than 0.1 to almost unity, further blurring the distinction
between atomic and molecular clouds.

In summary, CO is not exclusively a tracer of dense gas and its
formation rate may be enhanced by the gas dynamics. Moreover, molecular gas
exists without associated CO emission. This may bring partial answers
to the persistent questions raised on the CO--to--\HH\ ratio.

\subsection{The CO--to--\HH\ ratio}

The first major achievement of CO observers has been the determination
of the well known CO--to--\HH\ conversion factor, $X_{CO}= N(\HH)/W(\rm
CO)$ that provides the only way to measure the amount of molecular
hydrogen (which cannot be observed directly in cold environments, see
Section 6.1.1) from the integrated \twCO\ line emission, $W(\rm
CO)$. Estimates of total gas masses (atomic and molecular) were then
made possible, an essential step forward to unravel the physics of
molecular clouds.  The standard value $X_{\rm CO}= 2 \times 10^{20}$
\cq/(\Kkms) was calibrated on X--ray and near--IR extinctions towards
the central regions of the Galaxy, therefore kpc--long sight lines
encompassing GMCs, dark and diffuse clouds.  The $X_{CO}$ factor was
inferred by subtracting the contribution of the atomic gas to the
total extinctions.

Two recent space experiments {\it Fermi/LAT} in the $\gamma$--ray
domain (Abdo et al. 2010, Ackermann et al. 2011) and {\it Planck} in
the submillimeter domain (Planck collaboration et al. 2011) independently
confirm what was already found by the first $\gamma$--ray space
mission, {\it COS B} (Lebrun et al. 1983): there are significant
amounts of molecular gas in the Galaxy that are not traced by
\twCO\ and would be atomic gas rich in \HH\ without detectable CO
emission (e.g. Grenier et al. 2005).  This CO--quiet \HH\ gas lies at
the periphery of known bright molecular complexes, a fact consistent
with model predictions (Wolfire et al. 2010). In the case of {\it
  Fermi/LAT}, excesses of $\gamma$--ray emission are observed in these
regions that cannot be ascribed to higher densities of cosmic rays
because these are accurately measured.  In the case of {\it Planck},
the dust optical depth excess above that expected from known HI and CO
emission occurs in regions where $8 \times 10^{20}$\cq\ $< N_{\rm H}<
5 \times 10^{21}$ \cq, typical of the diffuse molecular gas.  The mass
of CO--quiet gas in the Solar Neighborhood is estimated to be about
30\% of its atomic gas and 100\% of the CO--bright gas.

Both groups, however, discuss the large uncertainties due to
underestimates of the HI optical depth in molecular clouds,
\ie\ overestimation of the spin temperature. Indeed, atomic hydrogen
at very low temperatures is found in molecular clouds (Li \& Goldsmith
2003).  If the amount of HI has been underestimated in the calibration
of the $X_{\rm CO}$ factor in the Galaxy, \HH\ would be overestimated
and $X_{\rm CO}$ is indeed too low.  This would be the origin of the
subsequent underestimate of \HH\ associated to a given CO observation.

A fact that has not always been fully appreciated is that CO emission
from diffuse gas leads to the same CO--to--\HH\ conversion factor
as in dense clouds (Liszt \& Pety 2012).  This can be understood in
the framework of radiative transfer: the radiation efficiency of CO
per molecule in low--density gas is much higher, because collisional
de--excitation is weak and each collisional excitation is followed by a
radiative de--excitation.

In the following sections, we will see that molecular clouds do not
have the same characteristics whether they are traced by \twCO(1--0)
emission, \thCO(1--0) emission (almost 100 times less optically thick)
or by emission of high dipole moment species (such as CS, CN...) that
are specific high density tracers ($n_{\rm H} > 10^{5}$ \cc).  The
diffuse component discussed above enters the concept of molecular
clouds because it is locally CO--rich, contributes to the
\twCO\ emission of molecular clouds at a significant level, and
because it comprises a large fraction of their mass. Finally, it is a
highly dynamic component that plays a key role, as will be seen, in
the formation of dense molecular clouds.

\subsection{Link between molecular clouds and their large scale environment}

The connection between molecular clouds and their surrounding ISM
in galaxies is an important issue because  molecular clouds
{\it (i)} presumably form out of their diffuse environment; and {\it (ii)}
evolve by interacting with it.

\subsubsection{The HI halos and the molecular cloud phases}
Many, if not all, molecular clouds are embedded into massive halos
detected in the HI 21~cm line, therefore containing a significant fraction
of atomic hydrogen (Hasegawa et al. 1983, Wannier et al. 1983,
Elmegreen \& Elmegreen 1987). Early on, the mass within these 
HI halos has early
been estimated to be comparable to (or even greater than) the mass
within the molecular clouds themselves.  While Hasegawa et al. (1983)
proposed that the HI halo constitutes the accretion reservoir of the
molecular clouds, Wannier et al. (1983) argue that the atomic hydrogen
around molecular clouds is warmer than in the rest of the Galaxy and
proposed that it is photo--dissociated \HH\ induced by the star
formation within the molecular clouds.

These observations remain difficult to perform in the Galaxy because
of the confusion along the line of sight, HI being ubiquitous.  
However, the velocity field allowed Elmegreen \& Elmegreen (1987) to
identify about 30 massive HI superclouds in the first quadrant of the
Galaxy, each surrounding groups of known GMCs detected in CO. These 
HI superclouds are 
found to be distributed along spiral arms with separations ($\sim$ 1.5
kpc) much larger than their size ($\sim 200$ pc).  Their mass ranges
between 10$^6$ and 4$\times 10^7$ \msol, the mean number density is $\sim
10$ \cc\ and they are self--gravitating.  Substantial effort has been
made  to conduct similar observations in external galaxies.

{\bf The Large Magellanic Cloud} (LMC), has the advantage of being not
too distant and  seen nearly face on, therefore limiting the
issue of the confusion. Wong et al. (2009) studied in detail the
correlation between the CO and the HI distribution.  In particular,
they stress that while the CO peaks are always associated to a peak in
HI, there are many HI peaks in which CO is not detected. The 
reason is not well understood yet: it may simply be a time--scale issue
or a consequence of different physical conditions.  This tends 
to support the view that molecular clouds form out of the HI instead
of HI being photo--dissociated H$_2$.

Whether the gas traced by HI is accreting onto molecular clouds is
still an open issue and Kawamura et al. (2009) address this question
in the LMC with the following approach.  They have classified 
clouds into three categories: (1)  
clouds in which massive stars are not observed, (2)
clouds with massive stars but no cluster, and (3) clouds with clusters. 
Kawamura et al. (2009) propose that this is an evolutionary
sequence and they estimate that each phase lasts, respectively, 6, 13 and
7  Myr. A possible key point is that they observe that the mean mass of
the third category is a few times larger than the
mean mass of category I and II.  If this evolutionary scenario is
confirmed, it would imply that GMCs in the LMC are indeed accreting.
Fukui et al. (2009) estimate the gas density in the HI halos to be of
the order of 10 cm$^{-3}$ with an accretion velocity less than 7 \kms,
\ie\ the rms velocity dispersion of the HI clouds. This leads them to
estimate the accretion rate to be at most 0.05 M$_\odot$ yr$^{-1}$.
The GMCs within the LMC appear to be similar to those within the
Galaxy (Hughes et al. 2010), therefore seemingly suggesting that the
processes are reasonably similar in these two galaxies. 

{\bf M33 } is a low-metallicity gas-rich nearby blue spiral galaxy
much further away from the Milky Way (840 kpc).  It has been mapped in
CO(2-1) and HI over a radius of 8.5 kpc (Gratier et al. 2010).  The HI
mass (1.4 $\times 10^9$ \msol) is five times larger than that of the
molecular gas inferred from CO.  At the resolution of the observations
$\sim 100$ pc, the correspondence between the CO and HI patterns is
excellent. Gratier et al. (2012) have analysed the molecular 
clouds in M33. They also distinguish three categories of clouds 
similar to the classification of Kawamura et al. (2009) and reach 
close conclusions.

\subsubsection{What is filling the volume between molecular clouds ?}

Quite early on, it was recognized that the mean density of the GMC
(about 10 \cc\ at the 100 pc scale) inferred from their CO column
density and size was much lower than the gas density, of the order of
10$^3$ \cc\ or more, required to excite the observed bright CO lines
(Blitz \& Thaddeus 1980, P\'erault et al. 1985). It was then inferred
that the filling factor of dense gas within GMCs is significantly
below unity, raising the question on the nature of the gas filling the
volume between the CO--bright structures, \ie\ the interclump medium
(ICM)\footnote{Here, as a tribute to the observers community, we
  introduce the poor and ill--defined concept of {\it clump}. Let us say
  that a clump has an average density larger (or even much larger)
  than that of the CNM at the ambient pressure, and, by contrast, the
  ICM has much lower densities.}.  Indeed, the density and temperature
of the ICM presumably controls the evolution time--scale of molecular
clouds and contributes to the formation, confinement, and survival of
the dense clumps that eventually form stars.

The recognition that dense gas in molecular clouds was filling only a
small volume, prompted the emergence of GMC models in which dense
fragments were virialized in the gravitational potential well of the
GMC (\eg\ Scalo \& Pumphrey 1982). The GMC lifetime was thus
considerably extended because the observed large velocity dispersion
of the GMC was no longer turbulent and supersonic (with respect to the
cold cloud material) and no longer expected to dissipate within one GMC turnover
time, $\sim 10^7$ yr. In this picture, the fragment boundaries were
sharp density fall--offs, set by the thermal instability between the
CNM and WNM (Falgarone \& Puget 1985) and their orbital velocities
were therefore transonic with respect to the WNM: only the rare
collisions among fragments were controlling the dissipative GMC
evolution.  Falgarone \& Puget (1986) refined this concept by
introducing the magnetic field lines connecting the fragments which
provided a mechanism to feed the internal turbulent energy of
fragments by pumping energy from their orbital motions. The
dissipation time scale of turbulence within fragments was thus
lengthened to that of the quasi--virialized stage of the large scale.
In this scenario, the low--density medium was the WNM, at $\sim$ 8000K,
heated mainly by ion--neutral friction caused by the field lines
entrained in the fragment motions. Such a scenario provides molecular
clouds with an internal structure that closely resembles that of the
diffuse ISM with both cold and warm material, but with the cold phase
reaching higher densities.

In an attempt to probe the nature of the ICM medium, Williams et
al. (1995) studied the Rosette molecular clouds and 
the correlation between CO and HI emission. They found that
both are clearly anti--correlated therefore showing their
association. From the HI line integrated emission, they inferred 
an ICM density, $\sim 4$ \cc, 100 times lower than that of the
clumps, a value somewhat plagued by the unknown geometry of the medium
since only the column density is measured. From the
HI linewidths, they inferred an ICM velocity dispersion 
of $\sim$  10 km s$^{-1}$, providing the ICM with a ram pressure,
$\rho \sigma^2$, comparable to the internal turbulent pressure of
the clumps. They proposed that the clumps be confined by the ICM ram
pressure.  Unfortunately, they could not place constraints on the
ICM temperature and assumed it to be $\sim$ 20 K.
Such a low temperature is unlikely because
it would imply that the gas thermal pressure would be about 
100 K cm$^{-3}$, roughly 30 times below the mean ISM pressure, 
allowing the external ISM to penetrate into the GMC. 
Part of the ICM is therefore likely
much warmer than 20 K and/or turbulent, providing molecular clouds with a 
structure that resembles that of the diffuse medium.
Magnetic fields are also contributing to the pressure of the ICM
and an alternative possibility is that the ICM 
is indeed cold, diffuse and highly magnetized.
In the next section, we describe how numerical simulations are presently able to 
tackle this huge range of densities and temperatures.

\subsection{A scenario for the formation of molecular clouds}

Numerical simulations are now able to capture many of the steps
leading to the formation of molecular clouds out of the diffuse ISM.
The first step has been the study of the non--linear development of
thermal instability triggered either by a converging flow of WNM
(Walder \& Folini 1998, Hennebelle \& P\'erault 1999, 2000,
S\'anchez--Salcedo et al. 2002, Audit \& Hennebelle 2005, 2010, Heitsch
et al. 2005, 2006, V\'azquez--Semadeni et al. 2006, Hennebelle \& Audit
2007), or by shocks propagating in the WNM (Koyama \& Inutsuka 00, 02,
Inoue et al. 2009) or by turbulent driving in the Fourier space
(Seifreid et al. 2011).  The conclusions common to all these studies
are: {\it (i)} the WNM flow quickly breaks--up into a multi--phase
medium with dense clumps (CNM) confined by the warm external medium
(WNM); {\it (ii)} the statistics of these clumps have strong
similarities with what has been inferred for the CO clumps (discussed
in Sect~\ref{co_clumps}); {\it (iii)} there is a significant fraction
of thermally unstable gas whose existence is made possible by the
turbulent fluctuations (\eg\ Gazol et al. 2001); {\it (iv)} the
various phases are tightly interspersed; {\it (v)} the cold phase is
maintained supersonic (with respect to its own sound speed) and with a
velocity dispersion equal to a fraction of the WNM sound speed.
Simulations including the magnetic field have also been performed.
Although magnetic fields definitely modify the fluid dynamics, the
above conclusions remain qualitatively similar.

\begin{figure}
\begin{minipage}{7cm}
  \includegraphics[width=\textwidth{},angle=0]{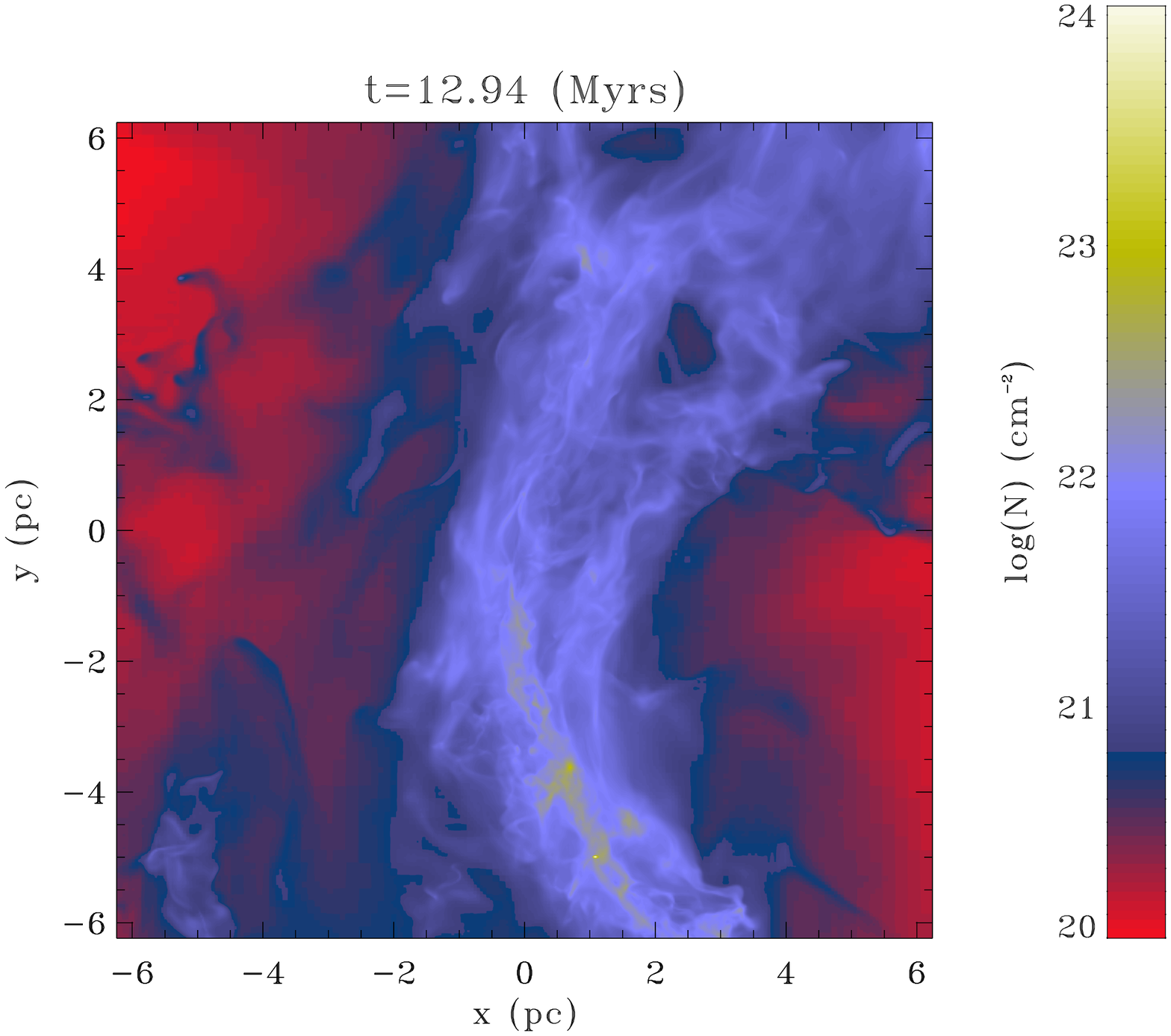}
  \includegraphics[width=\textwidth{},angle=0]{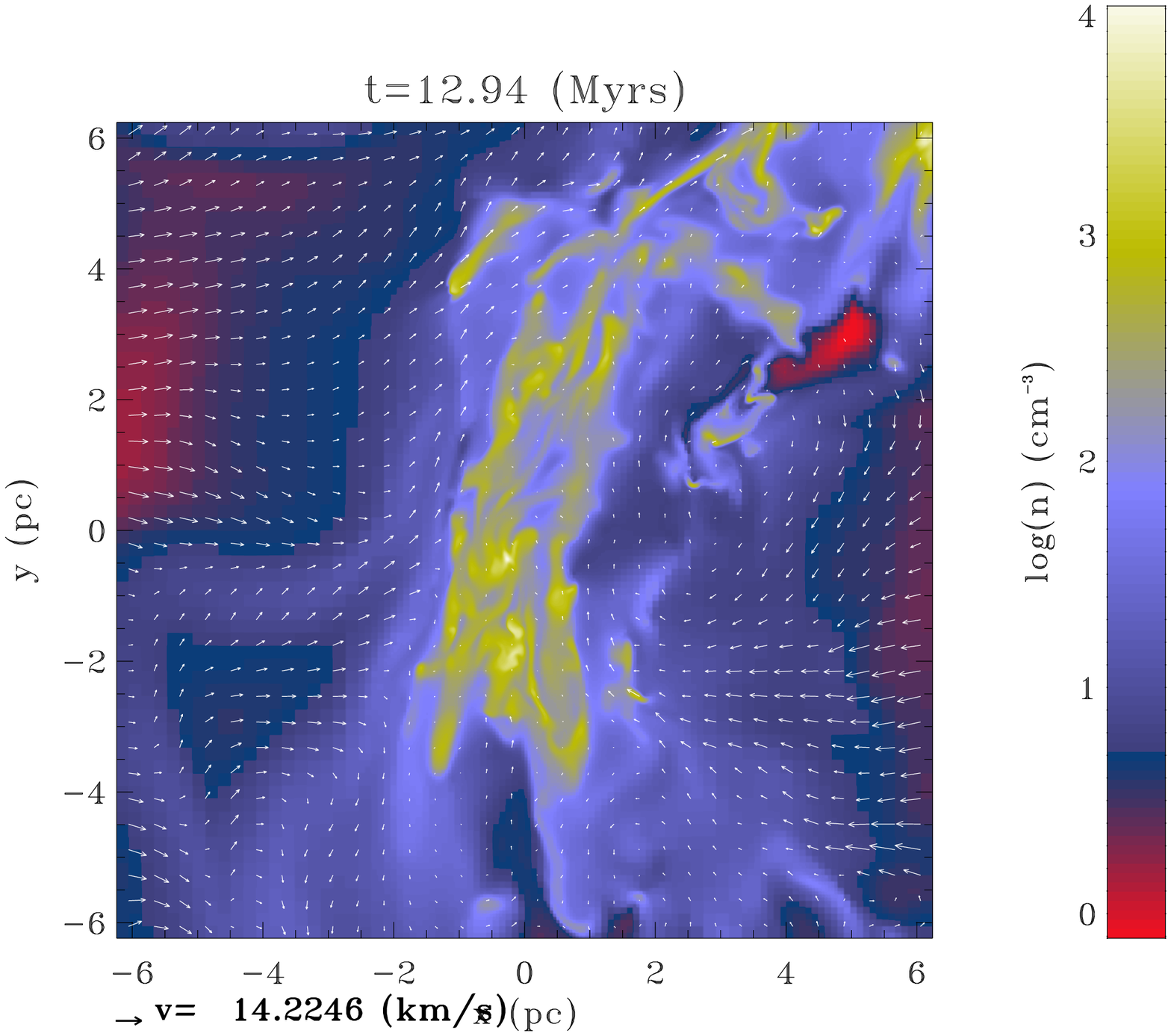}
\end{minipage}
\begin{minipage}{7cm}
  \includegraphics[width=\textwidth{},angle=0]{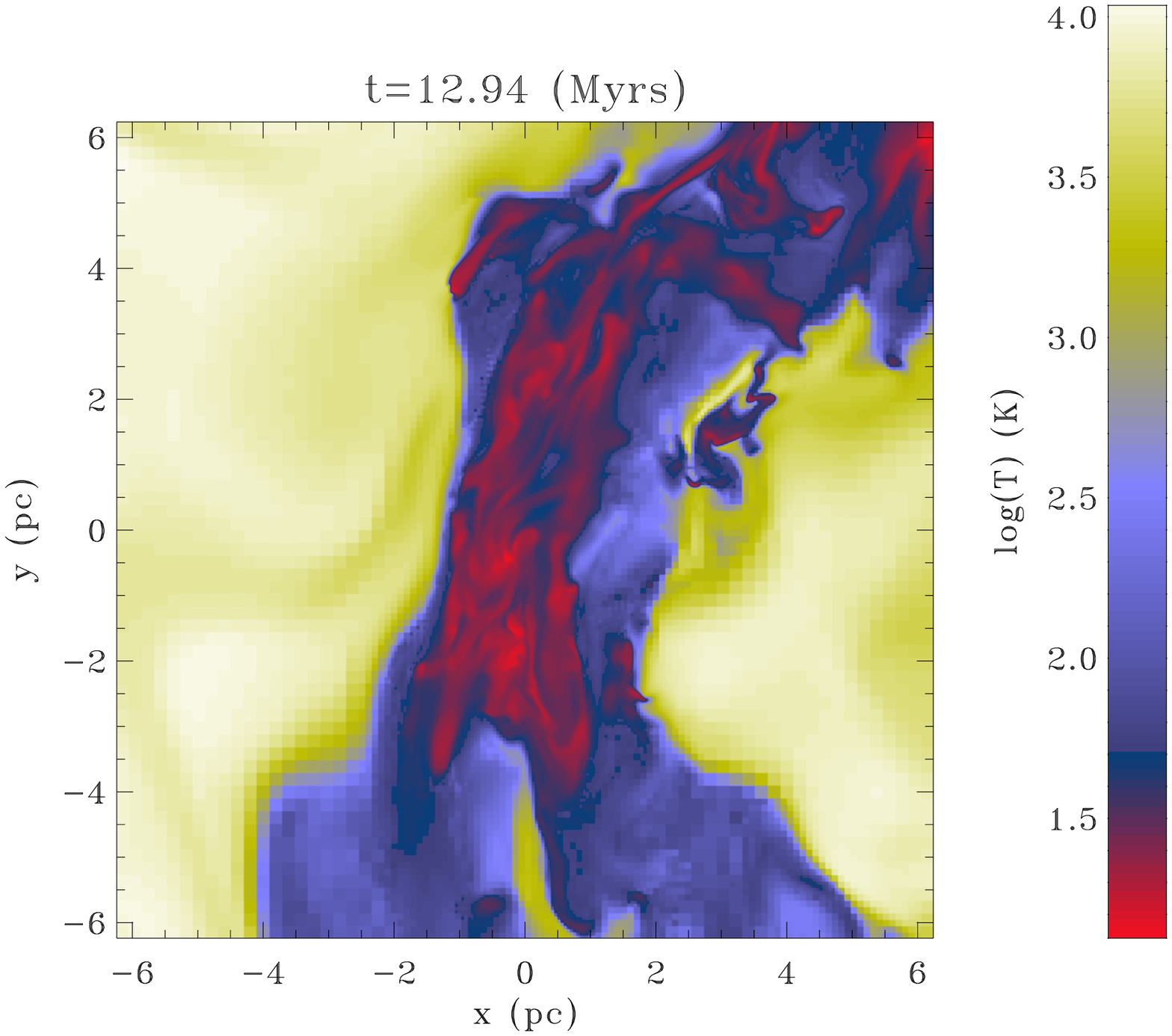}
  \includegraphics[width=\textwidth{},angle=0]{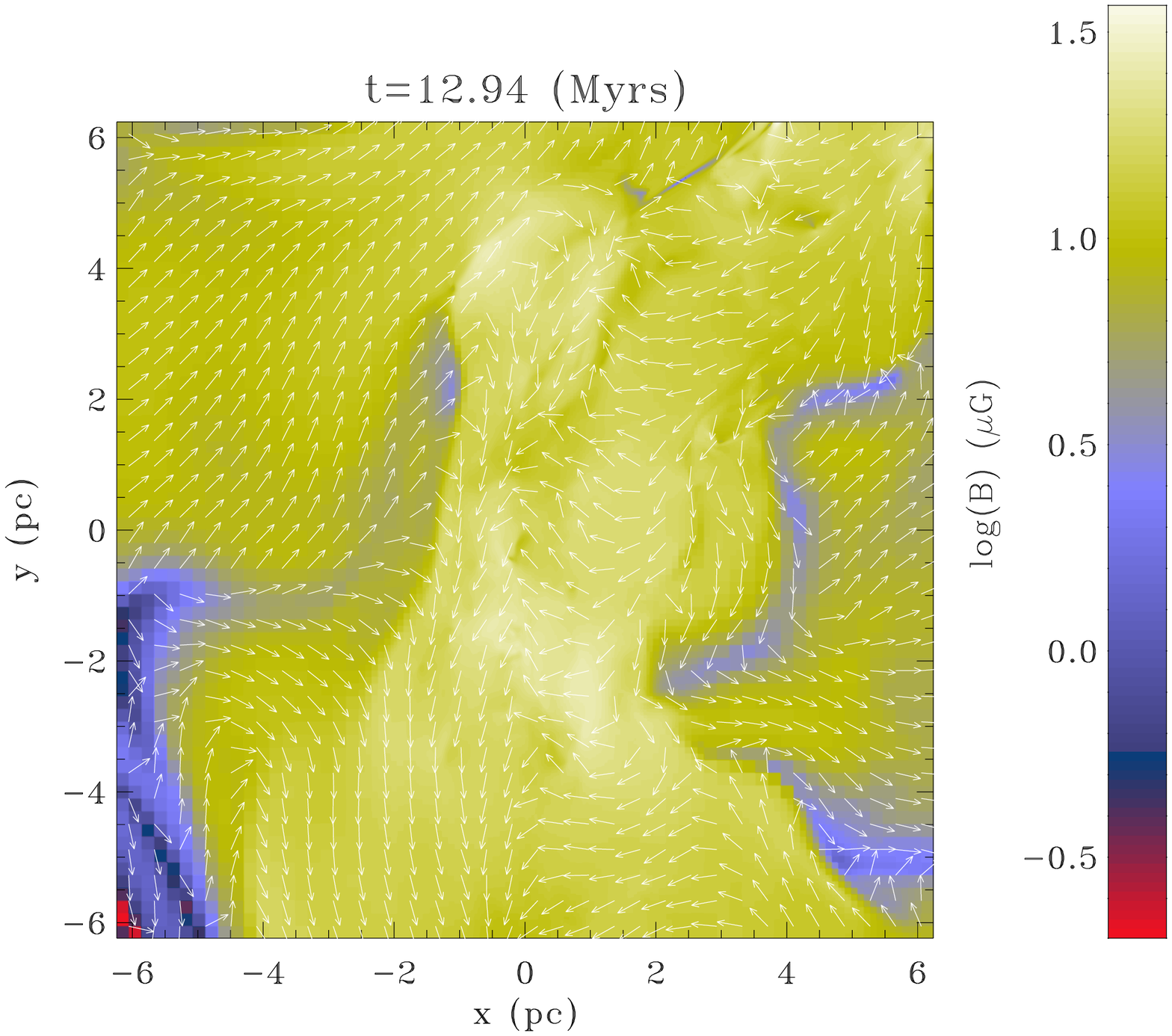}
\end{minipage}
  \caption{ Column density {\it (top left)}, and cuts of the 
 density with  projected velocity field {\it (bottom left)}, 
temperature {\it (top right)} and  magnetic field {\it (bottom right)} 
of a simulation snapshot of molecular
cloud formation (Hennebelle et al. 2008). 
While the total magnetic intensity 
is shown, the arrows indicate the direction of the field 
in the XY plane.}
  \label{multi-phase}
\end{figure}

In the second step, gravity has been introduced therefore letting the
condensations to proceed further (V\'azquez--Semadeni et al. 2008,
Heitsch et al. 2008, Hennebelle et al. 2008, Banerjee et al. 2009).
Fig.~\ref{multi-phase} shows a snapshot (column density, density,
temperature and magnetic field cut) of a colliding flow simulation
(similar to what is presented in Hennebelle et al. 2008 and Klessen \&
Hennebelle 2010) aiming at forming a molecular cloud. The dense gas
(\ie\ the clumps) has formed out of the diffuse gas.  The structure of
the cloud is made of dense and cold clumps embedded in a warmer and
more diffuse gas. Gravitational collapse has occurred in a few places.
As visible in the snapshot, at a few locations, the temperature
abruptly varies from $10^4$ to 10 K. The field shown in
Fig.~\ref{multi-phase} is part of a larger cloud  that contains
the  WNM shown here.
In simulations including gravity, the clumps share many
characteristics with the observed CO clumps (see
Section~\ref{co_clumps}) in terms of mean density, velocity dispersion
and mass spectrum (e.g. Banerjee et al. 2009).  The structure of the
cloud remains largely unchanged.  The salient difference with the non
self--gravitating case is that the PDF of the gas density extends towards much larger values
(see later discussion of the density PDF).  However, the bulk of the mass,
though, remains comprised in gas of density in the range $n_{\rm H}=
100-10^3$ \cc, even at late times, when densities as large as 10$^6$
\cc\ are reached.  \\
In a third step, UV-driven chemistry has been introduced either
directly during the simulation (\eg\ Glover \& MacLow 2007) or, at a
more sophisticated level, as a post-treatment of the WNM colliding
flow simulations (\eg\ Levrier et al. 2012).  This allows a proper
treatment of the combined influence of density and UV-shielding upon
chemistry (see Sect. 6) and of the cooling function. The results
provide a confirmation that the gas temperature is reasonably well
computed in the magneto-hydrodynamical (MHD) simulations.  
Levrier et al. (2012) beautifully
reproduce the observed behavior of molecules such as CH with column
densities of \HH\ (\eg\ Gerin et al. 2010) but, interestingly, fail to
reproduce (by almost a factor of 10) the observed CO abundances in
regions poorly shielded from the UV-field.  Several methods to follow
CO formation in turbulent simulations have been compared by Glover \&
Clark (2012): they find that, not unexpectedly, the gas dynamics is
not sensitive to the details of the chemistry models, but all the
adopted models so far fail to reproduce the CO abundances in such
regions (Shetty et al. 2011).

All these results are quantitatively in good agreement with the
two--phase picture of molecular clouds proposed above.  They illustrate
the importance of the connection between molecular clouds and their
surrounding low--density medium.  Moreover, they stress the major role
of large--scale velocity perturbations (shocks, turbulence) in the low
density gas upon the thermodynamical evolution of the whole medium.

In parallel to these works, which focus on the scale of a single
molecular cloud, a series of studies have been conducted to understand
the formation of molecular clouds at large scale.  A first type of
studies (e.g. V\'azquez--Semadeni et al.  1995, de Avillez \&
Breitschwerdt 2005, Joung \& Mac Low 2006, Koyama \& Ostriker 2009)
consider scales of about 1kpc with supernovae explosions as the main
driver of interstellar turbulence.  The second approach treats a whole
galactic disk including the gravity of stars either in a fixed
asymmetric potential (e.g. Tasker \& Tan 2009), or in a prescribed
spiral arm potential (e.g. Dobbs \& Bonnell 2007) or with self--consistent
dynamics of stars (e.g. Bournaud et al. 2010, Hopkins et al. 2011). 
Not surprisingly, it is
found that spiral arms play an active role in triggering molecular
cloud and star--formation.  These approaches are complementary to those
treating the formation of individual clouds since they provide a
consistent description of the cloud formation mechanisms. In
particular, statistics can be obtained such as the cloud mass spectrum
and internal velocity dispersion. Many of these observed properties
are satisfactorily reproduced, though the numerical resolution remains
limited, suggesting that the origin of molecular clouds is indeed due
to the interaction between turbulence and gravity.

\section{The turbulent velocity field}

The nature and properties of turbulence in the ISM are still a highly
debated and controversial issue in spite of dedicated observational
and numerical efforts.  This is due in part to the huge range of
scales separating those of the energy injection (at the Galaxy scale
and even beyond when infall is taken into account),
from those where it is dissipated, presumably below the milliparsec
scale.  It is also due to the fact that the turbulence is
compressible, magnetized and multi--phase. Unraveling the properties of
interstellar turbulence is essential, though, because along with the
magnetic fields,  it represents the main contributions to the 
pressure of the ISM
in equilibrium in the galactic gravitation potential field and  the
main support of molecular clouds against their
self--gravity. Turbulence dissipation is therefore a key process among
those leading to the formation of molecular clouds, star--formation,
and therefore Galaxy evolution (see the reviews of Elmegreen \& Scalo
2004 and Scalo \& Elmegreen 2004, MacLow \& Klessen 2004, McKee \&
Ostriker 2007).

\subsection{A few words on incompressible hydrodynamical turbulence}

The definition of turbulence is built on experiment.  Turbulence is an
instability of laminar flows that develops as soon as the inertial ${\bf v}
\cdot \nabla {\bf v}$ forces greatly exceed the viscous $\nu \Delta {\bf v}$
forces ($\nu$ is the kinematic viscosity), \ie\ when the Reynolds
number $Re=lv_l/\nu$, at a scale $l$ of characteristic velocity $v_l$,
exceeds a few hundreds. This instability at scale $l$ is at the origin
of an energy transfer to smaller scales, which eventually become
unstable too and transfer their kinetic energy to even smaller
scales, etc. This is the turbulent cascade that develops between the
integral scale, $L$, at which energy is injected, and the dissipation
scale $l_D$, close to the particle mean--free--path, where energy is
dissipated into heat due to the particle viscosity.  The time--scale for
the growth of this instability is of the order of the turnover time
$\tau_l=l/v_l$ at each length scale $l$.  Kolmogorov (1941) predicted
the self--similar behavior of the velocity field in incompressible
turbulence by postulating a dissipationless cascade characterized by a
transfer rate of kinetic energy independent of scale, $\epsilon
\propto v_l^2/\tau_l=v_l^3/l$, hence the well--known scaling $v_l
\propto l^{1/3}$. It is easy to demonstrate that this assumption leads
to an energy spectrum $E(k) = k^2 P_v(k) \propto k^{-5/3}$ known as the Kolmogorov
spectrum. $P_v(k) \propto k^{-11/3}$ is the power spectrum of the velocity.
$E(k)$ has the dimension of a kinetic energy per unit mass
{\it and} unit wavenumber because the average specific kinetic energy at
scale $l=2\pi/k$ is $\langle v_l^2 \rangle =\int_{k}^\infty E(k')
dk'$.  In Kolmogorov turbulence, the turnover time--scale $\tau_l$
therefore decreases towards small scales while the velocity gradient
$v_l/l \propto l^{-2/3}$ slowly increases.
 
\subsection{Intermittency of turbulence}  

For half a century now, turbulence has been recognized to be
intermittent, \ie\ the smaller the scale, the larger the
spatio-temporal velocity fluctuations, relative to their average
value.  Turbulent energy is not evenly distributed in space and time
by the turbulent cascade: at each step of the cascade, the active
sub--scales do not fill space so that the subset of space on which the
active scales are distributed has a multifractal geometry (see the
review of Anselmet et al. 2001 and the book by Frisch 1996).  The
statistical properties of the velocity fluctuations have been widely
studied experimentally in laboratory and atmospheric flows: in all
cases, the statistics of {\it velocity derivative and increment}
signals are found to be non--Gaussian, with large departures from the
average more frequent than for a Gaussian distribution.  The PDF of
the turbulent velocity field (projections and modulus), in turn,
remains Gaussian.  Moreover, the departure of the {\it velocity
  increments} PDFs from a Gaussian distribution increases as the lag
over which the increments are measured decreases.  All the functions
of the velocity involving a spatial derivative have therefore
non--Gaussian PDFs: the velocity gradients ($\partial_iv_i$) and shears
($\partial_jv_i$) and, accordingly, the rate--of--strain
$S_{ij}=\partial_jv_i +\partial_iv_j$ and the dissipation rate
$\epsilon_D = {\nu \over 2} \Sigma_{ij} (\partial_jv_i
+\partial_iv_j)^2$, with non--Gaussian wings more pronounced at small
scale.
 
The quantitative signature of intermittency appears in the behavior of
the high--order structure functions of the longitudinal velocity field
measured over a lag $l$, $\langle[\delta v_x(l)]^p\rangle \propto
l^{\zeta_p}$. This relation is statistical, not deterministic, and the
brackets hold for an average over an ``appropriate'', (\ie\ large
enough) volume with respect to $l^3$.  The anomalous scaling of the
exponents $\zeta_p \neq p/3$ characterizes the degree of intermittency
and provides the multifractal dimension of the most singular
structures, \ie\ that of the subset of space where the smallest active
regions, and turbulent dissipation, are distributed (Anselmet et
al. 2001).  In incompressible turbulence, the so--called {\it active
  small scales} are those of largest vorticity or velocity shear.

Various models have been proposed to explain the values 
of $\zeta_p$ (e.g. Frisch 1995). 
The most successful is certainly that  
of She \& Lev\^eque (1994) who infer the relation 
\begin{eqnarray}
\zeta_p = {-\gamma+1 \over 3} p + \gamma {1 - \beta^{p/3} \over 1 - \beta}, 
\label{she-lev}
\end{eqnarray}
where $\gamma$ is the exponent of the dissipation rate scaling in 
regions containing the most intermittent structures and 
$\beta$ is related to the codimension $C$ of the most intense  
dissipation structures as $\gamma / (1-\beta) = C$. 
It is now well established that this relation arises from
Poisson statistics (Dubrulle 1994, She \& Waymire 1995, Pan et al. 2008). 
The physical underlying idea is that the dissipative 
rates at two scales are related by a multiplicative factor
which is itself the result of dissipative events and 
defects (\ie\ lack of dissipative events). The latter follows Poisson statistics.
For incompressible hydrodynamical turbulence, the most dissipative 
structures tend to be filamentary so $C=3-1=2$ while 
$\gamma =2/3$  is assumed. 

Anticipating slightly on the next sections,  
Eq.~(\ref{she-lev}) has been used in the context of 
supersonic turbulence (Boldyrev et al. 2002) 
and magnetized turbulence (Politano \& Pouquet 1995, 
M\"uller \& Biskamp 2000). For instance, for supersonic
turbulence, it is expected that the dissipative structures
are sheets rather than filaments which leads to a 
codimension $C=1$ instead of 2 and therefore a parameter 
$\beta=1/3$.

\begin{figure}
 \includegraphics[width=8cm,angle=0]{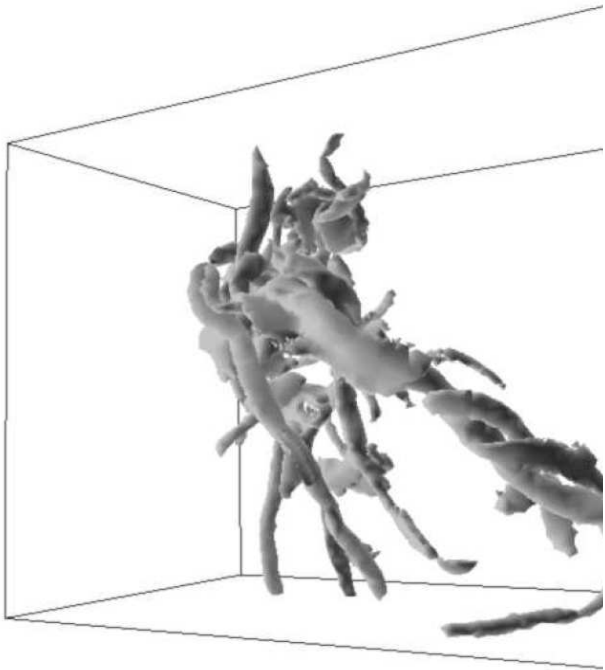} 
 \caption{Structure of most intense vorticity found in numerical
   simulations of 3-dimensional incompressible hydrodynamical
   turbulence. The pattern shown, consisting of a large tube
   surrounded by smaller ones wrapped around it, is frequently found
   (from Moisy \& Jim\'enez 2004).  In MHD
   turbulence, structures in which turbulent dissipation is
   concentrated appear to be more sheet-like (Politano \& Pouquet
   1995).}
 \label{moisy}
\end{figure}

Another essential facet of turbulent intermittency is that the active
small scales are not randomly distributed in space but are organized
into coherent structures -- {\it the sinews of turbulence}, as
qualified by Moffatt et al. (1994).  In incompressible turbulence, the
structures of largest vorticity tend to be filamentary as shown in
Fig. \ref{moisy}, while those of highest rate--of--strain and
dissipation are rather in the form of sheets or ribbons, following the
analysis of numerical simulations by Moisy \& Jim\'enez (2004).  These
structures are remarkable in the sense that they are both large scale,
\ie\ their length is comparable to the integral scale of turbulence,
and small scale structures, \ie\ they have substructure down to the
dissipation scale.  The coupling between large scales and small scales
operated by turbulence is repeatedly found in simulations. Mininni et
al. (2006) find that small--scale intermittency is more pronounced in
turbulent fields where the large--scale shear is larger.

\subsection{Incompressible magnetized turbulence}

While the simple Kolmogorov dimensional scaling relation, described
above, has proven to be very robust, MHD flows appear to be much more
difficult to understand.  Indeed, despite more than 35 years of
analytical, numerical and observational investigations, the energy
spectrum of MHD turbulence remains a subject of controversy.  The
first attempt to establish such a spectrum has been done by Iroshnikov
(1963) and Kraichnan (1965). In 
incompressible MHD turbulence, any function ${\bf v} \pm {\bf b}( {\bf
  r} \pm {\bf v}_A t)$ of ${\bf v}$ and ${\bf b}$, the velocity and 
magnetic field ${\bf B}$ fluctuations,  
is a solution of the MHD equations 
(here ${\bf v}_A= {\bf B}/\sqrt(4 \pi \rho)$ 
is the Alfv\'en velocity). This  implies
that Alfv\'en wave packets traveling in the same direction along the
magnetic field are not interacting. Thus, the interaction between
wavepackets occurs only between wavepackets moving in opposite
directions. The collision time $t$ being about $(kv_A)^{-1}$, the 
change in the velocity $\delta v_l$  is estimated from the  
non-linear advection term $\delta v_l/ t \simeq v_l^2 k $, so that one gets
the fractional change in the velocity, $\delta v_l / v_l$,  
proportional to $v _l / v_A$. 
Consequently, the number
of collisions, $N$, needed for a wavepacket to be significantly
modified is $N \simeq (v_A / v_l)^2$.  This is because the phase of
the wavepacket is changing randomly during a collision.  Thus the
cascade time is about $\tau \simeq N / (v_A  k)$.  Assuming that the
energy flux is the same at all scale, one gets $v_l^2 / \tau \simeq
\epsilon$, which leads to $v_l ^4 \propto k^{-1}$ from
which one infers  $v_l \propto
l^{1/4}$ and $E(k) = k^2 P_v(k) \propto k^{-3/2}$.  The energy spectrum
$E(k) \propto k^{-3/2}$ is thus slightly shallower than the Kolmogorov
one. An essential assumption of the Iroshnikov--Kraichnan approach is
that the eddies are isotropic, \ie\ have the same spatial extension in
the field--parallel and field--perpendicular directions.  However,
numerical and observational data accumulated for the last 30 years
indicate that in MHD turbulence the energy transfer occurs
predominantly in the direction perpendicular to the field (Biskamp
2003). This raises the question whether the picture proposed by
Iroshnikov and Kraichnan is grasping the essential physical
mechanisms.

An important progress has been performed by Goldreich \& Sridhar
(1995) who have developed a theory which takes into account the
anisotropy of the eddies in MHD. They suggested that as the energy
cascade proceeds to smaller scales, turbulent eddies progressively
become elongated along the large--scale field.  More precisely,
assuming that the Alfv\'en time--scale and the non-linear cascade
time-scale are in critical balance $k_z v_A \simeq v k_\perp$, and
that the cascade time in the perpendicular direction is still leading
to $v_\perp \propto k_\perp^{-1/3}$, they infer that the wave vector
along the z-axis is related to the wave vector $k_z \propto k_\perp
^{2/3}$.  As a consequence, they found that the energy transfer time
is different from the Iroshnikov--Kraichnan estimate, and identical to
the Kolmogorov one.  This leads  to a scaling for the
field-perpendicular energy spectrum, $E(k_\perp) \propto k_\perp
^{-5/3}$.  More recently, this issue has been investigated further in
various analytical and numerical studies (e.g. Cho et al. 2002,
Boldyrev 2005, Lee et al. 2010, Beresnyak 2011, Mason et al. 2012,
Wan et al. 2012) and still appears to be a matter of debate. Indeed,
even the question of the universality of the energy spectrum in
incompressible MHD turbulence remains unsolved.

\subsection{Supersonic turbulence}

In the ISM, turbulence is observed to be highly supersonic with
respect to the cold gas.  Given the complexity of the problem, it is
not surprising that very few results have been rigorously established
analytically (see nonetheless the recent work by Galtier \& Banerjee
2011 who attempt to obtain a rigorous expression for the two-point
correlation function) and most of our understanding comes from
numerical simulations performed during the last two decades.  The
biggest simulation performed to date is that by Kritsuk et al. (2007)
with an effective resolution of 2048$^3$ computing cells.  In the one
dimensional case (\ie\ Burgers turbulence, see Bec \& Frisch 2000 for
a review), one expects the power spectrum of the velocity $v$ to be
$P_v(k) \propto k^{-2}$. The reason is that a Fourier transform of an
Heavyside function, which is a good representation for shocks, is
proportional to $k ^{-1}$.
Thus, the energy spectrum is slightly stiffer than the Kolmogorov
spectrum. In  high resolution 3D numerical simulations, the power
spectrum of $v$ is $P_v(k) \propto k^{-3.95}$ (Kritsuk et
al. 2007). Thus, in 3D, the velocity power spectrum exponent $\alpha$
is bracketed by that of incompressible Kolmogorov turbulence
($\alpha$=11/3=3.67) and that of Burgers turbulence ($\alpha$=4) which
is fully compressible.  This is so because even highly supersonic
flows tend to have a large energy fraction in the solenoidal (or
incompressible) modes (e.g. Porter et al. 2002, Kritsuk et al. 2007,
Federrath et al. 2010), in particular when the flow is significantly
magnetized (Vestuto et al. 2003). 

An interesting issue, explored by Kritsuk et al. (2007) concerns  the
power spectrum of the density weighted velocity, $\rho^{1/3} v$. This
quantity stems from the fact that the energy flux, which for
incompressible turbulence is simply $\propto v_l^3/l$, becomes $\rho
v_l^3/l$ for compressible fluids. Thus $\rho v_l^3 / l \simeq
\epsilon$.  Interestingly, Kritsuk et al. (2007) find that the power
spectrum of this quantity has an exponent much closer to 11/3 than the
power spectrum of $v$. This raises the question as to whether the
ideas of Kolmogorov regarding the invariance of the energy transfer
rate, can be generalized and applied to compressible flows (see also
Galtier \& Banerjee 2011). Indeed, observations have provided
answers to these questions.  We revisit them in the next section.

\subsection{Revisited scaling laws and mass spectrum of molecular clouds}
\label{co_clumps}

Interstellar turbulence has been advocated a long time ago (von
Weizs$\ddot{\rm a}$cker 1951) on the  basis of estimated large
Reynolds number $Re>>10^7$ but formal proofs are hard to gather.  The
linewidth of a set of molecular clouds in the Galaxy was shown to
increase as a power of their size, with a spectral index, $p=0.38$,
reminiscent of the scaling of velocity fluctuations in incompressible
turbulence (Larson 1981).  This pioneering study has now been extended
to a variety of molecular clouds and tracers (Solomon et al. 1987,
Falgarone et al. 1992, Heyer \& Brunt 2004) and extended to external
galaxies.  This scaling has been often challenged because it is difficult
to recognize it on data samples that have too small a dynamic range.
In addition, the scaling laws derived from \twCO($J$=1--0) emission differ from
those inferred from other tracers (\thCO, CS, CN, \amm, ...).

\begin{figure}
\includegraphics[width=0.4\textwidth{},angle=-90]{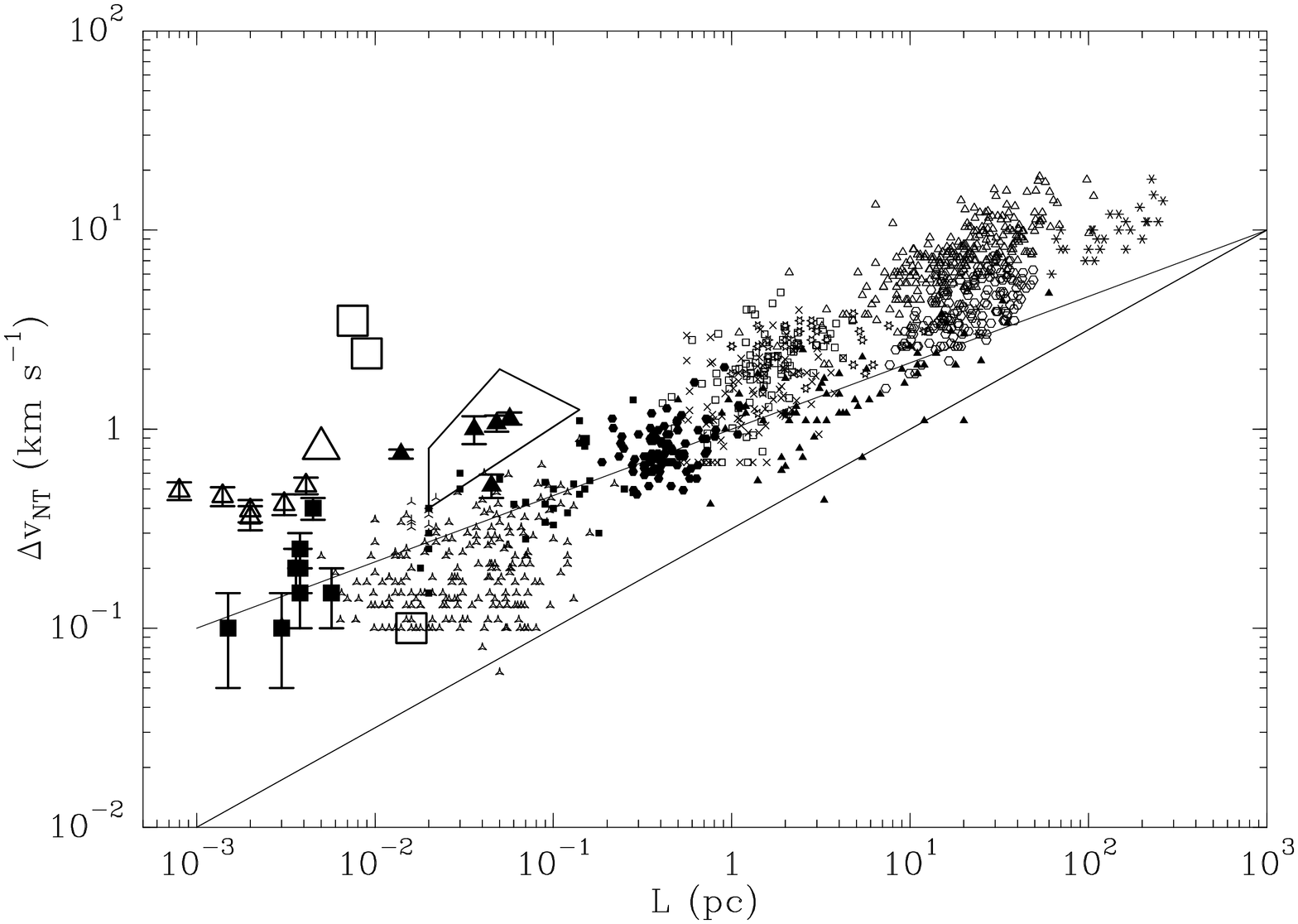}
\includegraphics[width=0.4\textwidth{},angle=-90]{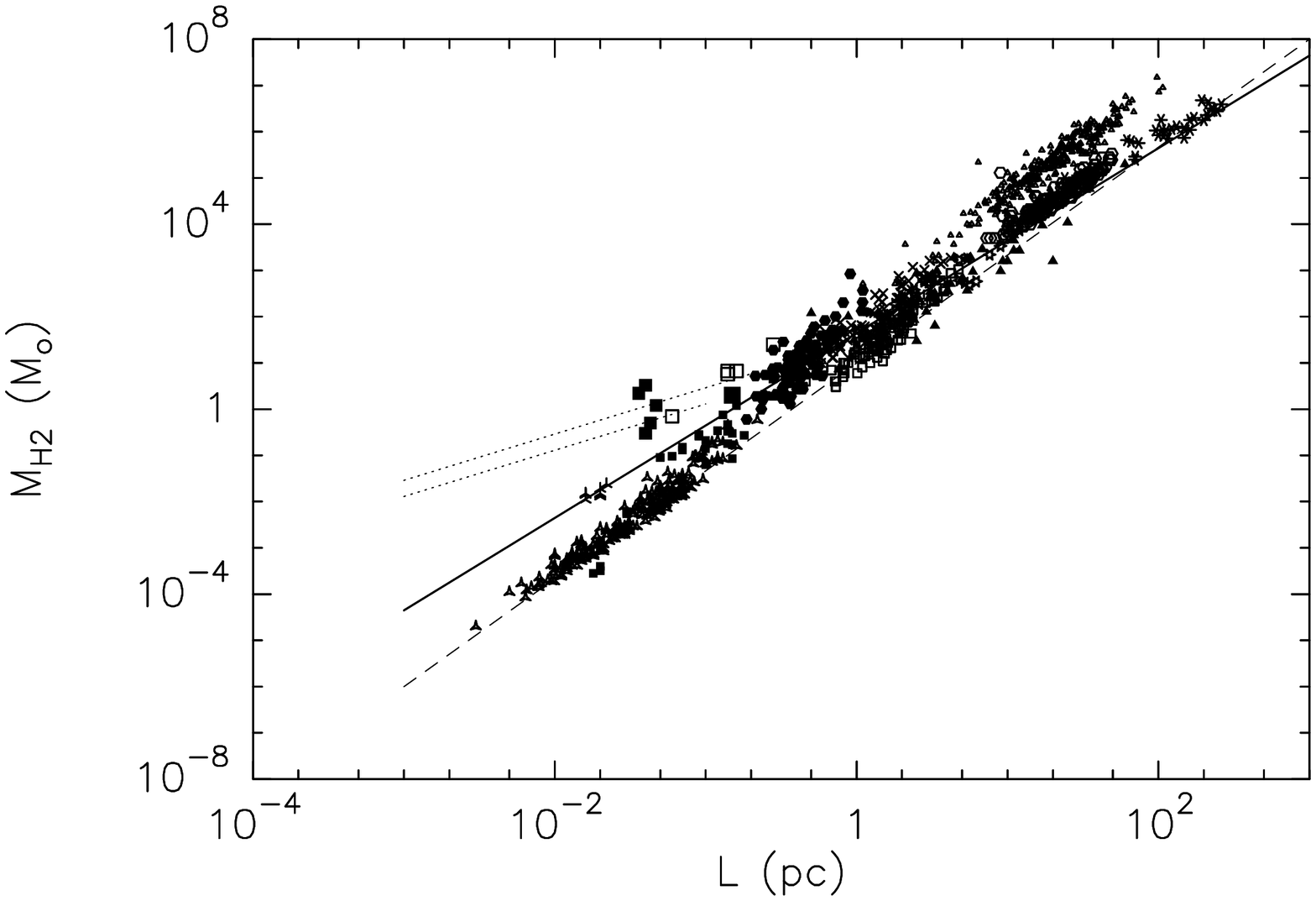}
\caption{Scalings of the masses and velocity dispersions with sizes of 
a large sample of molecular clouds.  The cloud properties are inferred from 
their \twCO\Jone\ line emission.
{\it(Left)} Non--thermal velocity dispersion versus size scale. 
The solid lines show the slopes $p=1/3$ and 1/2 of the scaling law  
$\Delta v_{NT} \propto L_{pc}^p$ (see references in Falgarone et al. 2009). 
{\it(Right)} Gas mass  
versus size scale. Different scalings,  
$\gamma=2$ (solid line), $\gamma=2.3$ (dashed line) defined by 
$M(L) \propto L^\gamma$ are shown. The scaling of isothermal self-gravitating polytropes
with size ($\gamma=1$) is
also shown  for two gas temperatures 10 K and 20 K (dotted lines).
The largest symbols (open and solid squares) refer to low-mass dense cores from the 
samples of Ladd et al. (1994) and Lada et al. (1997). }
\label{mr-vntr}
\end{figure}

CO structures are defined as connected structures identified in the
position--velocity space of \twCO\Jone\ maps.  Size, internal velocity
dispersion and column density (therefore mass) have been computed for
each of these structures.  Figure~\ref{mr-vntr} is an illustration of
the mass--size and linewidth--size scalings of such CO structures
(see references in Falgarone et al. 2004, 2009). The  method 
of structure extraction has long been criticized on
the justified basis that similar (or adjacent) projected velocities do not imply
actual spatial connection of the structures 
identified in position--velocity
space (Ostriker et al. 2001). However, a similar linewidth--size
scaling law has been obtained between 0.05 pc and 20 pc with a
different method (Ossenkopf \& Mac Low 2002).

Figure~\ref{mr-vntr} prompts four comments: {\it (i)} the
linewidth--size relation has quite a large dispersion (about a factor
of 10) about $\Delta v_{NT} \sim 1\kms\ L_{pc}^{0.5}$. {\it (ii)}
There is even a larger scatter (a factor of 50-100) below 0.1 pc: the
data points correspond to the small-scale structures of Heithausen
(2002, 2004, 2006), Sakamoto \& Sunada (2003) and Falgarone et
al. (2009).  {\it (iii)} When drawn over almost five orders of
magnitude in size scale, as in Fig.~\ref{mr-vntr}, the mass-size
scaling law cannot be restricted to one single scaling $M(L) \propto
L^\gamma$. Values between $\gamma= 2$ (solid line) at large scales and
$\gamma \sim 2.3$ (dashed line) at scales smaller than $\sim$ 0.5 pc
may be more relevant. Note that using clouds extracted from the
\thCO(1-0) Galaxy Ring Survey, Roman-Duval et al. (2010) find a single
(and very accurate) scaling with $\gamma=2.36 \pm 0.04$.  The range of
size scales involved is only 0.7 -- 30 pc, which shows the sensitivity
of such slope determinations to the sample dynamic coverage.  {\it
  (iv)} At scales smaller than $\sim 0.5$ pc, the mass in a given size
scale varies by orders of magnitude between self-gravitating dense
cores and much less massive structures of the same size.  Therefore
there is no such density--size scaling law $\rho \propto L^{-1}$ that
would be relevant to all small scale structures, as already noted by
Falgarone et al. (1992).

\begin{figure}
\includegraphics[width=0.4\textwidth{},angle=-90]{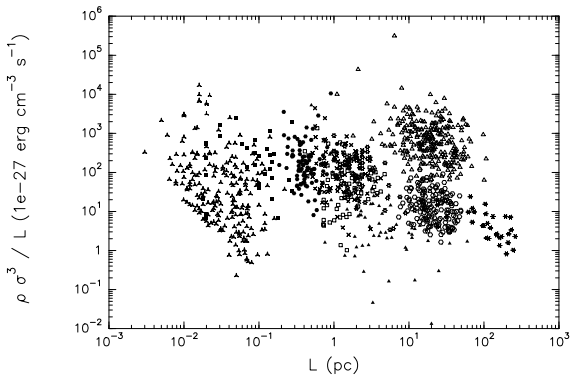}
\caption{Kinetic energy transfer rate $\rho\sigma^3/L$ versus size scale. Note that 
the value of this rate  inferred from HI in
the Solar Neighborhood is $ \sim 10^{-25}$ \eccs\ (see Table 1). }
\label{surf-eps}
\end{figure}
 
The linewidth--size (LWS) relation has often been used to argue that
molecular clouds are in virial balance between internal energy and
self-gravity, with their mass $M$, radius $R=L/2$ and 1-dimensional
internal velocity dispersion $\sigma$ related by $5 \sigma^2 = GM/R$,
where the geometric factor holds for spherical clouds (Bertoldi \&
McKee 1992).  In this virialized LWS relation, the coefficient between
the linewidth and the size depends on the cloud mass (or surface
density $\Sigma = M/\pi R^2$) so that $\sigma/R^{1/2} \propto
\Sigma^{1/2}$ (McKee, Li \& Klein, 2010).  Falgarone \& McKee (2013,
in preparation) show that molecular clouds traced by \twCO\ in the
Galaxy are virialized only above a surface density threshold $ \Sigma
\sim 100 \msol {\rm pc}^{-2}$.  The GMCs, as traced by CO, are
virialized {\it and} follow the same LWS relation as lower mass
clouds: below this threshold, the internal velocity dispersion (and
its large scatter) is independent of $\Sigma$, $\sigma\propto
R^{1/2}$, and this LWS relation may be interpreted as a manifestation
of the turbulent nature of the cloud internal motions.  In their
analysis of the Galactic Ring survey in \thCO(1--0) emission, Heyer et
al. (2009) find that $\sigma/R^{1/2} \propto \Sigma^{1/2}$, and that
the molecular clouds sampled by this survey are in virial balance. As
a consequence, they rule out the interpretation linked to turbulent
scaling laws. Since \thCO(1--0) line is more optically thin than the
\twCO(1-0) transition, the survey does not probe surface densities as
low a $\sim 1 \msol {\rm pc}^{-2}$, which \twCO\ emission does, and
the change of behavior at $\Sigma \sim 100 \msol {\rm pc}^{-2}$ has
been missed because of the intrinsic scatter of the data points.  The
same limitation appears to apply to the discussion of Field et
al. (2011) that relies on this data set.

The two different LWS relations may be reconciled with a more general
 expression of the virial theorem that takes surface terms (or an
 external pressure) into account:
\begin{equation}
\ddot I= 2M {\cal E } - \gamma GM^2/R - 4 \pi P_{ext} R^3 = 0,
\end{equation}
where ${\cal E } \propto \sigma^2$ is the specific internal 
energy due to turbulence and $\gamma \sim 1$
depends on the degree of the gas concentration at each scale.  Keto \&
Myers (1986) had already found that diffuse clouds are not
gravitationally bound and interpreted the departure of
$\sigma/R^{1/2}$ from the $\Sigma^{1/2}$ dependence as the cloud being in
quasi-static equilibrium with an intercloud medium exerting a
confining pressure $P_{ext}/k \approx 10^{3.5}$-$10^{4.5}$ K \cc. From
the above virial balance equation, in the regime of non-gravitationally bound 
clouds, $\sigma/R^{1/2}$ scales as
$\Sigma^{-1/2}$, a behavior absent in the \twCO(1-0) data set. 

The independence of $\sigma^2/R$ on $\Sigma$ is obtained if
$P_{ext}$ and $\Sigma$ have the same scaling with size. Indeed, this is
what is found in the above \twCO(1-0) cloud sample where
 the kinetic pressure increases with scale as
${1 \over 2}\rho \sigma^2 \sim 10^{-11} R_{pc}^{\delta}$ erg \cc, with
$\delta \sim$ 0.25 -- 0.5. If this non uniform pressure acts as a
confining pressure, then at each scale, $P_{ext} \propto R^{\delta}$ 
which is close to the scaling  
$\Sigma \propto R^{0.3}$ inferred from the observed $M \propto R^{2.3}$.
Thus, we would have $P_{ext} / \Sigma$ being nearly constant
as suggested by $\sigma \propto R^{1/2}$. The exact meaning of 
this relation remains however a little elusive since $\sigma^2$ 
certainly contributes to the internal pressure as well.

This scenario is reminiscent of the
hierarchy of clouds in dynamical equilibrium at all scales, with
$P_{ext}=\Pi(R)$, as proposed by Chi\`eze (1987). However, one has to 
understand why the 
turbulent pressure, at each scale, would act as an {\it isotropic}
stabilizing factor (as it is in the case of isothermal
self-gravitation polytropes bound by an external thermal pressure).
Chi\`eze (1987) ended his paper by
asking what is the physical nature of the process which tends to
maintain a constant energy density in molecular matter over a wide
range of linear scales. We now know that $\Pi(R)$ is not exactly constant,
and that the process that controls the energy density across 
the scales is called turbulence.

These  results shed light on the question ``What shapes the
structure of molecular clouds: turbulence or gravity?'' (Field et
al. 2008, 2011, Kritsuk \& Norman 2011). As these latter authors
show, the scaling laws can, to a large extent, be interpreted as a
signature of supersonic turbulence. The above discussion  suggests however 
that gravity certainly modifies the picture.
 The gas surface density threshold above which clouds are in
virial balance between self-gravity and internal energy is comparable to
the  stellar surface density within 1 kpc of the Sun, $\sim 90
\msol {\rm pc}^{-2}$, and one order of magnitude larger than that of
the gas, on average, $ \Sigma_{gas} \sim 10 \msol {\rm pc}^{-2}$. In
their numerical simulations, Koyama \& Ostriker (2009) find that the
mass-weighted midplane mean pressure is about one order of magnitude
larger than that of the midplane mean gas pressure. They argue that
self-gravity concentrates gas and increases the pressure within
molecular clouds without raising the ambient pressure. They also show
that the correlation of the molecular gas fraction with the
midplane pressure observed  in external galaxies by Blitz \& Rosolowsky (2004,
2006) is retrieved only if the epicyclic frequency $\kappa$ and
$\Sigma$ are kept proportional.  
This fact introduces  galactic rotation as an additional 
parameters driving the evolution of molecular clouds, a picture which 
has  started to emerge from recent observations 
(Swinbank et al. 2010, Herrera et al. 2011, 2012).

Last, the kinetic energy transfer rate $2\epsilon= \rho \sigma^3/L$ is
displayed as a function of the size in Fig. \ref{surf-eps}. It is
remarkable that the kinetic energy transfer rate in the whole
population of molecular clouds traced by \twCO\ shows no trend of
variation from structures of $\simeq$0.01 pc to GMCs of $\simeq$ 100
pc (the same large scatter of $\epsilon$ is observed at all scales)
and has the value observed in the HI gas (see Table 1).  Although the
dispersion of $\epsilon$ is large at each scale, this result suggests
that $\epsilon$ is indeed an invariant of the hierarchy of {\it
  molecular clouds traced by \twCO(1-0)} and that these clouds are
part of the same turbulent cascade as the atomic ISM. This conclusion
is in remarkable agreement with Kritsuk et al. (2007), generalizing
the ideas of Kolmogorov to compressible turbulence.
 
\begin{table}
\caption[]{Characteristics of the turbulence observed in 
various components of the ISM (Solar Neighborhood). 
$\epsilon$ is expressed in
L$_\odot$/M$_\odot$ for comparison with the energy provided by stellar
radiation, 
${1 \over 2} {\overline \rho} v_l^3/l$ in erg cm$^{-3}$ s$^{-1}$
and $P_{turb}$ in erg cm$^{-3}$.}
\begin{tabular}{cccc}
\hline
  &  CNM+WNM & molecular clouds & low-mass dense cores \\
       \hline
 ${\overline n}${\small (cm$^{-3}$)} & 30 & 200 & 10$^4$ \\
 $l$ {\small (pc)} & 10 & 3 & 0.1 \\
$\sigma_l$ {\small (km s$^{-1}$)} &$\approx$ 3.5 & 1 & 0.1 \\
 $B$ {\small ($\mu$G)} & 10 & 20 & 100 \\
$v_A$  {\small (km s$^{-1}$)} & 3.4 & 2 & 1.4 \\
$\epsilon={1 \over 2} v_l^3/l$  &  10$^{-3}$ & 10$^{-4}$ & 10$^{-6}$ \\
${1 \over 2} {\overline \rho} v_l^3/l$ &
2$\times 10^{-25}$ &  1.7$\times 10^{-25}$ &  2.5$\times 10^{-25}$ \\
$P_{turb}={1 \over 3} {\overline \rho} v_l^2$   & 3$\times 10^{-11}$ &
$2\times 10^{-11}$ & $10^{-11}$ \\ 
        \hline
      \end{tabular}
  \end{table}

The same methods used to isolate connected CO structures in 
space-velocity space allow the construction of mass spectra, $dN/dM \propto
M^{-\alpha}$.  Different samples provide very close exponents: 
$\alpha= 1.83$ for $10^3
{\rm M}_\odot <M< 2 \times 10^6 {\rm M}_\odot$ in the early CO survey of
the inner galactic plane (Solomon et al. 1987), $\alpha= 1.80$ for
$500 {\rm M}_\odot <M< 10^6 {\rm M}_\odot$ from the CO survey of the
outer galaxy (Heyer et al. 2001) and $\alpha= 1.84$ over $10^{-3} {\rm
  M}_\odot <M< 50 {\rm M}_\odot$ in a high latitude cloud (Heithausen
et al. 1998). The slopes of the mass spectra {\it of structures identified from CO lines} 
are therefore found to be
the same over nine orders of magnitude in masses, in the inner and outer
Galaxy, in GMCs and cirrus clouds, in star forming and in inactive
regions.  Similar values, with larger error bars, have been found in a
number of smaller samples covering the mass range 10 to 10$^3$
M$_\odot$. 

\subsection{Source of turbulent energy in molecular clouds}
A longstanding issue is the origin of the observed 
turbulent velocity dispersion of molecular clouds.
Indeed, turbulent supersonic motions are  expected to
rapidly dissipate in shocks within a cloud crossing time, $\approx$
1~Myr, for clouds of a few pc and internal velocity dispersion
of a few \kms. This has been largely confirmed by 
numerical simulations (e.g. MacLow 1999, MacLow \& Klessen 2004)
which have demonstrated, despite early speculations, that 
the magnetic field is not changing this result significantly, \ie\ 
in a few crossing times most of the energy has decayed. 
Since turbulent energy is decaying fast and that molecular 
clouds remain turbulent, turbulence has to be   
continuously replenished. The question is how.

Again, observations are providing clues to this question.
First, turbulence in molecular clouds that are not
forming stars, as for instance the Maddalena cloud, is
comparable to that in more actively star forming
clouds (Williams et al. 1994). This is also found by Kawamura et al. (2009)
in the LMC: the velocity dispersions they report for
 two categories (\ie\ clouds without and with massive stars)
are very similar (\eg\ their Table 4).  Second, there is no evidence
for a characteristic scale in the velocity field. Ossenkopf \& MacLow
(2002) using various methods conclude that it is consistent with a
forcing at a scale larger or equal to the size of the clouds they study. 
This  is also found by Brunt et al. (2009), who confront
synthetic molecular lines to real observations.  This suggests 
that, at least for a large fraction of clouds,
the turbulent driving is external, \ie\  most of the energy
is injected from outside.

Klessen \& Hennebelle (2010) argue that turbulence is driven 
by accretion onto the molecular clouds. Indeed when a piece of fluid 
falls onto the cloud, it carries a certain amount of kinetic energy 
which can be used to sustain turbulence in the molecular cloud. 
The question is then whether the energy injection rate,  
$\dot{E}_{in} \simeq
S_c \times v_{in} \times {1 \over 2} \rho_{in} v_{in}^2$, where $S_c$ is 
the cloud surface, $v_{in}$ is the  infall velocity of the 
gas  and $\rho_{in}$  its density, can 
compensate for the turbulent dissipation rate $\dot{E}_{dis} \simeq 
{1 \over 2} M_c \sigma_c^2 / \tau_c \simeq {1 \over 2 \sqrt{3}} M_c \sigma_c^3 / L_c$
 where $M_c$, $L_c$ and $\sigma_c$ are the cloud mass, size and velocity
dispersion.  The difficulty to perform 
such an estimate is obviously the lack of empirical knowledge on the 
accretion rate. From their data Fukui et al. (2009)  infer
a rate on the order of a few $10^{-2}$ M$_\odot$ yr$^{-1}$
for typical GMCs, which is enough to reproduce the 
observed velocity dispersion according to Klessen \& Hennebelle (2010). 
A direct estimate using 
colliding flow simulations leads to a similar conclusion.  
Goldbaum et al. (2011) have performed a time-dependent analytical 
calculation of a molecular cloud evolution and conclude
 that turbulence can  efficiently be driven by accretion.
The accretion-driven turbulence model
implies that energy is injected in the ISM at large scales 
in the Galaxy, presumably by  a combination of 
supernovae/bubble expansion and galactic differential rotation, 
and {\it cascades} to smaller scales as the gas accumulates into 
the GMC. This mechanism emphasizes the 
link between molecular clouds and their environment
as described earlier in the review. 

When enough stars have formed inside molecular clouds, stellar feedback
presumably becomes important.  Matzner (2002) computes the impact of
stellar winds, supernovae and HII regions onto cloud turbulence and find
that HII regions are the most efficient to
sustain turbulence. Following Williams \& McKee (1997), Matzner
concludes that the photo-evaporation induced by the HII regions
eventually destroys the cloud.  Numerical simulations of the impact of
the ionizing radiation from O stars have been performed by
Gritschneder et al. (2009, see also Tremblin et al. 2012). 
They find that it can induce the formation
of pillars of dense gas and inject turbulence within molecular clouds.
 The inclusion of the feedback from protostellar jets
has been considered by Nakamura \& Li (2007) and  Wang
et al. (2010).  These simulations suggest that protostellar jets can
efficiently trigger turbulence at the scale of a stellar cluster 
(say at a scale of about 1 pc) 
once enough stars have formed.  On the other hand, Maury et al. (2010)
estimate the support that the observed protostellar outflows are 
providing to the
star forming clump NGC2264-c and find that it is not sufficient to
resist gravitational collapse.

These qualitative estimates show how uncertain the energy injection
budget within molecular clouds is. One does not know either when the
internal feedback becomes dominant over the external feeding. In their
analytical model, Goldbaum et al. (2011) estimate that the total
contribution of external and internal sources is roughly comparable
over the cloud life time but their model is hampered by large uncertainties.

\subsection{Signatures of the intermittency of turbulence 
in molecular clouds}

\subsubsection{Parsec-scale coherent structures of intense velocity-shear}
\label{shear-fil}
Identifying regions of intermittency in interstellar turbulence is
challenging for several independent reasons.  First, these regions
are non-space filling and correspond to rare events in time and space: 
 finding them requires the analysis of large homogeneous statistical
samples of the velocity field.  This means observations at high
spectral and spatial resolution of large interstellar regions unperturbed 
by star--formation. 
Second, observations do not provide the full velocity field but only
its line-of-sight ({\it los}) projection provided by the Doppler-shift
of a molecular line. Then, statistical analyses similar to those
performed in laboratory flows or in the solar wind are not
possible. 
Third, only spatial variations of the {\it los} velocity in
the plane-of-the-sky ({\it pos}) are provided by observations.
Therefore, the velocity variations are by essence cross-variations
$\partial_jv_i$, \ie\ velocity shears.  Finally, the line emission being integrated along a
{\it los}, the velocity information at a given position is the full line
profile and its moments, the first moment being the line centroid
velocity. The statistics of the velocity increments are built 
using the line centroid velocity increments (CVI) measured between
two positions separated by a lag $l$ in the {\it pos} (e.g. Lis et al. 1996). 
The \twCO\ lines
have turned out to be a most useful tool for this search
because their large optical depth makes them sensitive tracers of rare events, 
in particular of the gas that emits at velocities in the far line--wings
(\ie\ extreme dynamic events).

\begin{figure}
\begin{minipage}{7cm}
  \includegraphics[width=0.6\textwidth{},angle=-90]{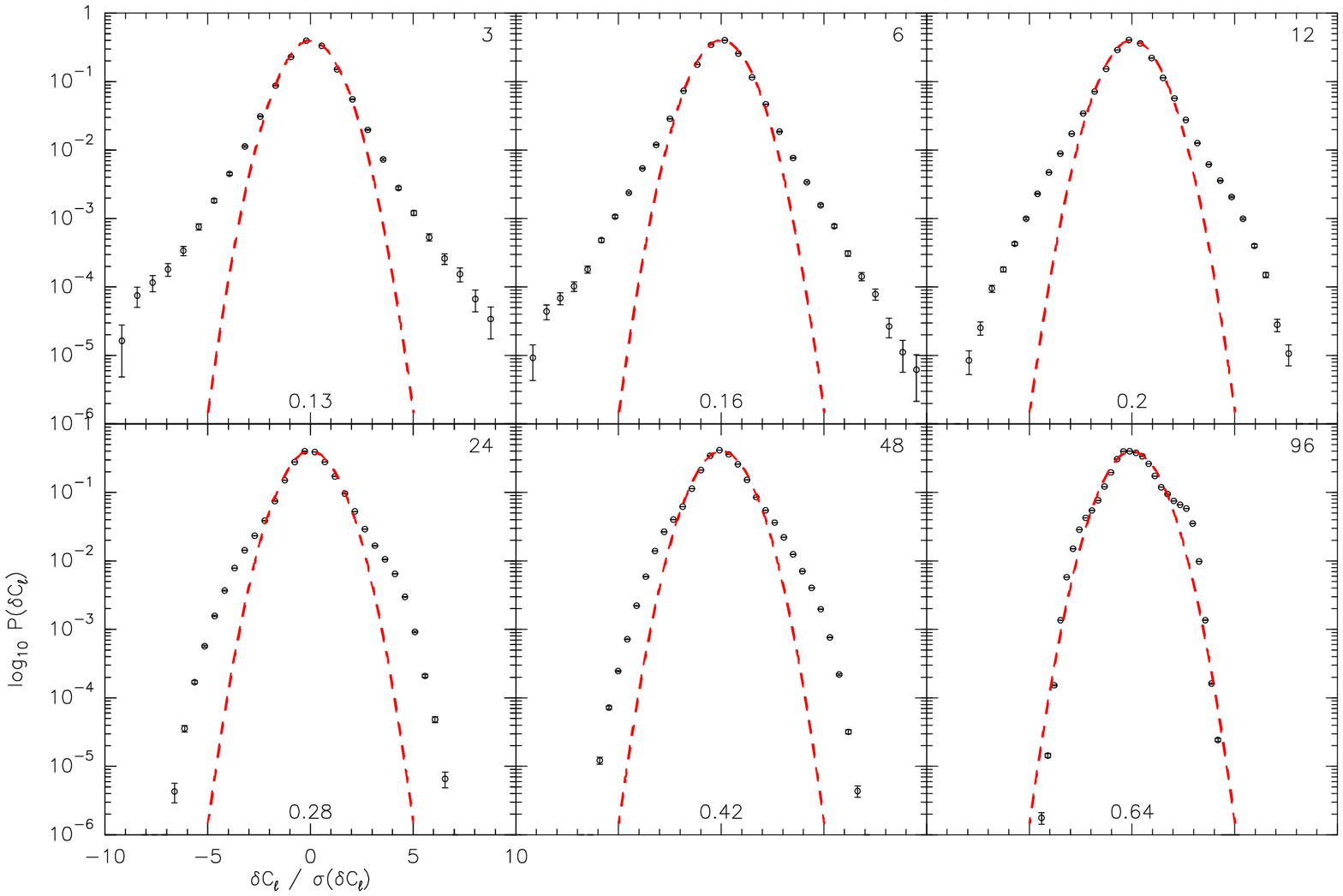}
\end{minipage}
\begin{minipage}{7cm}
  \includegraphics[width=\textwidth{},angle=0]{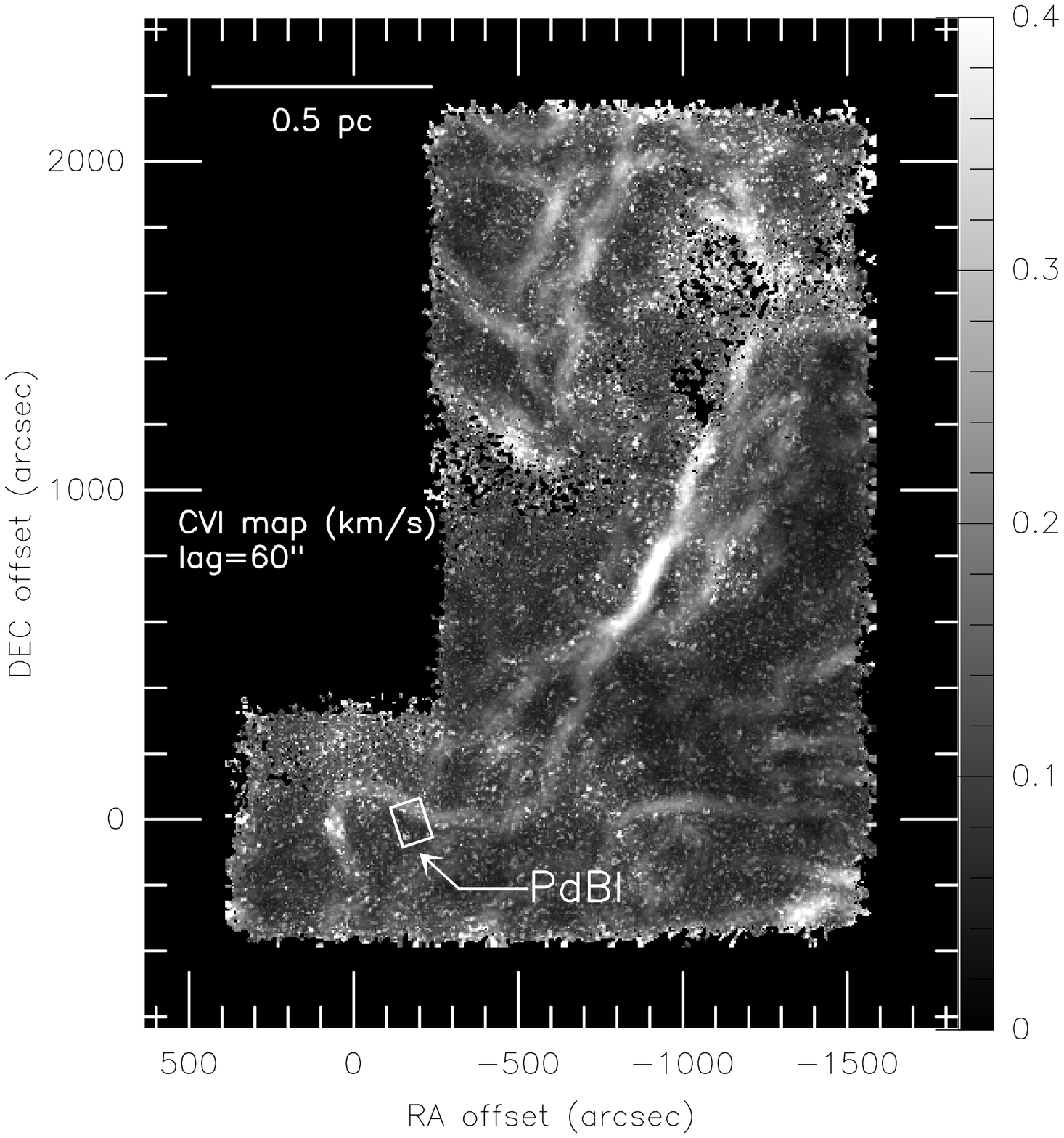}
\end{minipage}
  \caption{ {\bf Left:} Normalized PDFs of line centroid velocity increments (CVI) 
measured over variable lags, expressed in units of 15 arcsec (upper right corners), and 
computed within the field of Fig. 6.
The Gaussian of the same dispersion $\sigma(\delta C_l)$ 
(given in \kms\ at the bottom of each panel) 
are also drawn. The non-Gaussian wings of the PDFs increase as the lag decreases. 
{\bf Right:} In the same field, locus of the positions 
populating the non-Gaussian wings of 
the PDF for a lag of 60 arcsec. 
The wedge is in \kms. 
The rectangle is the area observed with the IRAM-PdBI
(From Hily-Blant et al. 2008).}
  \label{pdf-cvi}
\end{figure}

The most recent maps carried out at the IRAM-30m telescope in the
\twCO(2-1) line, produce up to $\sim 10^5$ independent spectra
 homogeneously sampling turbulence in the diffuse molecular gas 
(Hily-Blant \& Falgarone  2009).  
The CVI-PDFs in that field have the anticipated
non-Gaussian wings that increase as the lag over which the increment
is measured decreases (Fig.~\ref{pdf-cvi}, left).  The locus of
the extreme CVIs (the E-CVIs) that contribute the non-Gaussian wings of
the PDFs is an ensemble of elongated narrow structures ($\sim 0.03$~pc
thick) (Fig.~\ref{pdf-cvi},
right).  As expected, the lane of largest E-CVIs coincides with the
region where the velocity-shears are the largest, 40~\kmspc\ 
(Hily-Blant \& Falgarone 2009). Most interestingly, it also coincides 
with a lane of weak
\twCO(2-1) emission and one of the weakest filaments of dust thermal
emission detected at $250 \mu$m in that field with Herschel/SPIRE
(see Section 4). Last, two
low-mass dense cores in the field lie at the South-East tip of the
E-CVIs locus (Heithausen et al. 2002), suggesting a {\it causal link}
between intense turbulent dissipation, the formation of
CO molecule, the formation of tenuous dense filaments and, eventually, that of
low-mass dense cores.  

A similar analysis has been performed in a less turbulent field 
(specific kinetic energy four times smaller at the parsec-scale)
with the same
total hydrogen column density as that in the Polaris Flare.
The CVI-PDFs show departures from a Gaussian
distribution with an amplitude 2.5 times smaller than in the Polaris
Flare (Hily-Blant et al. 2008), 
in agreement with the theoretical predictions of Mininni et al. (2006), 
\ie\ the enstrophy density
at small scales increases with the amplitude 
of the large-scale shear feeding turbulence.

This ensemble of properties, {\it (i)} the increasing departure of 
CVI-PDFs from a Gaussian distribution as the lag decreases; {\it (ii)} 
the spatial 
coherence of E-CVIs structures and {\it (iii)} the link between the large-scale
properties of turbulence and the magnitude of the small-scale
E-CVIs, suggests that the \twCO(2-1) E-CVIs  trace the intermittency of
turbulence in diffuse molecular clouds.  
Intermittency is also 
probed by the non-linear dependence of $\zeta_p$
with $p$ up to $p=6$ discussed in Hily-Blant et al. (2008), 
a dependence now confirmed, up to higher orders, 
by the analysis of this larger data set. Unexpectedly (because these exponents are computed 
on projected velocities), the $\zeta_p$ 
dependence with $p$ in that field agrees, within the error bars, with 
the She \& Lev\^eque (1994) predictions for incompressible turbulence.

\subsubsection{Milliparsec-scale observations: approaching the dissipation scales}

A step further towards small-scales is provided by the IRAM Plateau de
Bure Interferometer (IRAM-PdBI) \twCO(1-0) line observations of the
field shown in Fig.~\ref{pdf-cvi} (right) at a resolution of $\sim 4$
arcsec or 3 milli-pc (Falgarone et al. 2009).  These observations are unique
so far because the lines detected are very weak. The spatial dynamic
range of the map is large enough to allow the detection of 8 elongated
structures with thickness as small as $\approx$ 3 milli-pc (600 AU) and
length up to 70 milli-pc.
These are {\it not filaments} 
but appear to be the sharp edges of extended CO emission.  Six, out of
eight filaments, form pairs of quasi-parallel structures at different
velocities.
Velocity-shears estimated for the three pairs 
include the largest values ever measured in non-star-forming regions,
up to 780 \kmspc.
Finally, the PdBI-structures are almost straight and their different position angles 
cover the same broad range of values as the projection of the magnetic
field inferred from absorption towards field stars in that area. This
suggests an interesting alignment of the {\it pos} projections of
magnetic fields and velocity shear-layers in these structures.  In
Section 6, we discuss the new routes that the intermittency of
interstellar turbulence opens for molecule formation.

\subsection{Anisotropy of turbulence}
The presence of large-scale pervasive magnetic fields in the galactic ISM 
is anticipated to induce anisotropy in the turbulence of molecular clouds.
Such an anisotropy has been sought in a field of the Taurus
molecular cloud where polarization of starlight reveals aligned and
ordered magnetic fields, associated with strong striations parallel to
the field in the \twCO(1-0) emission (Heyer et al. 2008).  These authors
 develop
an axis-constrained Principal Component Analysis method
that has been tested against
numerical simulations of MHD turbulence. In the real data, in spite of
projection effects that would tend to reduce it, Heyer et al. find an
anisotropy: there is  more power at 
small scales in the  spectrum of velocity fluctuations in the
direction perpendicular to the striations rather than along them.
Interestingly, Heyer \& Brunt (2012) have extended this analysis to the whole 
Taurus molecular cloud finding that the velocity anisotropy is limited
to the cloud edges and absent in the opaque regions of large column density.
The authors argue that these results, compared to MHD simulations,  suggest 
sub-Alfv\'enic turbulence in the edges and opposite in the opaque regions.

\section{Density structure}
In the context of structure formation, the
density distribution within molecular clouds
must be regarded as a fundamental quantity. In this section 
we review our knowledge and understanding of the origin 
 of the density distribution.

\subsection{The density distribution within molecular clouds}
A number of processes can shape the density field of a compressible 
flow. In the ISM compressible 
turbulence and gravity are likely to be largely responsible for the 
gas density distribution. Other important aspects are the equation 
of state that describes how pressure and density are related as well as
the forcing of the turbulence, \ie\ the way by which 
the turbulent energy is maintained, which  determines
the relative importance of solenoidal and compressive modes
and not surprisingly turns out to have a significant impact
on the density distribution. The dissipation of turbulence 
may also drive density fluctuations by triggering chemistry, 
modifying in turn the cooling rates and the equation of state (see Section 6).

\subsubsection{The role of supersonic isothermal turbulence}
As isothermal supersonic turbulence constitutes a simplified
and natural framework, most of the studies have been performed
under this assumption.
Since the pioneering works of V\'azquez-Semadeni (1994, see also
Blaisdell et al. 1993) and 
Padoan et al. (1997),  various simulations  of hydrodynamic supersonic 
turbulence, have established that the  density PDF is well 
represented  by a log-normal form (e.g. Kritsuk et al. 2007, 
Federrath et al. 2008, 2010),
\begin{eqnarray}
\label{Pr0}
{\cal P}(\delta) &=& {1 \over \sqrt{2 \pi \sigma_0^2}} 
\exp\left(- { (\delta - \bar{\delta})^2 \over 2 \sigma_0 ^2} \right) , \;
 \delta = \ln (\rho/ \rho_0 ), \\
 \bar{\delta}&=&-\sigma_0^2/2 \;
,  \; \sigma_0^2=\ln (1 + b^2 {\cal M}^2),
\nonumber
\end{eqnarray}
where ${\cal M}$ is the Mach number,  $\rho_0$ is the mean density
and $b \simeq 0.5-1$. 

An exact derivation of the density PDF is still pending but
qualitatively, at least, the origin of this distribution can be
understood as a consequence of the fluid particles being repetitively
and randomly compressed by the turbulent velocity fluctuations. The
key is that density enhancement is a multiplicative process. To
illustrate this better, let us consider a fluid particle of density
$\rho_0$ which is compressed by a first shock at mach number ${\cal
  M}_1$.  The density is then multiplied by ${\cal M}_1^2$ and becomes
$\rho_1 = \rho_0 \times {\cal M}_1^2$. Then the same fluid particle is
compressed by another shock which brings its density to $\rho_2 =
\rho_1 \times {\cal M}_2^2 = \rho_0 \times {\cal M}_1^2 {\cal M}_2^2$,
hence the density is {\it multiplied} each time a new shock
is passing. A complementary point of view is obtained by considering
the quantity $\ln ({\cal M}_i^2)$ which is {\it added} to the
logarithm of the density. By virtue of the central limit theorem, one
can expect that the distribution of $\ln(\rho)$ must be a Gaussian and
therefore that the density must have a log-normal distribution. This
argument is the essence of the explanation provided by Kevlahan \&
Pudritz (2009) and constitutes a nice qualitative
explanation. However, it is no more than qualitative for various
reasons.  First, as shown in Kevlahan \& Pudritz (2009), the result
depends on the number of shocks which implies that the average number
of shocks should be determined.  Moreover, this approach does not take
into account the reexpansion of the compressed layer between the
shocks.  Second, as demonstrated by Federrath et al. (2008, 2010), the
density PDF is really close to a log-normal only when the forcing is
solenoidal, i.e. when the force used to stir the turbulence
contains only incompressible modes. When this is not the case,
Federrath et al. (2008) show that the density PDF presents significant
deviation from a log-normal distribution, in particular at high
density.

Beyond the exact shape of the PDF, the dependences of $\bar{\delta}$
and $\sigma_0$ given in Eq.~(\ref{Pr0}) are also interesting.  The
expression of the former is a simple consequence of mass conservation,
$\int _{-\infty} ^{\infty} \exp(\delta) {\cal P} (\delta) d \delta =
1$ from which it is easy to show that $\bar{\delta} = \int _{-\infty}
^{\infty} \delta {\cal P} (\delta) d \delta = -\sigma_0^2 / 2$.  The
dependence of $\sigma_0$ on ${\cal M}$ is more empirical.  The
underlying physical idea (Padoan et al. 1997) is simply that the
density fluctuation induced by a shock is $\delta \rho / \rho_0 =
(\rho-\rho_0) / \rho_0 \simeq b^2{\cal M}^2$, while the variance of
the density field, $\sigma_\rho$ is by definition: $\sigma _\rho^2 =
\int (\rho-\rho_0)^2 {\cal P} (\rho) d \rho$.  However, the high
density regions are compressed and occupy a volume fraction $\propto
{\cal M}^{-2}$. Thus from a spatial integration of the density
variance, $\sigma _\rho^2 = (\int (\rho-\rho_0)^2 dV) / \bar{V}$, it
is inferred that $\sigma _\rho^2 \propto {\cal M}^4 / {\cal M}^2$.
Thus, $\sigma_ \rho$ has a linear dependence on the mach number,
$\sigma_ \rho \simeq b {\cal M}$.  Next, as pointed out by Federrath
et al. (2008), with a density probability distribution as stated by
Eq.~(\ref{Pr0}), it is an easy task to show that 
$\sigma_0^2 = \ln(1+\sigma_\rho ^2/\rho_0^2)$, by simply writing $\sigma _\rho^2 = \rho_0^2 \int
(\exp(\delta)-1)^2 {\cal P} (\delta) d \delta$.

Federrath et al. (2008) propose an heuristic model for the parameter
$b=1-2 \zeta / 3$, where $\zeta$ controls the importance of solenoidal
and compressible modes. For $\zeta=1$, the modes are purely solenoidal
while they are purely compressible for $\zeta=0$.  For $\zeta=0.5$,
the power in compressible modes is half of the power in solenoidal
modes which corresponds to energy equipartition between the modes (see
Federrath et al. 2010, Eq.~9). A comparison between a set of numerical
simulations and this approach shows a good agreement.  As the Mach
number increases, the width of the PDF increases. This is because the
shocks are stronger and generate larger density contrasts.

\subsubsection{Influence of non-isothermal equation of state}
When the gas is non-isothermal, the density PDF is not log-normal
anymore. Passot \& V\'azquez-Semadeni (1998) performed 1D simulations
for various values of $\gamma$.  In particular, they found that for
$\gamma<1$ at high densities, the distribution becomes a power-law
whose slope index varies with $\gamma$.  For $\gamma=0.5$, which is
not too far from the value $\gamma=0.7$ typical for densities of the
order of $10^2-10^4$ cm$^{-3}$, they infer an index of about -1.2.
Audit \& Hennebelle (2010) have presented 3D high resolution
simulations of colliding flows in the isothermal case and also in the
polytropic case, $\gamma=0.7$. They confirm the power-law behaviour
found in 1D by Passot \& V\'azquez-Semadeni (1998).

The two-phase case has also been investigated by various teams
(e.g. V\'azquez-Semadeni et al. 2006, Gazol et al. 2005, Audit \&
Hennebelle 2010).  It has generally been found that the density PDF
presents two peaks, one at low density (a few particle per cm$^{-3}$)
and one at high density (a few hundreds of particle per cm$^{-3}$).
The thermally unstable region which is continuously filled by the
turbulent motions comprises relatively less mass than in the vicinity of the two
branches of equilibria.  Note that it appears particularly difficult
to obtain numerical convergence at high densities (10$^3$ cm$^{-3}$
and above) for this type of simulations. For example, Hennebelle \&
Audit (2007) warn that numerical convergence was probably not reached
even in their bidimensional $10^4 \times 10^4$ cell simulations.

The influence of the magnetic field on the density PDF has also been
investigated by various authors both in the isothermal case (Ostriker
et al. 2001, Lemaster \& Stone 2008, Molina et al. 2012) and in the
two--phase case (e.g. Hennebelle et al. 2008) generally concluding that
it has a weak impact.  As discussed below, this conclusion accords
well with the fact that the gas which is not self-gravitating
essentially flows along the field lines.

\subsubsection{The role of gravity}
As gravity leads to collapse it is clear that it must have a strong
impact on the density PDF. Numerical simulations indeed have revealed
that as gravity takes over, a high density tail develops precisely
because some gas is collapsing (e.g. Klessen 2000, Slyz et al. 2005,
Hennebelle et al. 2008, Kritsuk et al. 2011).  Moreover it has been
found that this tail is typically a power-law with an exponent of
about -3/2 (Kritsuk et al. 2011).  The latter is interpreted as a
consequence of the $\rho(r) \propto r ^{-2}$ profile developed by a
collapsing sphere (e.g. Shu 1977). Indeed, the volume of the gas whose
density lies between $\rho$ and $\rho + \delta \rho$, is $4 \pi r^2
dr$.  The exponent simply follows from $r \propto \rho^{-1/2}$, $dV
\propto \rho^{-3/2} d \log \rho $.

This gravitational tail of the density PDF develops
with time. It first affects only the highest density 
part of the PDF and then becomes  gradually important at
smaller densities as collapse  proceeds. 

\subsubsection{Observed column density distribution}

\begin{figure}
 \includegraphics[angle=270,width=0.7\textwidth]{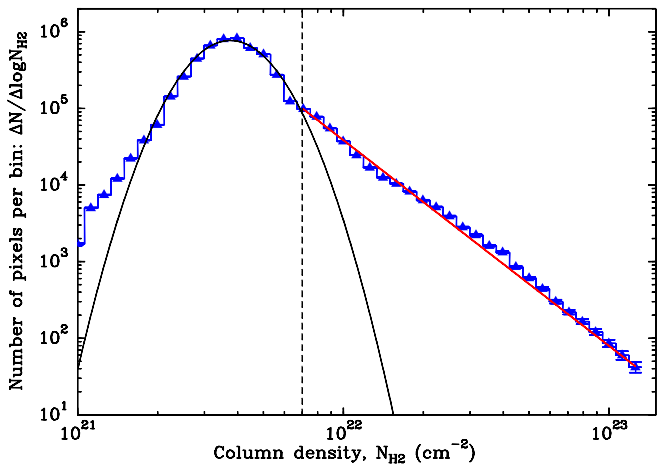} 
 \caption{Column density PDF of filaments in the Aquila star-forming 
molecular cloud (Andr\'e et al 2011). The $N_{\rm H}$ values are inferred 
from the submillimeter dust emission measured by {\it Herschel}/SPIRE. 
Note the log-normal and power-law parts. The vertical dashed line
marks the extinction that separates the two regimes at visual extinction,  $A_V=3.9$.}
\label{PDFaquila}
\end{figure}

\begin{figure}
 \includegraphics[width=0.8\textwidth,angle=0]{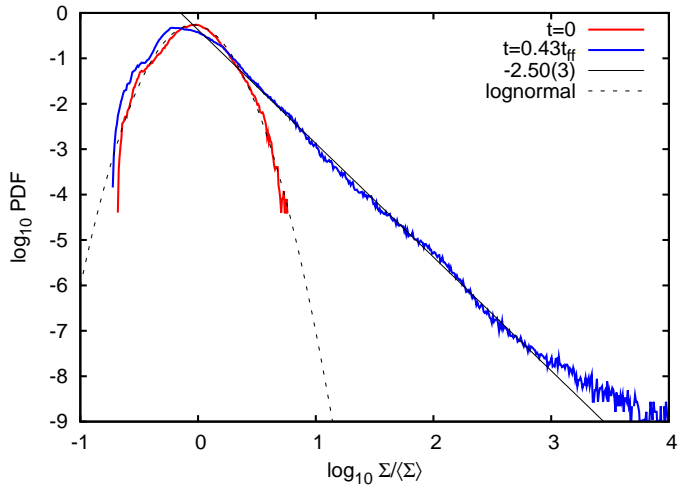} 
 \caption{Column density PDF within the simulation by 
Kritsuk et al. (2011). At $t=0$ the distribution is
log-normal. At later times, a high density power-law tail 
develops.}
\label{PDFkrit}
\end{figure}

The density field cannot be directly inferred from observations.
Owing to the integration along the line of sight, column densities are
usually obtained.  Column density PDF within molecular clouds have
been observed for the first time by Kainulainen et al. (2009) who
determined the extinction from the 2MASS archive and more recently by
Andr\'e et al. (2011) using the thermal dust emission (see
Fig.~\ref{PDFaquila}).  Kainulainen et al. (2009)
obtain the column density of 23 nearby molecular clouds.
They find that the cloud which are not forming stars,
have a log-normal distribution up to 
extinctions comparable to, or slightly below the cloud mean value. 
The clouds that are forming stars have also a log-normal distribution
but only up to some extinction above which the PDF is 
more compatible with a power-law and resembles, even quantitatively, the
high column density tail found in numerical simulations (see
Fig.~\ref{PDFkrit} from Kritsuk et al. 2011).  As shown by Kritsuk et
al. (2011), this result can be understood by the following argument.
In projection the surface of a sphere lying between radius $r$ and
$r+dr$ is $dS \propto r dr$. Assuming that the density is proportional
to $r^{-2}$, the column density $\Sigma \simeq r \rho$ is thus
proportional to $r^{-1}$ and thus $d r \propto d \Sigma / \Sigma
^{2}$.  Thus, we get $dS \propto \Sigma ^{-2} d \log \Sigma$.

Turbulence, however, is not the unique explanation for having 
a log-normal PDF as argued by Tassis et al. (2010). Therefore,
it should not be used as a direct proof in favor of the existence 
of supersonic turbulence but as a fact compatible with it.
Finally, we note that in some cases,
 at low visual extinction, the PDF 
 deviates significantly from the log-normal distribution 
(see Fig.~\ref{PDFaquila})
and appears to be reminiscent of the column 
density inferred in two--phase flows (e.g. Hennebelle et al. 2008).

\subsection{Density higher order statistics}
Beside the density PDF, which does not entail spatial correlations in
the flow, higher order statistics have been investigated both in
simulations and observations.

\subsubsection{Density power spectra: theory}
One of the most fundamental quantity in turbulence is the 
power spectrum, which,  in the case of the velocity field, 
describes how energy is distributed
 and a lot of effort has been made 
to infer the power spectrum of the density field. 

In the limit of weakly compressible flows, analytical expansions have
been performed by Zank \& Matthaeus (1990) and Bayly et al. (1992).
They conclude that the result is not universal and depends on the
thermal properties of the flow. When the pressure, density and
temperature fluctuations are all of the same order, they predict that
density should behave like pressure and the power spectrum should be
proportional to $P_\rho(k)k^2 \propto k^{-7/3}$.  On the other hand,
when the temperature and density fluctuations are large with
respect to the pressure ones, the density is close to a passive scalar
which in turn implies that the power spectrum should be $P_\rho(k)k^2
\propto k^{-5/3}$. Note that it is necessary to go through their
analytical expansions to reproduce these results.  As molecular clouds
are very compressible, it is unclear that these analytical predictions
are relevant.

When the flow is very compressible, it is not possible to apply these
analytical techniques and simulations must be used.  Kim \& Ryu (2005)
perform a series of isothermal simulations at various Mach
numbers. They find that at low Mach numbers (say ${\cal M} \simeq 1$),
the density power spectrum  is close to the
theoretical predictions described in the previous paragraph, \ie\ is
roughly proportional to $k^{-5/3}$.  At higher Mach numbers, it
gradually becomes shallower with exponents that could become as small
as $-0.7$. The reason seems to be due to the term $\rho \nabla . {\bf
  v}$ in the continuity equation because the density power spectrum is
then related to the power spectrum of this term. Indeed, in the high
Mach number case, the strong shocks create thin shocked layers in the
flow which resemble Dirac functions.
As their power spectra  are 
 $\propto k^{0}$, it is expected to get exponent power spectra 
close to this value and thus much shallower than the subsonic case. 

On the other hand, when the equation of continuity is formulated using
$s=\log \rho$ instead of $\rho$, it simply becomes $d_t s= - \nabla
. {\bf v}$ which suggests that the variable $s$ should be less
affected by this effect. Indeed various authors have computed the
power spectrum of $s$, both in the isothermal and two--phase cases
(Beresnyak et al. 2005, Schmidt et al. 2009, Audit \& Hennebelle 2010)
and they all find that the exponent of the power spectrum of $s$ is
always not too far from the Kolmogorov exponent although no rigorous
analytical prediction has been made so far.

\subsubsection{Power spectra: observations}

\begin{figure}
        \includegraphics[width=0.6\textwidth,angle=-90]{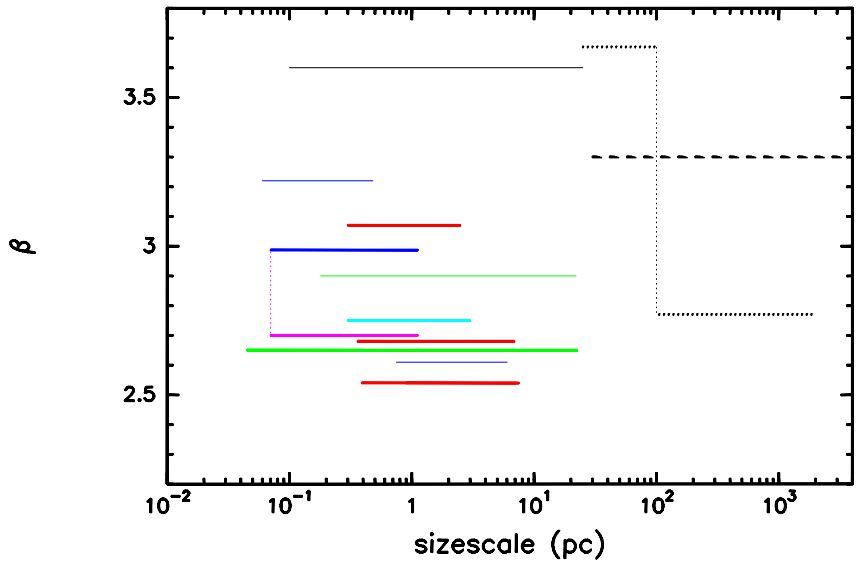}
        \caption{ 
Spectral indices $\beta$ of emission maps   
versus the range of size scales where the power-law $P_I(k) \propto
k^{-\beta}$ is observed. Indices for 
dust emission (green), \twCO\ (blue and purple) 
and \thCO\ (red) line emission, HI emission (black) and absorption (turquoise) are  
obtained with various methods (see text for details).
 The error bars (not shown for clarity) on all
the values of $\beta$ vary between 0.01 and 0.3.}
\label{beta_size}
\end{figure}

Figure~\ref{beta_size} displays the power spectral indices $\beta$ of
an ensemble of intensity ($I$) maps versus the range of size scales
over which the power-law $P_I(k) \propto k^{-\beta}$ is observed.
Various methods are used.  The $\Delta$--variance method, introduced
by Stutzki et al. (1998), has been used on maps of line integrated
areas (\ie\ close to column density maps) of CO emission in molecular
clouds (Bensch et al. 2001): \thCO(1-0) in star-forming regions (red
lines), \twCO(1-0) and \thCO(1-0) in the Polaris Flare (blue thin
lines).  In the \twCO(2-1) line, two values are shown (connected by a
dotted line): for the whole line emission (thick blue) and for the
line-wing emission only (purple) (Hily-Blant private communication).

\begin{figure}
\begin{minipage}{7cm}
  \includegraphics[width=\textwidth{},angle=0]{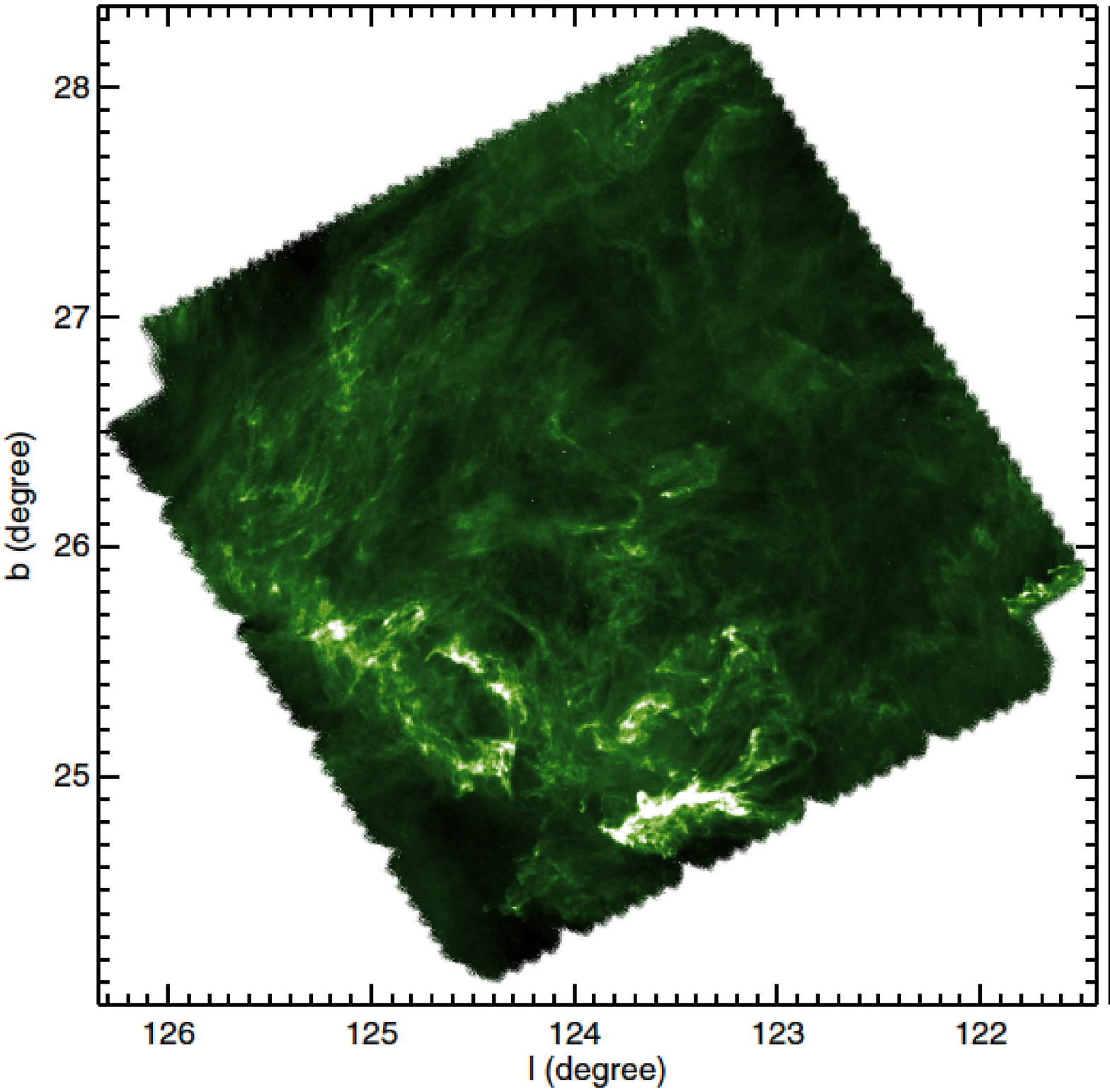}
  \includegraphics[width=\textwidth{},angle=0]{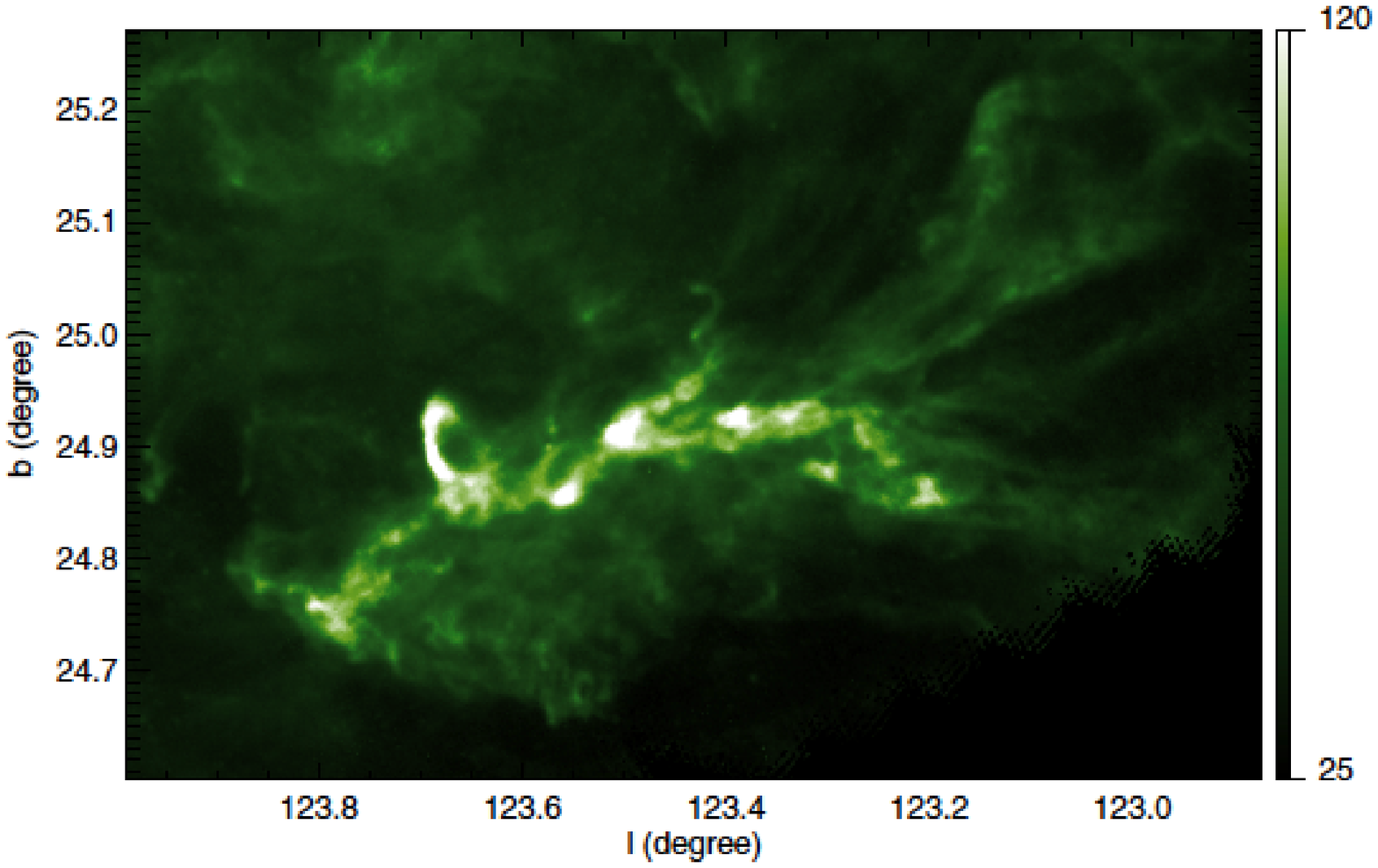}
\end{minipage}
\begin{minipage}{7cm}
  \includegraphics[width=\textwidth{},angle=0]{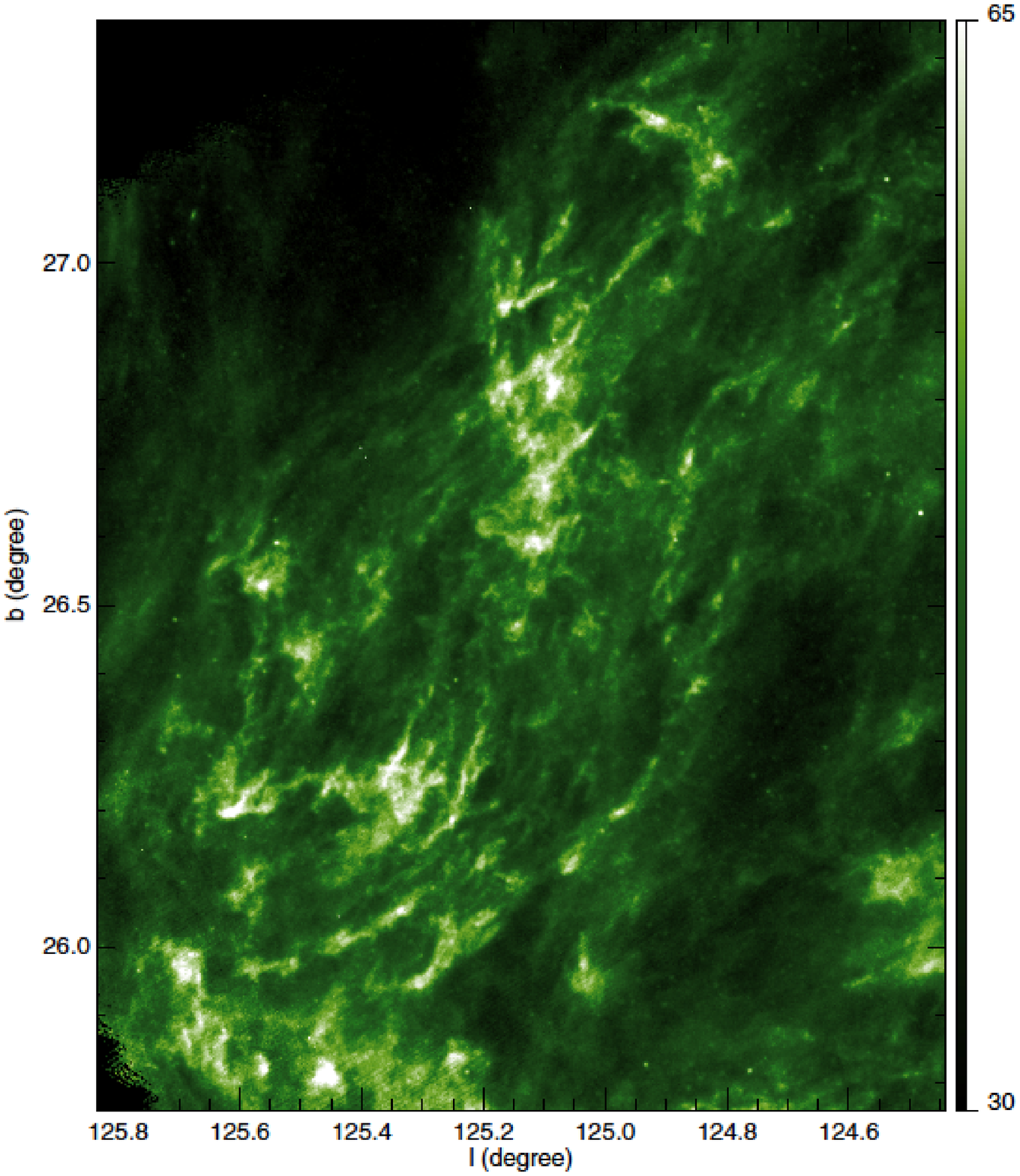}
\end{minipage}
  \caption{ {\bf Left:} {\it Herschel}/SPIRE 250$\mu$m map of the Polaris Flare over $\sim 7$pc 
$\times 7$pc. Units are MJy sr$^{-1}$ {\it (top)}. 
Subfield of the region of largest column density discussed in Sect. 3.6 {\it (bottom)}.
{\bf Right:} Another subset of regions of lowest column density. Note the brightness difference.
(from Miville-Desch\^enes et al.  2010).}
  \label{polarisSPIRE}
\end{figure}

Slopes of dust emission power spectra
are shown in green: IRAS 100$\mu$m emission of the high galactic
latitude sky (thin) (Gautier et al. 1992) and {\it Herschel}/SPIRE 250$\mu$m
emission combined to IRAS 100$\mu$m in the Polaris Flare (thick)
(Miville-Desch\^enes et al. 2010).  The constancy of $\beta$ over
almost three orders of magnitude in scales obtained with the SPIRE
data (Figure~\ref{beta_size}) is impressive.  It hides, however, the
intrinsic diversity of small scale structures illustrated in the two
subsets of Fig.~\ref{polarisSPIRE}.

Slopes of HI emission power spectra are displayed in black: Large
Magellanic Cloud (thick dotted) (Elmegreen et al. 2001), Small
Magellanic Cloud (thick dashed) (Stanimirovic \& Lazarian
2001). Galactic HI in emission: North Polar Spur (thick solid)
(Miville-Desch\^enes et al. 2003a). The slope of the power spectrum of
Galactic HI in absorption is in turquoise: VLBI maps of HI absorption
optical depth against Cas A (Deshpande et al. 2000).  Values derived
from HI emission are predominantly WNM values, those from HI in
absorption relate to the CNM.  In the LMC, a slope discontinuity is
observed, and two values of $\beta$ (connected by a dotted line) are
displayed over each relevant range of scales.  This change of slope is
ascribed to the transition from 2-- to 3--dimensional turbulence at
about 100 pc, the estimated thickness of the HI layer in the LMC
according to the criterion of Lazarian \& Pogosyan (2000).

Figure~\ref{beta_size} provides quantitative insights on the link
between turbulence in the diffuse medium and that in molecular clouds.
First, the values of $\beta=3.6$ obtained from HI in emission are
significantly larger than those obtained from HI in absorption
$\beta=2.75$. Second, a value of $\beta=2.75$ is similar to the values
inferred from dust emission at similar scales.  Cold dust emission at
small scale ($\leq 0.5$ pc), however, has a significantly flatter
spectrum ($\beta=2.65$) than that of \twCO(2-1) ($\beta=3$), but is
remarkably similar to that of CO line--wing emission ($\beta=2.7$).
This illustrates the fact that at small scales, CO emission of
molecular clouds appears almost featureless in maps of integrated
intensity, due to projection (not optical depth) effects.  Structure
is detectable only in the velocity field, supporting a close link
between gas dynamics and the formation of structures.  Indices even
higher than $\beta=3$ have been obtained in the Perseus complex by Sun
et al. (2006) with a small dynamic range of scales.

These results are qualitatively in good agreement with the trends
inferred in the previous section. In particular, the HI seen in
emission (predominantly WNM), which is not very supersonic on large
scales, presents an index $\beta$ close to 11/3. The CO and the HI in
absorption, which both trace colder supersonic media, have power
spectra that are much shallower and reminiscent of what is obtained
in numerical simulations for high Mach number flows.  The trend of the
index being higher at smaller spatial scales may also signal the
transition from the supersonic regime to the subsonic one. Indeed, the
sonic length is typically of the order of 0.1 pc, a scale around which
the value of $\beta$ seems to increase significantly.

Alternatively, the increase of $\beta$ toward small scale, 
found from the values derived from CO maps,
may be seen as a manifestation of the multi--fractal
facet of CO molecular clouds. As shown by Stutzki et al. (1998), there is
a relation $\beta=\gamma(3-\alpha)$ between the slope  $\beta$ of the power spectrum 
of a fractal Brownian motion (fBm) structure, the index 
$\gamma$ of the mass-size relation $M \propto l^{\gamma}$, and the slope of
the mass spectrum $\alpha$.  Since $\alpha$ does not vary (see
previous section) among molecular clouds, the increase of $\beta$ from 2.2 to 3.2 is
nothing more than the steady increase of the fractal dimension of the
cold ISM as the size scale decreases, \ie\
a decrease of lacunarity in the maps.
However, it cannot be ruled out that 
the variation of $\beta$ is not gradual but abrupt 
and, alike HI, is a manifestation of the dimensionality of molecular clouds: 
the increase of $\beta$ at about 1 pc would suggest that
molecular clouds (and CNM clouds) are sheet-like structures of thickness $\sim 1$ pc. \\

Incidentally, the spectral index $\beta=3.6$ of the power spectrum of
the HI integrated emission (\ie\ the HI column density) in the North
Polar Spur is the same as that found for the line velocity centroid
maps. Miville-Desch\^enes et al. (2003b) have shown, by using fBm
simulations, that in the case of an optically thin medium, the former
is the spectral index of the 3--dimensional density field and the
latter, that of the velocity field. The value $\beta=3.6$ being
consistent with the Kolmogorov scaling of incompressible turbulence,
the galactic HI value in emission (also found in other samples)
provides one more example of Kolmogorov scaling found in the
compressible interstellar turbulence (see Federrath et al. 2008).

\subsubsection{Mass-size relation}

Some authors have attempted to characterise the mass-size relation in
numerical simulations using various techniques (see e.g. Federrath et
al. 2009). The so-called mass size method has been applied by Kritsuk
et al. (2007) and Federrath et al. (2009). It consists in considering
a series of boxes of size $l$ centered around all density peaks and
counting the mass, $M(l)$ within the box. The recovered $M(l)$ has the
form of a power-law, $M(l) \propto l^\gamma$.  The exponent $\gamma$
has been measured to be $2.39$ according to Kritsuk et al. (2007) and
2.11 with solenoidal forcing and 2.03 for compressible forcing
according to Federrath et al. (2009) who count only the mass above
half the value of the local maximum.

Another way to estimate a mass-size relation which is closer to
observations, though still significantly different, is to clip the
density field choosing a density threshold and to reconstruct the
structures using a friend of friend algorithm, i.e. 
gathering all connected cells. Then, the mass of each identified
structure can be easily obtained. The size can be estimated either by
taking the most distant cells or by computing and diagonalizing the
inertia matrix and taking the largest
eigenvalue.  Plotting $M$ against $l$, Audit \& Hennebelle (2010) find
that the relation is again close to a power-law $M \propto
l^{2.3-2.5}$ both for isothermal and two--phase flows.

Considering that gravity is not included in these simulations, the
physical origin of the mass-size relation can only be attributed to
turbulence.  Yet, no quantitative explanation has been inferred
neither for the power-law behaviour nor the measured exponent. A
phenomenological model has been proposed by von Weizs{\"a}cker (1951)
and Fleck (1996) who postulate a hierarchy of density fluctuations at
successive scales given by
\begin{eqnarray}
{\rho_\nu \over \rho_{\nu -1}} = 
\left( { l_\nu\over l_{\nu-1}} \right)^{-3 \alpha}. 
\end{eqnarray}
This leads to a mass-size relation $M \propto l^{3-3 \alpha}$ which
suggests that $\alpha \simeq 0.1-0.2$.  The interesting point is that, 
assuming the phenomenology of Kolmogorov still holds in the
compressible case (\ie\ $\rho v^3 \propto l$), one gets $v
\propto (l/\rho)^{1/3}$, which leads to $v \propto
l^{1/3+\alpha}$ and the power spectrum of $v$, $P_v(k)$ is thus
proportional to $k^{-5/3 -2 \alpha}$ (since $<v^2> \propto
\int_{k_v}^\infty P_v(k) k^2 dk$).  This offers a valuable explanation
for a velocity field power spectrum  being steeper in the
compressible case than in the incompressible case, since the exponent is 
on the order of -2.  Using this expression, Kritsuk et al. (2007) measure the exponent
to be $\alpha \simeq 0.15$.

As shown in Fig.~\ref{mr-vntr}, the mass size relation inferred from
observations is also a power-law with an exponent $\gamma \simeq
2-2.3$ and therefore close to what is measured in numerical
simulations. Although this apparent agreement is interesting, we
recall that the CO clumps are determined using both spatial and
kinematic information, \ie\ in the position-position-velocity (PPV)
space while the structures described above are defined using the 3D
density field, independently of the velocity. Whether the two
approaches select the same structures is not clear at this
stage. Using periodic boxes, Ostriker et al. (2001) show that the
structures selected in the PPV-space are sometimes not spatially
continuous.  On the other hand, in the context of colliding flows,
Heitsch et al. (2009) use the PPV methods to identify the structures
and find a good correspondence with the structures identified using 3D
spatial information.  This issue certainly requires further careful
investigation.

\subsection{Mass spectra of clumps and cores}

\subsubsection{Some observational results}
To date, the mass spectra of two types of cloud have been
convincingly inferred from observations, namely the so-called CO
clumps and the prestellar cores.  As recalled in
section~\ref{co_clumps}, the mass spectrum of the CO clumps identified
in the PPV-space obeys the same power-law $dN/dM \propto M^{-1.8}$,
over about 7-8 orders of magnitude in mass.

The core mass spectrum (CMF) has been first obtained by Motte et
al. (1998), Testi \& Sargent (1998), Johnstone et al. (2000) and more
recently by Alv\'es et al. (2007) and Andr\'e et al. (2010). The
derivations did not make use of the velocity information but are based
only on the spatial distribution.  It has been pointed out that at
masses larger than $\simeq 1 M_\odot$ the dense core distribution is
compatible with a power-law significantly steeper than that of the CO
clump power spectrum with an exponent close to the value 2.3 inferred
by Salpeter (1955) for the initial mass function, $dN/dM \propto
M^{-2.3}$ (or equivalently $dN/dlogM \propto M^{-1.3}$).  At lower
core mass, the most recent studies claim to see a change in the
distribution with a pronounced flattening and possibly a peak around
$0.5-1 M_\odot$ (Andr\'e et al. 2010).  In general, it is found that
the CMF is very similar in shape to the IMF, but shifted towards
larger masses by a factor of about 3.

It is not easy to find a theory which accounts for all these facts
simultaneously. In particular, we would like to account for the fact
that $i)$ the relation $dN/dM \propto M^{-\alpha}$ with $\alpha \simeq
1.8$ extends over 7-8 orders of magnitude and applies both to CO
clumps which are self-gravitating (e.g. $M > 10^3 M_\odot$) and CO
clumps which are not; $ii)$ while the dense cores are self-gravitating
entities, their mass distribution is different from the most massive
clumps which are also self-gravitating.  Different approaches recently
developed lead to a possible explanation.  A strong hint in this
direction comes from the work of Peretto \& Fuller (2010) who have
shown that while the mass distribution within the infrared dark clouds
(IRDC) resembles the mass spectrum of the CO clumps, the mass
distribution of their fragments, \ie\ the density fluctuations within
the IRDC resembles the mass spectrum of the dense cores.  This clearly
calls for a unifying interpretation.

\subsubsection{Mass spectra of turbulent non self-gravitating clouds}
\label{turb clump}
Let us start with elementary considerations which suggest that indeed
$dN/dM \propto M^{-2}$ is a natural power spectrum in a 3D space. 
In the Fourier space, the volume occupied by fluctuations  
of wavenumber between $k$ and $k+dk$ is $4 \pi k^2 dk$. 
It seems reasonable to assume  that for a broad variety of situations, 
$dN(k) \propto k^2 dk$. Now, in 3D we expect that the mass
of a clump can be simply  proportional to its volume, that is
$M \propto R^3$. Therefore $dM \propto k^{-4} dk$ and thus 
$dN/dM \propto k^{6} \propto M^{-2}$. It is interesting to 
stress that the power spectrum of the CO clumps and the 
high-mass part of the dense core distributions have exponents
close to -2. 
The exponent -2 is also the value for which the mass is equally
distributed between cores of low and high mass since
$\int _{M_{min}} ^{M_{max}} dN/dM M dM \propto \ln(M_{max}/M_{min})$.
Beyond these elementary considerations, 
it is clearly necessary to infer the mass spectrum of the different
types of cloud from the specific physical processes that drive 
their evolution.

As CO clumps are presumably density fluctuations, it seems
natural to relate their mass spectrum to the statistical properties of
the density field, likely a consequence of supersonic turbulence.  The
first attempt along this line has been made by Stutzki et al. (1998)
who relate the column density power spectrum index, $\beta$, the
mass-size relation index, $\gamma$ and the clump mass spectrum index,
$\alpha$ through the relation: $\beta=(3-\alpha) \gamma$.  This
follows from calculating the column density power spectrum as a sum of
Gaussian clumps.  It suggests that the mass-size relation and the
distribution of clumps are related (see also Elmegreen \& Falgarone
1996).  More recently, Hennebelle \& Chabrier (2008) have performed a
multi-scale analysis following the approach developed in cosmology by
Press \& Schechter (1974). They assume that the density PDF is
log-normal and that the power spectrum exponent of $s=\log \rho$ is
close to the Kolmogorov exponent, $P_s(k) \propto k^{-n^{\prime}}$
with $n^\prime \simeq 11/3$.  In this way, they obtain the relation
$\alpha=3-n^\prime/3$ or equivalently $n^\prime = (3-\alpha) \times 3
$, a relation quite similar to that obtained by Stutzki et al.  (1998)
with some important differences. First, 
while $\beta$ is the power spectrum index of the {\it column density
  field} (or equivalently the density field), $n^\prime$ is the power
spectrum index of $\log(\rho)$. Second, while the coefficient $\gamma$
must be specified in the approach of Stutzki et al. (1998), Hennebelle
\& Chabrier (2008) do not introduce this parameter which can be
deduced from the calculations as emphasized in Chabrier \& Hennebelle
(2010).  Note that the power spectrum indices of $\rho$ and
$\log(\rho)$ are different especially at high Mach numbers where
density contrasts are high and the gas is organised in clumps that do
not fill the volume. Indeed, these two expressions for $\alpha$ can be
combined leading to $n'=3 \beta / \gamma $. This in turn suggests that
for highly supersonic flows, the smaller values of $\beta$ result in a
more lacunary medium.

The mass spectra of non self-gravitating clumps, \ie\ clumps induced
by pure turbulence, have been obtained numerically by Hennebelle \&
Audit (2007), Heitsch et al. (2008), Dib et al. (2008), Audit \&
Hennebelle (2010) and Inoue \& Inutsuka (2012).  These authors have
simulated various types of flow including isothermal and two--phase
medium as well as periodic boxes and colliding flows. In each case the
mass spectrum has been inferred to be compatible with a power-law with
$\alpha=$1.7-1.8 similar to the mass spectrum of the CO clumps and to
the analytical prediction, $\alpha = 3 - n'/3$, mentioned
above. Considering that both the input physics and the boundary
conditions are different in these studies, the lack of variations
for the measured $\alpha$ strongly suggests that it is the very nature
of turbulence that determines almost entirely the value of $\alpha$
while cooling, instabilities and forcing play a secondary role.

\subsubsection{Mass spectra of self-gravitating clouds}
The mass spectrum of self-gravitating condensations has been inferred
by Padoan et al. (1997) assuming a log-normal density distribution and
considering thermal support only.  Hennebelle \& Chabrier (2008, see
also Hennebelle 2012) include the turbulent support that resists
gravity and show that it is crucial to predict the mass spectrum.
They use the same formalism than for the turbulent fluctuation case
described previously, but instead of adopting a simple uniform density
criterion, require that the parcel of gas should be dominated by its
own gravity. More precisely they perform a multi-scale analysis and
select the mass enclosed in the smallest scale at which a given piece
of gas is self-gravitating.  The distribution they infer is very
similar in shape to the core mass function (e.g. Andr\'e et al. 2010)
and to the initial mass function (e.g. Kroupa 2002, Chabrier 2003).
In particular the slope at high masses is similar to the Salpeter
(1955) estimate $dN/d {\rm log}M \propto M^{-x}$, $x\simeq 1.35$.  At
low masses the distribution presents a peak and its overall shape is
close to a log-normal distribution.

\begin{figure}
 \includegraphics[width=8cm,angle=0]{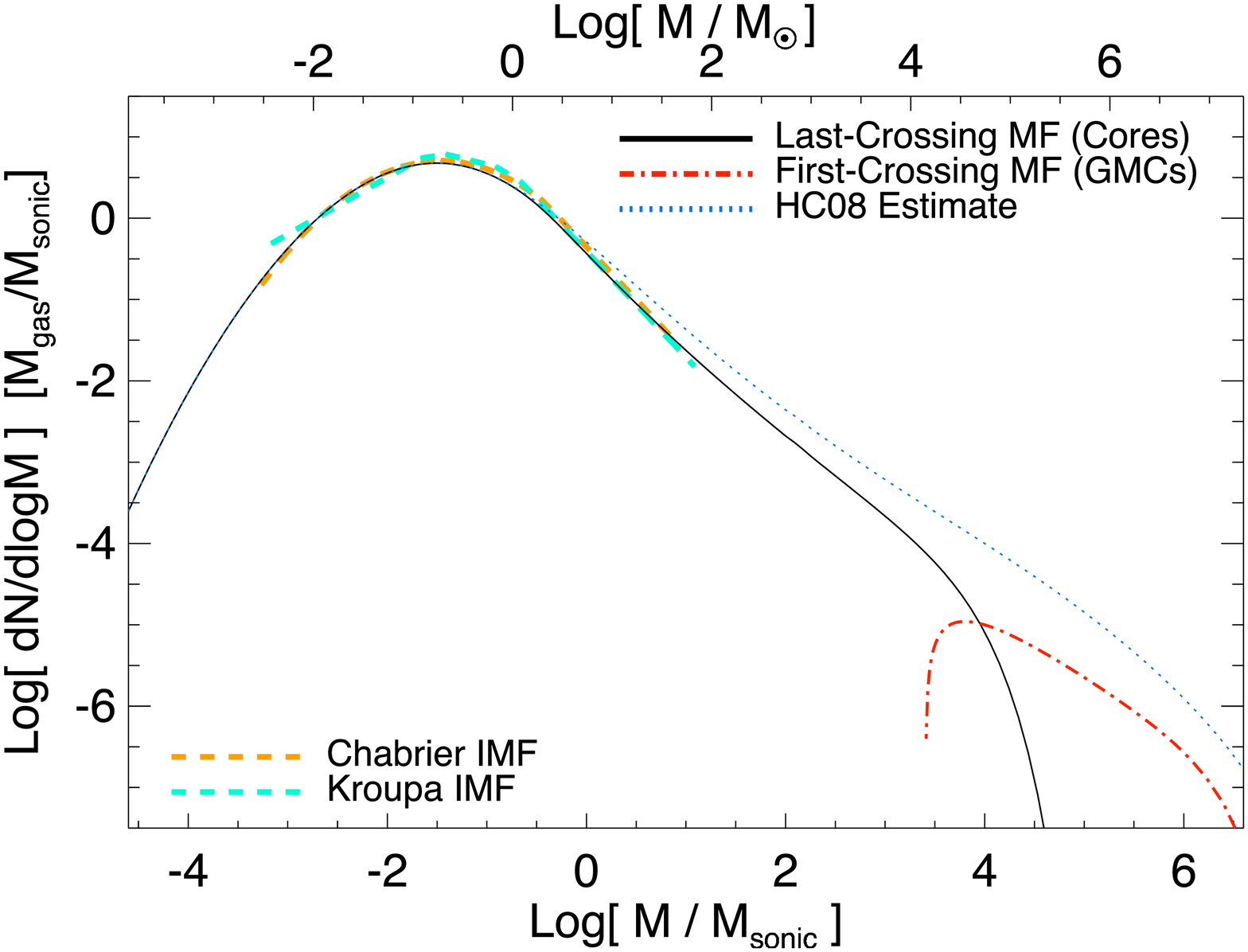} 
 \caption{Core mass spectrum and GMC mass
spectrum as calculated by Hopkins (2012b). The blue dotted
curve shows the core mass function inferred by Hennebelle \& Chabrier
(2008) while the solid and dot-dashed lines show structures 
defined by the last and first crossing of the self-gravitating
barrier.}
 \label{hopkins}
\end{figure}

Recently, Hopkins (2012a,b) improves on this approach by applying the
excursion set theory (Bond et al. 1991) which consists in making a
random walk in the Fourier space of the density field and counting the
density fluctuations that cross the requested barrier, \ie\ reach a
requested density threshold.  Hopkins presents a global model that
includes the spatial scales larger than the molecular clouds
appropriate for a description of the whole galactic disc. An
interesting result is that the clouds defined as entities which cross
first the requested self-gravitating barrier have a mass spectrum
which is slightly shallower than $M^{-2}$ while the condensations
defined by the last crossing of the barrier have a mass spectrum almost 
identical to that by Hennebelle \& Chabrier (2008) except at large
masses.  This is illustrated in Fig.~\ref{hopkins}.
 An obvious interpretation is that the first type of structures likely
 represents the massive molecular clouds that are not themselves
 embedded into a larger self-gravitating cloud while the second type
 can possibly be associated to the prestellar core progenitor.  Indeed
 it suggests that, while large scale self-gravitating clumps should
 have a mass spectrum close to $M^{-1.8-2}$, the mass spectrum of the
 smallest self-gravitating fluctuations, likely to be the core
 progenitors, should be close to the core mass function.

Interestingly, the mass spectrum for the purely turbulent clumps as
inferred by Hennebelle \& Chabrier (2008, see sect.~\ref{turb clump})
and that of the largest self-gravitating clumps obtained by Hopkins
(2012b) are both power-laws with very similar exponents. This is an
appealing explanation for the mass spectrum of the CO clumps
apparently constant over almost 9 orders of magnitude in mass
(e.g. Heithausen et al. 1998, Kramer et al. 1998) even though the most
massive ones are self-gravitating while the least massive are not, with
the transition occurring at $10^2-10^3$ M$_\odot$.  A possible route
for the least massive self-gravitating ones is that they originated by turbulent decay
within non self-gravitating clumps and that gravity eventually took
over during the condensation process through cooling and energy
dissipation. In this case, the part of the CO clump mass spectrum
which is due to turbulence could be extended towards higher masses.
On the other hand, the formation of the most massive clumps may have
been initially triggered by their own self-gravity.  It is probably
difficult at this stage to estimate where the possible transition
between the two regimes appears.

It must be noted that CMF in good agreement with the observed
distributions have also been constructed using different approach and
assumptions.  For example Myers (2009a) suggests a model based on gas
accretion in dense cores with a distribution of infall durations. Kunz
\& Mouschovias (2009) develop a model based on magnetically dominated
cores assuming a log-normal distribution of magnetic fluxes.

Finally, the CMF has also been inferred from various numerical
simulations. Using hydrodynamical and MHD simulations without gravity,
Padoan et al. (2007) identify the cores as the density fluctuations
induced by turbulence that would have been
self-gravitating. Considering pure thermal support (as in Padoan et
al. 1997), they find that the CMF produced is too steep. 
When thermal plus magnetic support is taken into account,
the core mass function is found to have a Salpeter-like slope.
Schmidt et al. (2010) perform similar hydrodynamical simulations and
analysis but include the turbulent support in the core
definition. They find that the CMF is a power-law at high
masses whose index is close to the Salpeter exponent while it is too
steep when only thermal support is included in the core
definition. Performing quantitative comparisons with the predictions
of Hennebelle \& Chabrier (2008), they infer a good agreement. Smith et
al. (2009) carry out simulations that include both turbulence and
self-gravity and define the cores using the potential wells.  They
also find a CMF close to that observed. All these results underscore
the importance of non-thermal support to reproduce the observed CMF.

\subsection{Filaments}
With the increased sensitivity of CO observations and the availability
of larger maps away from the brightest spots coinciding with star
forming regions, the evidence for the filamentary structure of the CO
emission became clear (e.g. Ungerechts \&
Thaddeus 1987, Bally et al. 1987).  At about the same time, the far-IR all-sky
IRAS survey (Low et al. 1984) revealed the ubiquitous filamentary
structure of the ISM, and discovered the cirrus clouds, \ie\ the
filamentary structure of the diffuse ISM.  This suggested a link
between filaments in the diffuse medium and more opaque filaments.
Abergel et al. (1994) showed quite convincingly that the filaments of
cold dust were those of largest column density ($N_{\HH} \geq 10^{22}$
\cq) seen in \thCO\ only and invisible in the \twCO\ maps because of
the high opacity of \twCO(1-0) in diffuse molecular gas.

Subsequently, several mid-infrared surveys of the inner Galaxy ({\it
  ISO}, P\'erault et al. 1996, {\it MSX}, Price et al. 2001, {\it
  Spitzer/MIPS}, Benjamin et al. 2003) revealed the existence of
massive filamentary structures of cold dust, seen as dark lanes
against the bright galactic background emission.  These regions are
now known as infrared dark clouds (IRDC). One of the most spectacular
IRDC is the Nessie nebula with a linear extent of 80 pc long and a
width of only 0.5 pc thick. The linear mass density is still poorly
constrained (Jackson et al. 2010).

\subsubsection{Self-gravitating filaments}
The presence of filaments in molecular clouds 
 and their active role in the star
formation process has long been recognized and is particularly
suggested by the presence of dense cores often elongated in the same
direction (Tachihara et al. 2000) and sometimes periodically
distributed (e.g. Dutrey et al. 1991). Moreover, in the case of the
Taurus molecular cloud, Hartmann (2002) shows that the protostars
distribution closely follows the filament network identified by Mizuno
et al. (1995). The association of filaments and cores suggests
that the latter likely form from the gravitational fragmentation of
the filaments, a process supported by analytical studies
(e.g. Stodolkiewicz 1963, Ostriker 1964, Larson 1985, Inutsuka \&
Miyama 1992, Fiege \& Pudritz 2000a, Curry 2000), which in particular
show that in the isothermal case there is a critical line mass (mass
per unit length) equal to $2 c_S^2 / G$ for which equilibrium is
possible leading to a density profile
\begin{eqnarray}
\rho(r)= {\rho_0 \over (1+r^2/l_0^2)^2}, \;
 l_0=\left({2 c_S^2 \over \pi G \rho_0} \right)^{1/2}. 
\end{eqnarray}
where $c_S$ is the sound speed.
This equilibrium solution is, however, subject 
to gravitational instabilities. The growing fluctuations would 
give birth to cores  separated by a few 
Jeans lengths, which corresponds to the typical wavelength 
of the fastest unstable mode.

The formation mechanism of these self-gravitating filaments is still
not well understood yet but a few theoretical arguments are worth
considering. The most important point is that gravity tends to amplify
any pre-existing anisotropy, as emphasized for example by Lin et
al. (1965).  This is simply because the force is the gradient of the
gravitational potential and is thus higher along the shortest
direction.  This implies that a gravitational structure which would be
slightly elongated, for example because of the way it was assembled,
will quickly become very elongated. Another way for filament formation
is by gravitational instabilities of self-gravitating layers, as shown
for example by Spitzer (1978) and Nagai et al. (1998). The expression
of the density profile of the layers is given by
\begin{eqnarray}
\rho(z)= {\rho_0 \over {\rm ch}(z/l_0)^2}, \; l_0=\left( {c_S^2 \over 2 \pi G \rho_0} \right)^{1/2}.
\end{eqnarray}
 Again, filaments will be periodically distributed with separation of
 a few Jeans lengths. Interestingly, Nagai et al. (1998), who
 performed linear stability analysis of a magnetized self-gravitating
 layer, show that the orientation of the most unstable mode tends to
 be correlated with the magnetic field direction. The result depends
 on the external pressure that determines the scale height, $z_b$ at
 which the solution is truncated. If $z_b \gg l_0$ then the fastest
 growing mode is aligned with the magnetic field, resulting in
 filaments which are perpendicular to the field direction.  The
 physical reason is that, since the width is large relative to the
 Jeans length, the layer is compressible and density fluctuations are
 easier to develop along the magnetic field.  On the other hand when
 $z_b \ll l_0$, the fastest growing mode is perpendicular to the
 magnetic field and the filaments are aligned with it.  This is
 because the layer is almost incompressible (since the scale height is
 smaller than the Jeans length), thus the instability develops through
 the bending of the layer. As perturbations whose wave vectors are
 perpendicular to the magnetic field do not bend the field lines,
 these perturbations develop more easily.  A nice bidimensional
 solution which describes the continuous and quasi-static evolution
 from the self-gravitating layer to the self-gravitating filament has
 been derived by Schmid-Burgk (1967). It is given by
\begin{eqnarray}
\rho(x,z) = 
{ \rho_0 (1-A^2) \over \left( {\rm ch}(z/l_0)-A \cos(x/l_0) \right)^2}. 
\end{eqnarray}
When $A=0$, it is simply reduced to the self-gravitating layer while
when $A \rightarrow 1$, it tends to a set of discrete periodic
filaments. Recently, this solution has been used by Myers (2009b) to
model the hub filament structures that are often observed around young
stellar groups. In this model, a self-gravitating layer forms first,
then it fragments in filaments as predicted theoretically.  Owing to
the presence of local overdensities, the filaments converge towards
each other forming a denser and more massive region, which eventually
give birth to a star cluster.

\subsubsection{A continuum of filaments}
Thanks to its unprecedented sensitivity and large-scale mapping
capabilities, Herschel/SPIRE has now provided a unique view of these
filamentary structures of cold dust (e.g. Miville-Desch\^enes et al.
2010, Ward-Thompson et al. 2010, Andr\'e et al. 2010). One of the main
findings of the SPIRE observations is the very large range of column
densities -- a factor of 100 between the most tenuous ($N_{\HH} =2
\times 10^{20}$ \cq) and most opaque ($N_{\HH} \sim 10^{23}$ \cq) of
the observed filaments in several fields (Fig. \ref{PDFaquila})
contrasting with a narrow range of filament width (between 0.03 and
0.2 pc) barely increasing with the central column density (Arzoumanian
et al. 2011).

Understanding this unexpected result is challenging.  First, one would
naively expect that self-gravitating filaments would have a thickness
comparable to the Jeans length at odds with the observations (see
figure 7 of Arzoumanian et al. 2011). Second the very existence of
such thin structures in a turbulent medium is not at all
straightforward. One possibility is that it corresponds to the sonic
length. Indeed, if filaments are produced in shocks, then their
density, $\rho_f$, should be linked to the background density,
$\rho_0$ by the Rankine-Hugoniot relation: $\rho_f = \rho_0 {\cal
  M}^2$, where ${\cal M}$ is the Mach number, ${\cal M}= v / c_S$.
The velocity on the other hand is linked to the scale as $v(L) \simeq
v_0 (L/1 {\rm pc})^{\eta}$ which is simply the Larson relation
discussed above and $v_0 \simeq 0.8 \; {\rm km \; s^{-1}}$ while $\eta
\simeq 0.4-0.5$.  As the size of the shocked layer is simply given by
$L_f = L \rho_0 / \rho_f$, we get $L_f = (c_S/v_0)^2 \times L (L/1
       {\rm pc})^{-2 \eta}$.  In particular assuming that $\eta=0.5$,
       the shocked layer becomes independent of the fluctuation size
       and with $c_S \simeq 0.2$ km s$^{-1}$, we get $L_f \simeq 0.07$
       pc which is close to the thickness inferred by Arzoumanian et
       al. (2011).  However, there are a couple of intriguing issues
       with this explanation. First, the derived scale is the typical
       thickness of the shocked layers and not of the filaments, thus
       one step is missing to go from the layer to the filament.
       Second, it has assumed that the gas is isothermal.  Thus once
       the velocity fluctuation has vanished (typically in a few
       crossing times $L/v$), the filament re-expands due to the
       over-pressure relative to the ambient medium. In a few sound
       crossing times $L_f/c_S$, the thickness broadens significantly
       therefore leading presumably to a wide thickness distribution.
       This scenario is somehow similar to that proposed by Padoan et
       al. (2001) which show that, in their simulations, the filaments
       correspond to low velocity regions and are often found at the
       intersection between shocked layers.

Recently, an alternative explanation has  been proposed by 
Fischera \& Martin (2012) who argue that the characteristic size 
of the filaments is simply the result of mechanical 
equilibrium in the radial direction. Assuming that the filaments 
are pressure bounded, they  find that the equilibrium of the isothermal 
gas, between thermal pressure and gravity leads to a diameter 
of about 0.1 pc with a  weak dependence on the column density. 
We must stress that no detailed analysis,
however, on the exact nature of filaments ubiquitously found in 
numerical simulations has been performed to date. There is, however, 
a general visual trend of MHD simulations that appear more filamentary than 
hydrodynamical simulations (e.g. Padoan et al. 2007, Hennebelle et al. 2008).
If confirmed by quantitative arguments, this trend could suggest that 
shocks are only part of the process. Indeed, it is likely the case that 
the shear which is at the very heart of turbulence should also 
play a role in filament formation.

In this regard, the dynamic origin of filaments is observationally supported 
by the remarkable coincidence 
between one of the faintest filaments detected by SPIRE in the Polaris Flare 
(Fig \ref{polarisSPIRE}, bottom left), also visible as a weak CO filament,  and the 
parsec-scale coherent structure of intense velocity-shear discussed in 
 section~\ref{shear-fil}
(Hily-Blant  \& Falgarone 2009). 
Whether the velocity-shear is associated with a true compression
due to a shock propagating in the plane of the sky cannot be decided because of 
projection effects.

\subsubsection{The most tenuous filaments}

Tenuous filaments are interesting structures since
they may shed light on the origin of filaments and a possible growth
scenario, but they are difficult to study for obvious reasons.  A
tenuous filament connected to a low-mass dense core of the Taurus
cloud has been thoroughly studied in different \twCO\ transitions and
isotopes (Falgarone et al. 2001).  It is a spinning filament, with a
density drop-off as steep as $n_{\HH}(r) \propto r^{-4}$.  Periodic
oscillations of its azimuthal velocity have been found.  Its average
inner pressure, including a turbulent contribution, exceeds by a factor
$\sim 10$ the surface pressure, even including the contribution due to
the self-gravity of the Taurus complex.  It has a low mass-to-virial mass
ratio $m/m_{vir}$ (here, the mass is the linear mass), hence it is not
held together by self-gravity. 
It is possibly confined by helical magnetic fields 
because the periods of the observed wavelike behaviors of the velocity field 
are reminiscent of the instability signatures computed by
Fiege \& Pudritz (2000ab) for filaments threaded by such
fields.  In this framework, the fragment separation observed along
this tenuous filament suggests a toroidal magnetic flux five times
larger than poloidal.

\section{Magnetic fields in molecular clouds}
\subsection{Theoretical considerations}
Early on, magnetic fields have been  proposed to provide 
an important mechanical support to the gas (see e.g. Shu et al. 1987),
possibly explaining the low star--formation efficiency within the 
Milky Way. This suggestion has contributed to trigger 
studies of interstellar magnetic fields. Although, as described below, the actual 
trend indicated by measurements of the magnetic intensity
is that magnetic fields, although dynamically important, 
may be not dominant, we recall below the theoretical elements 
that were put forward in support of the magnetically regulated star 
formation theory.

\subsubsection{Magnetic support}
Magnetic forces are highly non-isotropic.  An easy way to estimate the
strength of magnetic support is to compute the ratio of the magnetic
over gravitational energies. For simplicity, let us consider a
spherical and uniform cloud of mass $M$, volume $V$, radius $R$,
threaded by a uniform magnetic field, of strength $B$.  The magnetic
flux within the cloud, $\Phi$ is equal to $\Phi= \pi R^2 B$. As long
as the magnetic field remains well coupled to the gas, the magnetic
flux threading the cloud will remain constant with time.  The ratio of
magnetic over gravitational energies for uniform density cloud
threaded by a uniform magnetic field, is
\begin{eqnarray}
{ E_ {\rm mag} \over E_{\rm grav}} = {B^2 V \over 8 \pi} \times {2  R \over 5 G M^2}
\propto {B^2 R^4 \over M^2} \propto \left( {\Phi \over M } \right)^2.
\label{emag}
\end{eqnarray}
Remarkably, the ratio of magnetic over gravitational energies is
independent of the cloud radius. This implies that if the cloud
contracts or expands, the relative importance of these two energies
remains the same. This is unlike the thermal energy of an isothermal
gas, which becomes smaller and smaller compared to the gravitational
energy as the cloud collapses.  It is clear from Eq.~\ref{emag}, that
there is a critical value of the magnetic intensity for which the
gravitational collapse is impeded even if the cloud was strongly
compressed.  Mouschovias \& Spitzer (1976) have calculated accurately
the critical value of the mass-to-flux ratio using the virial theorem
and numerical calculations of the cloud bidimensional equilibrium.  A
cloud which has a mass-to-flux ratio smaller  than this
critical value cannot collapse and is called subcritical.  
It is called supercritical when the mass-to-flux is larger than the 
critical value.
It is usual to define $\mu = (M/\Phi) / (M/\Phi)
_{\rm crit}$. Large values of $\mu$ correspond to small magnetic
fields and thus supercritical clouds.

\subsubsection{Ambipolar diffusion}
One of the problems with  magnetic support is that when ideal MHD
applies, the field remains strong and prevents collapse.  Mestel \&
Spitzer (1956) have first considered the possibility of leakage of the
magnetic flux which would reduce the strength of the field.  At
microscopic scales, the neutrals are not experiencing the Lorentz
force which applies only on charged particles. Strictly speaking, this
implies that at least two fluids should be considered, the neutrals
and the ions (in different contexts more than one fluid of charged
particles must be considered), to treat the problem properly. Since
the two fluids are coupled to each other by collisions, the neutrals
are nevertheless influenced by the magnetic field if the gas is
sufficiently ionized. Since the ionization in molecular clouds is
usually of the order of $10^{-7}$, the density of the ions is much
smaller than the density of the neutrals.  It is thus possible to
neglect the inertia of the ions. A strong assumption however is that
of the equilibrium between the Lorentz force and the drag force
exerted on the ions. This leads to
\begin{equation}
{(\nabla \times {\bf B}) \times {\bf B} \over 4 \pi} = 
\gamma \rho \rho_i ( {\bf v}_i - {\bf v}),
\label{coupling}
\end{equation}
where $\rho_i$ and {\bf$v_i$} are the ion density and velocity
respectively, $\gamma \simeq 3.5 \times 10^{13}$ cm$^3$ g$^{-1}$
s$^{-1}$ is the drag coefficient (Mouschovias \& Paleologou 1981).
From Eq.~\ref{coupling}, the ion velocity can easily be expressed as a
function of the neutral velocity and the Lorentz force. Considering
now the induction equation, which involves the velocity of the ions,
and using Eq.~\ref{coupling}, we obtain
\begin{equation}
\partial _t {\bf B}  + \nabla \times  ({\bf B} \times {\bf v}) = 
\nabla \times \left( {1 \over 4 \pi \gamma \rho \rho_i}  ( (\nabla \times {\bf B}) \times {\bf B}) \times {\bf B} 
 \right). 
\label{induc}
\end{equation}
The left part of this equation is identical to the induction equation except that the  
 velocity of the  neutrals appears instead of the velocity of the ions. The right term 
 is directly responsible for the slip between the neutrals and the magnetic field. Although 
it is of the second order, it is not rigorously speaking a diffusion term. 
From this equation, there can be easily  inferred a typical time--scale 
for  ambipolar diffusion
\begin{eqnarray}
\tau_{\rm ad} \simeq {4 \pi \gamma \rho \rho_i L ^2 \over B^2 },  
\label{time}
\end{eqnarray}
where $L$ is the typical spatial scale relevant for the problem. In the context of 
star--formation, $L$ could be the size of the cores, $R$. Ionization equilibrium 
allows to estimate that the ions density is about $\rho_i = C \sqrt{\rho}$, where 
$C=3 \times 10^{-16}$ cm$^{-3/2}$ g$^{1/2}$. 

If a dense core is initially subcritical (therefore magnetically supported), the diffusion 
of the field will progressively reduce the magnetic flux within the cloud. So after a
few ambipolar diffusion times, the cloud  becomes supercritical and 
the magnetic field cannot prevent the collapse any more. 
The important and interesting question at this stage is  
``How much is the collapse   delayed by this process?'' 
In order to estimate this time, it is usually  assumed that the cloud is in virial 
equilibrium, $B ^2 / 4 \pi   \simeq M \rho G / R$ (within a factor of a few) 
and to compute 
the ratio of $\tau_{\rm ad}$ and $\tau_{\rm ff}$, the free--fall time (Shu et al. 1987).
This leads to:
\begin{eqnarray}
{ \tau _{\rm ad} \over \tau_{\rm ff}} \simeq 8.
\label{time_ratio}
\end{eqnarray}
 The implication is that ambipolar diffusion can 
 reduce the star--formation rate significantly making it closer to the
 observed value (e.g. Shu et al. 1987).  
Numerical simulations of magnetized collapse
 controlled by ambipolar diffusion have been performed (e.g. Basu \&
 Mouschovias 1995).  These simulations are generally one dimensional
 and assume a thin disk geometry.  When the dense core is very
 subcritical, with values of $\mu$ as low as $0.1$, Basu \&
 Mouschovias (1995) found that the collapse is significantly delayed
 and occurs in about 15 free--fall times. For nearly critical cores,
 $\mu \simeq 1$, the collapse occurs after about $\simeq$3 free--fall
 times.

\subsubsection{Turbulence in  magnetized clouds with ambipolar diffusion}
In Sect.~3.2 we have described the properties of MHD turbulence. 
Here we focus on the studies which have  
 investigated simultaneously the effect of turbulence and 
ambipolar diffusion in molecular clouds.

Basu \& Ciolek (2004) and Li \&
Nakamura (2005) have presented 2D simulations combining  magnetic support
(magnetic field lines are parallel to the cloud) and  turbulent
motions. Since the clouds are initially subcritical, ambipolar
diffusion plays an important role as it allows to reduce locally the
magnetic flux. Interestingly, it has been found (Li \& Nakamura 2005,
Li et al. 2008) that in this context, turbulence tends to accelerate
the star--formation. This is because turbulence creates stiff
gradients, due to the formation of shocks, in which the ambipolar
diffusion takes place quickly. Indeed, Eq.~\ref{time} shows that the
ambipolar diffusion time decreases with the spatial scale, $L$. A very
interesting results is that these simulations are able to reproduce
the low star--formation efficiency observed in the Milky Way provided
the initial value of $\mu$ is small enough (typically $\mu \simeq 1$)
and  turbulence is sufficiently strong (typically $v_{\rm rms}
\simeq 10 \times c_S$). More recently, these results have been extended
to the 3D and multi-phase case by V\'azquez-Semadeni et al. (2011).

In 3D simulations including turbulence, gravity and supercritical
magnetic fields, it is found that the star--formation rate is reduced
by a factor of a few, i.e. the star--formation rate is
smaller in a cloud that has a supercritical magnetic field than in the
otherwise identical non magnetized cloud (Price \& Bate 2008, Padoan
\& Nordlund 2011), although the exact physical mechanism by which this
happens is not well understood.

Last, a series of recent 3D simulations of supersonic turbulence with
ambipolar diffusion (Downes \& o'Sullivan 2009, Li et al. 2008, 2012,
Tilley \& Balsara 2010) have emphasized the impact that ambipolar
diffusion has on the turbulence properties. They conclude that, as
expected, ambipolar diffusion enhances the rate of turbulent
dissipation.  Also, the power spectra of density, velocity and
magnetic fields tend to be steeper. Finally, the synthetic
spectra of the neutral and ionic species show that the latter tend to
be narrower than the former. The larger velocity dispersion of the
neutrals is certainly a consequence of their partial decoupling from
the magnetic field.
In principle, these effects  could be observed.

\subsubsection{Magnetic intensity versus density relation}
The relation between the magnetic intensity and the gas density 
is another important input to  understand   
star--formation, as  it bears the signature of the gas motions relative
to the field lines.

Before describing the results inferred from numerical simulations, it
is worth to recall the different behaviors that can be expected.  If
the contraction occurs along the field lines, then the magnetic field
is not amplified and $B \propto \rho^{\kappa}$ with $\kappa =0$. If
the motion is perpendicular to the field lines, then it is easy to
show that $\rho / B$ stays constant (combining the continuity and
induction equations) and thus $\kappa=1$.  Note that in this
configuration the magnetic pressure is proportional to $\rho^2$ and
therefore quickly halts any contraction. If the contraction is
spherical, then the mass enclosed is simply $\propto \rho R^3$ while
the magnetic flux is $\propto B R^2$ thus leading to $B \propto
\rho^{2/3}$. It is, however, unlikely that a contracting cloud remains
spherical, especially if the magnetic field is not negligible. In this
case, it is expected that an equilibrium along the field lines settles
leading to $c_S^2 \simeq \phi$, where $\phi$ is the gravitational
potential.  The Poisson equation leads $\phi \propto \rho h^2$ where
$h$ is the thickness of the cloud along the field lines. Then, as the
mass enclosed is now $\propto \rho R^2 h$ while the magnetic flux is
still $\propto B R^2$, we get $B \propto c_S \rho^{1/2}$.  Basu (2000)
has compared the data provided by Crutcher (1999) with this expression
and has obtained a good agreement, which improves if the velocity
dispersion $\sigma$ instead of $c_S$ is used.  Another even simpler
interpretation of this relation is energy equipartition between
magnetic and kinetic energy, $B^2/(4 \pi) \propto \rho \sigma^2$.

Various simulations of 3D ideal MHD turbulence tend to show that
without gravity (e.g. Padoan \& Nordlund 1999, Hennebelle et al. 2008,
Banerjee et al. 2009), the magnetic intensity weakly depends on the
density field. A weak correlation is found with typically $\kappa
\simeq 0.1-0.2$.  This has been interpreted in the context of
two--phase flows by Hennebelle \& P\'erault (2000) as a consequence of
the magnetic tension, which tends to unbend the magnetic field lines
and to align the magnetic and the velocity fields. This eventually
facilitates the gas contraction. In the context of polytropic flows,
the weakness of the correlation has been interpreted by Passot \&
V´azquez-Semadeni (2003) (see also Burkhart et al. 2009) in terms of
the fact that the various types of (nonlinear, or simple) MHD wave are
characterized by different scalings of the field strength with the
density.  Indeed while for fast waves, magnetic intensity and density
are correlated, they are anti-correlated for slow waves and not
correlated for Alfv\'en waves.  Thus, in a turbulent flow in which all
kinds of wave pass through a given point, the field strength is a
function of the history of wave passages, rather than of the local
density.  Moreover, the simulations which treat both self-gravity and
turbulence find that at high density the magnetic intensity is
$\propto \rho^{0.5}$ (e.g. Hennebelle et al.  2008, Banerjee et
al. 2009), which accords well with the analytical predictions deduced
above.

\subsection{Magnetic field measurements and interpretation}

\begin{figure}
\begin{minipage}{7cm}
  \includegraphics[width=\textwidth{},angle=0]{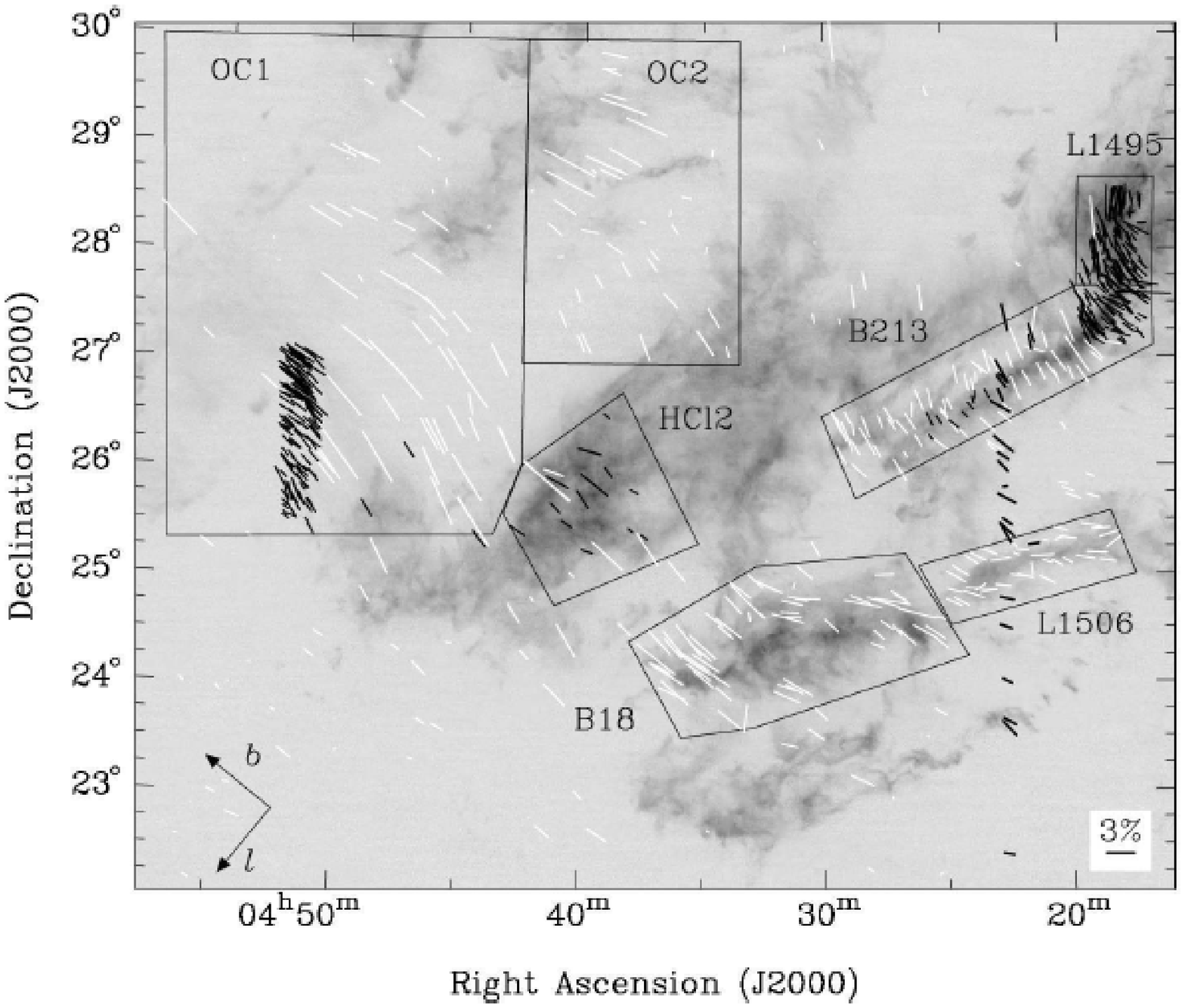}
\end{minipage}
\begin{minipage}{7cm}
  \includegraphics[width=\textwidth{},angle=0]{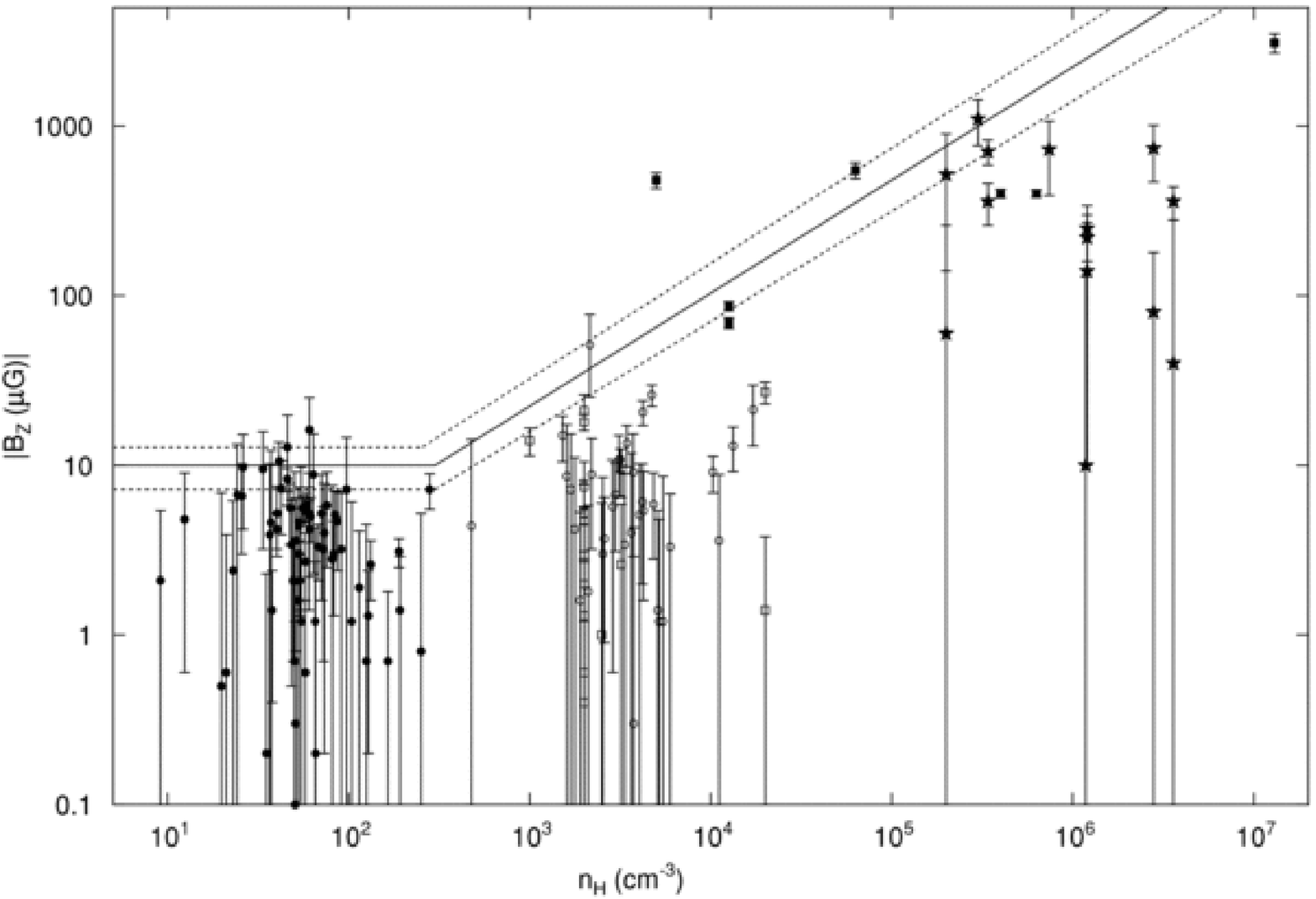}
\end{minipage}
  \caption{{\it (Left)} Polarization vectors of starlight (therefore
    parallel to magnetic field projection) overlaid on a map of
    \thCO(1-0) integrated emission of the Taurus complex (Chapman et
    al. 2011). White lines : Heiles 2000; black lines : Chapman et
    al. {\it (Right)} Magnetic field intensity inferred from Zeeman
    measurements in the HI, OH and CN lines as a function of gas total
    column density (from Crutcher et al. 2010). The solid line shows
    the critical mass-to-flux ratio.  }
  \label{polar}
\end{figure}

Dust polarization measurements in molecular clouds are making
impressive progress in the visible and near-IR ranges (polarization of
starlight) and in the submillimeter domain (polarization of dust
thermal emission).  In the former case, large sensitive cameras allow
the detection of thousands polarization vectors simultaneously.  An
example is given in Fig \ref{polar} (left) that displays the
measurements of Heiles (2000) throughout the field of the Taurus
molecular complex and recent measurements by Chapman et al. (2011)
that reveal the small scale coherence of the polarization vectors.  In
the latter case, sensitive bolometer arrays provide maps of polarized
emission of dust in dense cores and filaments. The {\it Planck}
all-sky survey will offer a global view of the dust polarized emission
in the Galaxy.

These measurements do not provide the magnetic field intensity, but
probe a clear decrease of the polarization degree in the interior of
dense cores and in the inner parts of the dense filaments (Matthews et
al. 2009).  There are several possible interpretations of the
depolarization, among which: {\it (i)} the decrease of the alignment
capability of dust grains if spinning sustained by radiative torques
drops due to extinction (Lazarian \& Hoang 2007), {\it (ii)} a helical
magnetic field threading cores and filaments (Fiege \& Pudritz 2000b).

The well-known method described by Chandrasekhar \& Fermi (1953) to
infer the magnetic field intensity from the fluctuations of the
polarization angle turns out to be less straightforward than foreseen,
although it has been tested in numerical simulations of MHD turbulence
(Ostriker et al. 2001).  The polarization due to the fluctuations of
the larger scale fields in the galactic ISM is contributing to that
observed and the separation between the two contributions relies on
modelization (see Houde et al. 2009, Hildebrandt et al. 2009). Li et
al. (2006) find a correlation of the magnetic field direction in
molecular clouds with that in the Galactic plane.

The only direct measurements of the magnetic field intensity are based
on the Zeeman effect.  Unfortunately, the line frequency shift between
left and right circularly polarized emissions is sensitive only to the
projection of the magnetic field along the line-of-sight, an effect
which is plagued by the signal integration along the line-of-sight
itself. Furthermore, given the low magnetic field intensities, there
are only a handful of species with large enough sensitivity to the
magnetic field to attempt its detection in molecular clouds. These are
species with non-zero magnetic moment due to unpaired electrons, for
instance (the elusive O$_2$, 10$^3$ less abundant than model
predictions, being one of them). So far, the best diagnostics have
been OH in diffuse molecular clouds and CN in star-forming
regions. All the available results are gathered in Crutcher et
al. (2010) and displayed in Fig. \ref{polar} (right). They include
Zeeman measurements that sample $B_{\parallel}$ in diffuse (mostly
atomic) clouds in HI from Heiles \& Troland (2005), in low-mass dense
cores and molecular clouds in OH and in star-forming regions from CN
(Crutcher 1999, Falgarone et al. 2008).  The results of the Bayesian
analysis performed on the ensemble of Zeeman measurements are the following,  \\ 
1 - There is no increase of the magnetic field intensity up to gas
densities $n_{\rm H,0} \sim 300$ \cc, and above this threshold the
observed values are consistent with an upper value increasing as
$B_{tot} \propto n_{\rm H}^{\kappa}$ with $\kappa \approx 0.65$. This
change of behavior occurs at $N_{\rm H} \sim 3 \times 10^{21}$ \cq,
where clouds start to be self-gravitating.  \\ 2 - The sharp increase
of the {\it upper limit of the magnetic field} with density above
$n_{\rm H,0} \sim 300$ \cc\ and the large scatter of the data have no
straightforward interpretation, given the small number of data points.
These results do not seem to support the model of star--formation
controlled by gravity-driven ambipolar diffusion. \\ 3 - The line
drawn on Fig. \ref{polar} (right) corresponds to the critical
mass-to-flux ratio $(M/ \Phi )_{crit} = 3.5 \times 10^{-21} N_{\rm
  H}/B_{tot}$. Therefore all the Zeeman measurements in molecular
cloud cores, whether they are low-mass dense cores (OH) or high-mass
dense cores and star-forming regions (CN), show that the mean
mass-to-flux ratio $M / \Phi \sim 2$ to $ 3 \times ( M /
\Phi)_{crit}$.  Thus, none of the molecular cloud cores observed so
far have subcritical $M/ \Phi $. \\ 4 - The turbulent and magnetic
energies in all these clouds, including star-forming regions, are in
approximate equipartition.

Sensitive OH Zeeman measurements performed in the core and envelope of
four dark clouds by Crutcher et al. (2009) provide a mass-to-flux
ratio smaller in the cores than in the envelopes, $(M/ \Phi
)_{core}<(M/ \Phi )_{env}$.  It is a provocative result, although
close to the detection limit, because it is the opposite of the
prediction of the ambipolar diffusion theory.  Lazarian et al. (2012)
propose that a process they term ``reconnexion diffusion'' efficiently
removes magnetic flux from the turbulent envelopes of molecular
clouds.  However, the uncertainties and difficulties at measuring the
magnetic field intensity make any definite conclusion likely
premature.

\section{Openings}

All along this review, emphasis has been put on the importance of
turbulence and magnetic fields on the physics of molecular clouds.
Here, we would like to address, as openings, two related issues for
which the dynamics of molecular cloud is crucial.  First, we address
the inter-relation between turbulence and chemistry. Indeed, while
turbulence can drive chemistry by inducing high temperatures in
dissipation bursts, chemistry helps to study the velocity field.
Second, we present some theories that have been proposed to explain
the star--formation rate in molecular clouds.

\subsection{Link between chemistry and turbulence}

\subsubsection{Warm \HH\ glitters of the cold ISM}
We have seen that the turbulent energy density of molecular clouds is
of the same order as the gravitational energy at large scale. One
question has not been addressed yet: how is this huge energy
dissipated, and where? Do we have any signpost of dissipation?  The
answer is probably positive: we may be witnessing turbulent
dissipation in the recently discovered \HH\ pure rotational emission
of diffuse molecular clouds and in their remarkable molecular
richness.

Cold \HH\ cannot be detected in emission due to the symmetry and small
moment of inertia of the molecule.  Its lowest lying rotational levels
have so large energies\footnote{The first excited levels of ortho- and
  para-\HH\ lie at 510K ($J=2$) and 1020K ($J=3$), respectively, above
  the ground states $J=0$ and $J=1$} that it precludes the detection
of even the first pure rotational transitions of \HH\ in environments
devoid of UV radiation or shock excitation.  By detecting the lowest
rotational lines of \HH\ at an unexpected level in cold diffuse
environments, {\it ISO-SWS} has challenged the traditional view of the ISM
and disclosed the transient existence of tiny fractions of warm gas,
disseminated within the cold ISM, but warm enough to excite these
lines (Falgarone et al. 2005).  The \HH\ line emission has since then
been observed by {\it Spitzer}/IRS in one edge of the Taurus molecular
complex (Goldsmith et al. 2010) and in high-latitude clouds (Ingalls
et al. 2011).  The detected amounts of warm \HH\ (\ie\ in rotational
levels $J\geq 2$ of the ground vibrational level) cannot be explained
by the known sources of excitation in these environments.  The
existence of a few percent of warm molecular gas mixed with the cold
diffuse ISM is not only inferred from these observations.
\HH\ absorption lines observed by FUSE, in the direction of late B
stars probe a similar fraction of warm \HH\ embedded in the cold
diffuse ISM (Gry et al. 2002, Lacour et al. 2005).  Similarly, the
excess of \HH\ rotational line emission in photo-dissociation regions
(PDRs) with low UV-irradiation (Habart et al. 2011) calls for an
additional heating mechanism.

A possible source of heating comes from the bursts of dissipation of
supersonic turbulence which pervades the whole ISM.  It is because the
dissipation of turbulence is intermittent that the {\it local} gas
heating is large enough to excite the \HH\ pure rotational lines by
collisions.  These space-time bursts are modeled as low-velocity MHD
shocks (Flower \& Pineau des For\^ets 1998, Lesaffre et al. 2012)
and/or thin coherent vortices temporarily heating a small fraction of
the gas ($\approx$ 1\%) to temperatures up to 10$^3$ K (Joulain et
al. 1998, Godard et al. 2009).  The heated gas eventually cools down once
the dissipation burst is over.

\subsubsection{Need for alternative routes for molecule formation 
in diffuse unshielded environments}

Molecular lines, and CO lines in particular, are major coolants of
molecular clouds at high densities (Goldsmith \& Langer 1978) while
the fine--structure line of \Cp\ (and to a lesser extent OI) dominates
the cooling of diffuse molecular gas. On average, in the galactic
plane, \Cp\ cooling exceeds by far all the other coolants, as first
shown by the {\it COBE} submillimeter all-sky survey (Bennett et
al. 1994). At high latitude there is an excellent correlation of the
\Cp\ cooling rate with neutral atomic hydrogen, $\Lambda(\Cp)=2.6
\times 10^{-26}$ erg s$^{-1}$ H$^{-1}$ .  Chemistry is therefore a key
issue in the understanding of the evolution of molecular clouds.  The
many unsolved issues, \ie\ the unexpected molecular richness of the
diffuse gas, its large (and highly fluctuating) CO abundances, to name
a few, are indicators that some aspects of the cloud chemistry have
been overlooked. \\
 
State-of-the-art models of photo-dominated regions (PDR) (Le Petit et
al. 2006) have met with success even with uniform density
distributions. Some species, like CH, have observed behaviours that
are fully consistent with models predictions. But noteworthy
discrepancies exist.  Recently,  outputs of numerical simulations of
MHD turbulence (Hennebelle et al. 2008) have been used to compute the
propagation of the UV radiation, the gas temperature and the chemistry
in a self-consistent manner (Levrier et al. 2012).  The average
molecular abundances of \HH, CO, CN, CS are increased by a factor of a
few, in better agreement with observations.  But CO abundances remain
below the observed values by about a factor of 10 for \HH\ column
densities in the range from a few $\times 10^{19}$ to 10$^{21}$ \cq.

In standard chemistry, CO forms rapidly by reaction of \Cp\ with OH
and \wat\ and is photo--dissociated by UV photons. As \HH, CO
self--shields because its photo--dissociation proceeds via absorption in
its own electronic transitions (van Dishoeck \& Black 1988).  OH and
\wat\ form via reactions of \Hthp\ with oxygen. \Hthp\ results from
the ionization of hydrogen by cosmic rays followed by reaction with
\HH. Quoting Glassgold \& Langer (1975), ``When \Cp\ recombines
in a gas rich in \HH, it does not produce neutral carbon but directly
CO''.  The computation of the \Cp/C/CO transition, as well as the
H/\HH\ transition, is still a challenge to theorists.  In 
steady-state chemistry, \Hthp\ plays a pivotal role.

There is another route to CO that proceeds via highly endoenergetic
reactions. This is why it has been overlooked in the chemistry of cold
molecular clouds.  It is initiated with the formation of \CHp\ via
\Cp\ + \HH\ ($\Delta E/k=4940$K), by rapid hydrogenations of
\CHp\ that produces \CHtp\ and \CHthp.  This  chain stops at
 \CHthp\ which becomes the pivotal cation of the ``warm''
chemistry by reacting with O to form \HCOp\, the  main
source of CO through dissociative recombination. \\

The TDR model (for Turbulent Dissipation Regions, Godard et al., 2009)
is based on the fact that turbulent dissipation that involves $ \nabla
\cdot {\bf v}$ and $\nabla \times {\bf v}$ is an intermittent
quantity.  The free parameters of the model are constrained by the
known large-scale properties of turbulence (Joulain et al. 1998).  In
this model, the dissipation is due to both viscous dissipation and
ion-neutral friction induced by the decoupling of the ionized and
neutral flows in the central regions.  Since the diffuse medium has a
low density, its chemical and thermal inertia are large. A random line
of sight through the medium therefore samples actively dissipating
regions, relaxation phases, and the ambient medium.  The chemical
relaxation times of molecular species cover a broad range, from 200 yr
for CH$^+$ up to 5$\times 10^4$ yr for CO.  This introduces a
potential complexity in the comparison of the observed abundances of
different species.

The main successes of the TDR model are: \\ 
- the agreement of \CHp\ and \SHp\ observations with model predictions. \SHp\ 
has even a larger formation
endothermicity than \CHp, via S$^+$ + \HH, ($\Delta E/k=9860$K)
(Menten et al. 2011, Godard et al. 2012);\\
- the scaling of \CHp\ abundances with the turbulent dissipation rate; \\
- the \HH\ pure rotational line emission of diffuse molecular gas 
(see previous section); \\
- the CO abundance of diffuse molecular gas.

An interesting properties of these models is that a fraction as small as a
few per cent of warm gas, heated by the dissipation of turbulent
energy, is sufficient to reproduce the observed \HH\ line intensities,
and their line ratios, as well as the abundances of specific
molecules. These tend to be in better agreement with low
rates-of-strain, \ie\ models in which dissipation is dominated by
ion-neutral friction.  Finally, models of low velocity C-shocks with
shock velocities in the range 5 to 8 \kms\ have very similar chemical
outputs (Lesaffre et al. 2012).

\subsection{Role of density PDF for the  star--formation rate (SFR)}
Below we emphasize the importance of the density PDF
regarding one of the most fundamental problems related to the
formation of stars namely the star--formation rate.  Various authors
(namely Krumholz \& McKee 2005, Padoan \& Nordlund 2011, Hennebelle \&
Chabrier 2011) have been recently proposing analytical theories that
predict the star--formation rate within molecular clouds.
The dimensionless {\it star--formation rate per free-fall time},
$SFR$, which has been introduced by Krumholz \& McKee (2005) is
the fraction of cloud mass converted into stars per
cloud {\it mean} free-fall time, $\tau_{ff}^0$, \ie\: $SFR =
{\dot{M}_*\over M_c} \, \tau_{ff}^0$. 
In this expression $\dot{M}_*$ denotes the
{\it total star--formation rate} arising from a cloud of mass $M_c$,
size $L_c$ and mean density $\rho_0$.

As the physical ideas behind these theories are similar and relatively
straightforward, though the results may differ significantly, we start
with a general description. Considering a cloud with a density
distribution, ${\cal P}$, close to the expression stated by
Eq.~(\ref{Pr0}): the densest parts will collapse in a few free-fall
times if they are dense enough.  This suggests that the star--formation
rate could be simply estimated by summing over the density PDF up to
some threshold, $\rho_{crit}$, and dividing by a few free-fall times,
$\phi_t \tau_{ff}$, and multiplied by some efficiency $\epsilon \simeq
1/3-1.2$ that takes into account the fact that only a fraction of the
mass initially within cores eventually ends up into stars (Matzner \&
McKee 2000, Ciardi \& Hennebelle 2010).  Thus, $SFR = \epsilon
\int ^\infty _{\delta_{crit}} (\tau_{ff}^0 / (\tau_{ff}(\rho) \phi_t))
(\rho/\rho_0) {\cal P}(\delta) d{\delta}$, where
$\delta_{crit}=\ln(\rho_{crit}/\rho_0)$. The differences between the
various estimates available in the literature reside mainly in the
choice of $\rho_{crit}$ and of the free-fall time.

The first theory along this line has been proposed by Krumholz \&
McKee (2005).  In their approach, $\rho_{crit}$ is determined from the
condition that the corresponding Jeans length must be equal to the
sonic length, $\lambda_s$, \ie\ the length at which the velocity
dispersion is equal to the sound speed.  The underlying assumption is
that turbulent support will be too efficient to enable star--formation
at scales larger than the sonic length. This yields ${\rho}_{crit, KM}
= \rho_0 (\phi_x \lambda_{J0}/\lambda_s)^2$, where $\phi_x$ is a
coefficient of order unity, $\lambda_{J0}$ is the Jeans length at the
mean cloud density. Furthermore, they assume that the relevant
free-fall time, $\tau_{ff}(\rho)$, is equal to the cloud mean
free-fall time $\tau_{ff}^0$. Thus, they obtain a star--formation rate
of the order of few percents.

Padoan \& Nordlund (2011) make different choices regarding the
critical density and the free-fall time. For the former they assume
that it is obtained by requiring that the corresponding Jeans length
is equal to the typical thickness of the shocked layer, inferred by
combining isothermal shock jump conditions and the turbulent velocity
scaling, $v \propto l^{0.5}$. This yields $\rho_{crit, PN} \simeq
0.067\, \theta^{-2} \alpha_{vir} {\cal M}^2 \rho_0$, where
$\theta\approx 0.35$ and $\alpha_{vir}$ is the virial parameter,
$\alpha_{vir}=2E_{\rm kin}/E_{\rm grav}=5 V_0^2 / (\pi G { \rho_0}
L_c^2)$, where $V_0$ is the rms velocity within the cloud,
representative of the level of turbulent vs. gravitational energy in
the cloud.  To estimate the free-fall time, they compute its value at
the critical density. This leads to a factor $\tau_{ff}^0 /
\tau_{ff}(\rho_{crit}) = (\rho_{crit}/\rho_0)^{1/2}$ which is {\it
  independent} of the density. The SFR obtained in this way is about
3-10 times higher than what has been inferred by Krumholz \&
McKee. Moreover the dependence of the SFR on the Mach number at a
given value of the virial parameter $\alpha$ is different. The SFR
increases with the Mach number in the Padoan \& Nordlund model, while
it decreases according to Krumholz \& McKee.

Hennebelle \& Chabrier (2011) proceed differently.  First, they infer
a star--formation rate by a direct integration over the core mass
spectrum derived by Hennebelle \& Chabrier weighted by $\phi_t
\tau_{ff}$, where $\tau_{ff}$ is the free-fall time associated to the
core. It can be shown that the typical crossing time, at the scale of
the core progenitor under the assumption that it is virialised and
irrespectively of the scale, is about 3 times the free-fall
times. Therefore $SFR = \rho_0^{-1} \int _0 ^{M_{cut}} \tau_{ff}^0 /
\tau_{ff} M dN/dM (M) dM $ where $M_{cut}$ is the core mass until
which the integration must be performed.  In the general case, this
integral cannot be converted into a density integral of the kind
obtained by Krumholz \& McKee (2005) and Padoan \& Nordlund (2011)
mainly because of the scale dependence of the density
PDF \footnote{More precisely, this is expressed by the second term of
  the right hand side of Eq.~(33) of Hennebelle \& Chabrier (2008)
  that we do not write here for conciseness}. The physical reason is
that an integral over the density only, does not take into account the
spatial distribution. Indeed, one could construct a medium with a
density PDF identical to the one observed in molecular clouds but with
a very clumpy gas spatial distribution such that all clumps would have
a mass smaller than the Jeans mass.  In this case, no star could form
by gravitational collapse.  However, if the gas distribution is
sufficiently regular, as it is the case for density structures
generated by turbulence (appendix B of Hennebelle \& Chabrier 2008),
then the SFR can be written as an integral of the density field and it
is not necessary to take the detail of the spatial distribution into
account.

As it is clear from the last SFR expression, the density PDF should be
weighted by the local free-fall time when performing the integration
over the density to get the SFR. Estimating the free-fall time at the
mean cloud density or at the critical density, leads to overestimate
the free-fall time in the densest regions and therefore underestimate
the SFR.  When the free-fall time is kept inside the integral, as it
should, the following expression for the SFR is obtained:
\begin{eqnarray}
\nonumber
SFR
 &=& \epsilon
\int ^\infty _{\delta_{crit}} {\tau_{ff}^0 \over \tau_{ff}(\rho) \phi_t} 
 {\rho \over \rho_0} {\cal P}(\delta) d{\delta} =
{\epsilon  \over  \phi_t}
\int ^\infty _{\delta_{crit}}
\left({\rho \over \rho_0} \right)^{3/2} {\cal P}(\delta) d\delta  \\
&=& {\epsilon  \over  2 \phi_t} \exp( 3 \sigma_0^2 / 8 ) 
\left[ 1 + {\rm erf} \left( { \sigma_0^2 \ln( \rho_{crit}/\rho_0 ) \over  
2^{1/2} \sigma_0 }  \right) \right],
\label{sfr_corrected}
\end{eqnarray}
where $\rho_{crit}$ is the critical density. 

Comparing the various models, Hennebelle \& Chabrier (2011)
 conclude (their figure 1) that 
the choice of Krumholz \& McKee (2005) to estimate 
the free-fall time as the mean cloud free-fall time, leads to underestimate
the SFR by almost one order of magnitude. On the other hand, the choice 
of Padoan \& Nordlund (2011) to estimate the free-fall time at the critical
density leads to underestimate the SFR by a factor 2-3. 
The choice of the critical density has also some impact
and leads to SFR that vary by a factor of the order of 3.
Direct comparisons with SFR inferred from 
observations (e.g. Heiderman et al. 2010, Kennicutt and Evans, 2012) have been performed but
given the relatively large scatter of the data, it seems 
hard to draw firm conclusions. Once the dependence 
on the free-fall time is properly taken into account,
all model predictions are in comparable 
agreement with the molecular cloud data. Moreover,
the theoretical predictions are all hampered by large 
uncertainties  due to the parameters $\epsilon$, $\phi_t$
but also the value of the critical density up to which the integration
is performed.

\section{The road ahead}

This review has stressed the role played by turbulence in the physics
of molecular clouds. As the whole field of turbulence, including
incompressible turbulence and ideal MHD turbulence, remains a
challenge to physicists, unraveling interstellar turbulence is an even
greater task because it is compressible and magnetized, and it
entails complicated non-ideal MHD processes, radiative cooling, and
interplay with self-gravity.  Moreover, the lesson drawn from
observations is that the range of scales coupled by interstellar
turbulence likely extends from AU-scales, close to the mean free path
of atoms and molecules, to galactic scales.  Two facets are therefore
particularly challenging for the future, on observational and
theoretical grounds: (1) a better knowledge of interstellar turbulence
and the nature of its dissipation processes; (2) the respective impact
of the large scale environment and the star--formation feedback upon
the evolution of star-forming clouds.

Several aspects of interstellar turbulence are particularly relevant
to our understanding of molecular clouds. First, the density PDF is
largely determining the SFR and the CMF. A better theoretical
knowledge of its dependence on the Mach number and on the adiabatic
index, would lead to more quantitative predictions of the SFR and CMF.
Second, the nature of the dissipation structures and the local heating
induced by turbulent dissipation, may be a clue to explain the
abundance of some chemical species.  In turn, the observations of
these species will provide reliable diagnostics to help constraining
the dynamics of the flows.  Third, the origin of the filaments
observed in molecular clouds and the role they play in their evolution
as well as in the star--formation process remain speculative. Finally,
the coupling between the magnetic field and the gas in presence of
turbulence is only partially understood (e.g. Santos-Lima et
al. 2012).

As amply emphasized, molecular clouds continue to accrete during a
long period of their lifetime.  At some stage, when the gas has formed
enough stars, stellar feedback becomes important and even dominant,
possibly disrupting the cloud and setting the star--formation
efficiency at the observed low level.  The different phases of this
scenario need to be assessed and quantified, in particular the
duration as well as the amount of energy which is released inside the
cloud from the various sources.  While it seems unavoidable that dense
cloud formation involves converging/colliding flows, the
nature of these flows remains largely open and important questions are
still unsolved such as: What drives them ?  What is the amount of mass
they deliver ?  What is their geometry ?  Are stars forming while the
cloud is still growing in mass?  Concerning feedback, it is not clear
which process is dominant and which scales are affected.

In conclusion, in spite of significant progress, the old controversy
of what is regulating star--formation within molecular clouds, magnetic
fields, turbulence or feedback, is still largely present and
unsolved. Likely the three factors play a significant role and
contribute to determine the outcome of star--formation.

\section{Acknowledgement}
This review is dedicated
  to the masters that opened the field in the early 1970s, Pat Thaddeus,
  Phil Solomon and Tom Phillips on the observations side and George Field
  on the theory side. 
We thank Alexei Kritsuk, Fred Lo, Fr\'ed\'eric Bournaud, Maryvonne Gerin,
Pierre Lesaffre and Leo Blitz for 
a critical reading of the manuscript and for 
suggestions which have improved it.

\end{document}